\documentclass[11pt,leqno]{article}
\usepackage{amssymb,amsmath,amsthm,thmtools}
\usepackage[left=1in,right=1in,top=1in,bottom=1in]{geometry}
\usepackage{amsfonts,bbm}
\usepackage{blindtext}
\usepackage{booktabs}
\usepackage[utf8]{inputenc}
\usepackage{bm}
\usepackage{caption}
\usepackage{comment} 
\usepackage[shortlabels]{enumitem}
\usepackage{float} 
\usepackage{graphicx}
\usepackage{bbm}
\graphicspath{ {figures/} }
\usepackage{indentfirst}
\usepackage{mathtools}
\usepackage{mathrsfs}
\usepackage{multirow}
\usepackage{upgreek}
\usepackage{xcolor}
\usepackage[bibstyle=numeric, citestyle=authoryear, natbib=true,doi=false, url=true, backend=biber, maxbibnames=6, maxcitenames=3, uniquelist=false, uniquename=false, sorting=nyt]{biblatex}
\renewbibmacro{in:}{}
\newcommand*{\MNWyearlabels}{%
  \iffieldequalstr{entrykey}{MNW-2023}{\csdef{abx@field@extradate}{1}}{}%
  \iffieldequalstr{entrykey}{mackinnon-nielsen-webb-2023fast}{\csdef{abx@field@extradate}{2}}{}%
}
\AtEveryCitekey{\MNWyearlabels}
\AtEveryBibitem{\MNWyearlabels}
\renewbibmacro*{date}{%
  \ifboolexpr{test {\iffieldequalstr{entrykey}{MNW-2023}} or
              test {\iffieldequalstr{entrykey}{mackinnon-nielsen-webb-2023fast}}}
    {\printdateextra}
    {\printdate}%
}

\usepackage{array}
\usepackage{rotating}
\usepackage{setspace}
\usepackage{subcaption}
\usepackage{tikz}
\usepackage[plainpages=false]{hyperref}
\usepackage{cleveref}
\crefname{enumi}{}{} 
\usepackage{dcolumn}
\usepackage{float}
\usepackage{longtable}
\usepackage{threeparttable}
\renewcommand{\arraystretch}{1}


\newtheorem{theorem}{Theorem}

\newtheorem{corollary}{Corollary}
\newtheorem{lemma}{Lemma}

\newtheorem*{runningexample}{Running Example}

\newtheorem{assumption}{Assumption}
\newtheorem{remark}{Remark}
\newtheorem{proposition}{Proposition}

\newtheorem*{expectilemapcondition}{Condition (Expectile-map regularity)}

\newtheorem{inneruassumption}{Assumption}
\newenvironment{namedassumption}[1]
{\renewcommand\theinneruassumption{#1}\inneruassumption}
{\endinneruassumption}
\newenvironment{modelassumption}[3]
{\renewcommand\theinneruassumption{#1 #2}\inneruassumption[#3]}
{\endinneruassumption}

\crefname{figure}{Figure}{Figures}
\crefname{assumption}{assumption}{assumptions}
\crefname{inneruassumption}{assumption}{assumptions}
\Crefname{inneruassumption}{Assumption}{Assumptions}
\crefname{lemma}{Lemma}{Lemmata}
\crefname{DGP}{DGP}{DGPs}
\crefname{table}{Table}{Tables}
\crefname{condition}{Condition}{Conditions}

\numberwithin{theorem}{section}
\numberwithin{corollary}{section}
\numberwithin{stheorem}{section}
\numberwithin{lemma}{section}
\numberwithin{result}{section}
\numberwithin{definition}{section}
\numberwithin{remark}{section}
\numberwithin{proposition}{section}
\numberwithin{condition}{section}
\numberwithin{equation}{section}

\newcommand{\E}{\mathbb{E}}

\newcommand{\G}{\mathcal{G}}

\newcommand{\Prob}{\mathbb{P}}
\newcommand{\Probbr}[1]{\Prob\!\left[#1\right]}

\newcommand{\Probstarbr}[1]{\Prob^*\!\left[#1\right]}

\newcommand{\R}{\mathbb{R}}
\newcommand{\T}{\mathcal{T}}
\newcommand{\V}{\mathbb{V}}
\newcommand{\Op}{\mathcal{O}_p}
\newcommand{\Opstar}{\mathcal{O}_{\Prob^*}}
\newcommand{\proofpart}[1]{%
  \par\medskip\noindent{\bfseries\itshape #1.}\enspace\ignorespaces
}
\newcommand{\proofstep}[2]{%
  \par\medskip\noindent{\bfseries\itshape Step #1: #2.}\enspace\ignorespaces
}

\newcommand{\indicator}[1]{\mathbbm{1}\{#1\}}

\newfloat{sfigure}{tbp}{sfig}
\DeclareNameAlias{default}{family-given/given-family}

\begin{document}
\hypersetup{pageanchor=false}
\begin{refsection}
\title{$Z$-Estimation under Unignorably Large Clusters}
	\date{\today }
	\author{Agossou Michel Yomakou\footnote{\href{mailto:michel.yomakou@um6p.ma}{michel.yomakou@um6p.ma}, AIRESS \& FGSES, Universit\'e Mohammed VI Polytechnique, Morocco.} \and Abderrahim Taamouti\footnote{\href{mailto:abderrahim.taamouti@liverpool.ac.uk}{abderrahim.taamouti@liverpool.ac.uk}, Department of Economics, University of Liverpool Management School.} \and Emmanuel S. Tsyawo\footnote{\href{mailto:estsyawo@gmail.com}{estsyawo@gmail.com}, Department of Economics, Finance and Legal Studies, Culverhouse College of Business, University of Alabama}}
	\maketitle
\thispagestyle{empty}
	\begin{abstract}\noindent
Cluster-size heterogeneity can invalidate conventional cluster-robust inference when a small number of clusters dominate observation-weighted estimating equations. This paper develops asymptotic theory for cluster-balanced \(Z\)-estimators under independent clusters, arbitrary realised cluster sizes, and unrestricted within-cluster dependence. Building on the size-adjusted reweighting of \citet{chiang-sasaki-wang-2025cluster}, the analysis distinguishes target definition, identification, and inference. A common cluster-level zero, imposed as an identifying condition or derived from individual-level exogeneity or design-based mechanisms, makes the structural target invariant to deterministic cluster reweighting; separation of the reweighted population moments then ensures identification. The pointwise results establish asymptotic normality while allowing the covariance sequence to vary along the cluster array. The results for functional parameters establish weak convergence and validity of a cluster multiplier bootstrap for uniform inference. The framework covers smooth and non-smooth scores with distribution, quantile, and expectile regressions as concrete applications. Monte Carlo experiments illustrate pointwise and uniform inference under heterogeneous cluster sizes, and a supplementary empirical application illustrates the consequences of cluster weighting in municipal-school data.
		
        \vspace{0.2cm}
		
		
        \end{abstract}
\clearpage
\setcounter{page}{1}
\hypersetup{pageanchor=true}
\section{Introduction}
Clustered sampling is ubiquitous in empirical research in economics and related fields, where observations within clusters, such as states, schools, or industries, may exhibit arbitrary within-cluster dependence. Ensuring the validity of inference under clustered sampling has hence become a \emph{sine qua non}. Inference procedures based on cluster-robust standard errors or cluster bootstrap methods, typically coupled with \(t\)- and Wald tests, are routinely employed to account for such dependence  \citep{white_1984,liang_longitudinal_1986,arellano_computing_1987,cameron_bootstrap-based_2008,cameron_practitioners_2015,MNW-2023}. Additionally, bootstrap methods are often preferred when the number of clusters is small and asymptotic approximations may be unreliable, or when the estimation of cluster-robust standard errors requires tuning parameters, as in regression quantile models \citep{hagemann-2017-cluster}. The validity of these inference procedures, however, ultimately relies on the asymptotic normality of the underlying estimator. This property can fail in the presence of extreme cluster size heterogeneity. Specifically, letting \(G\) denote the number of clusters, \(N\) the total sample size, and \(N_g\) the size of cluster \(g\in \{1,\ldots,G\}\), asymptotic normality may fail when $ \displaystyle \max_{1\leq g \leq G} N_g^2 / N \not\to 0$ as $N\rightarrow \infty $ \citep{hansen2019asymptotic,chiang-sasaki-wang-2025cluster}. Recent theoretical advances, including \citet{djogbenou_asymptotic_2019} and \citet{hansen2019asymptotic}, have relaxed restrictions on cluster growth and within-cluster dependence but continue to require the ignorability of cluster size heterogeneity.

As documented by \citet{chiang-sasaki-wang-2025cluster}, the cluster size heterogeneity condition $ \displaystyle \max_{1\leq g \leq G} N_g^2 / N \to 0$ likely fails in a wide range of empirical applications. Substantial cluster-size heterogeneity arises in settings such as schools, firms, and municipalities. The supplementary empirical illustration revisits the municipal-school design of \citet{Akhtari2022}, in which students are clustered within Brazilian municipalities of substantially unequal size. The \emph{Unignorably Large Cluster} (ULC) problem does not require a cluster to retain a positive limiting sample share: the conditions
\[
\max_{g\leq G}\frac{N_g}{N}\to0,
\qquad
\max_{g\leq G}\frac{N_g^2}{N}\not\to0
\]
can hold simultaneously. Under sufficiently strong within-cluster dependence,
cluster scores can then remain first-order influential, so consistency may
coexist with failure of conventional Gaussian approximations.
To assess the empirical relevance of this issue, \citet{chiang-sasaki-wang-2025cluster}, in a pioneering effort, examines 31 articles published in the \textit{American Economic Review} and \textit{Econometrica} between 2020 and 2021 that implement cluster-robust inference using publicly available data. Rejection of normality in 24 of the 31 studies (77\%) therein raises concerns about the validity of conventional inference in those studies.

\citet{chiang-sasaki-wang-2025cluster} highlights several theoretical challenges arising when cluster size heterogeneity is unignorable. Estimators may be inconsistent, may fail to be asymptotically normal, their conventional cluster-robust variance estimators may fail to consistently estimate asymptotic variances, standard \(t\)- and Wald tests may become invalid, and the wild bootstrap offers no general remedy. The formal asymptotic analysis in \citet{chiang-sasaki-wang-2025cluster} treats cluster sizes as random and potentially heavy-tailed, so that observation-weighted sums can have non-Gaussian stable limits. The present paper adopts a different, complementary asymptotic framework. Throughout the theoretical analysis, cluster sizes are treated as a realised triangular array, and the probabilistic arguments are finite-moment cluster-level weak law of large numbers (WLLN), central limit theorem (CLT), and empirical-process arguments in the tradition of \citet{hansen2019asymptotic}. Thus, this extension borrows and generalises the size-adjusted cluster-robust (SACR) reweighting device of \citet{chiang-sasaki-wang-2025cluster} without extending the stable-law asymptotic theory therein.\footnote{The SACR proposal discussed here is from \citet{chiang-sasaki-wang-2025cluster}, the 27 January 2025 version, arXiv:2308.10138v6. The 5 October 2025 revision (v8) focuses on a cluster-score bootstrap and omits the SACR proposal.} Its application to \(Z\)-estimating equations is termed \(Z\)-SACR.

Cluster reweighting, as proposed by \citet{chiang-sasaki-wang-2025cluster},
replaces observation-weighted averages in the estimating equations with
averages of cluster-level scores. Conventional observation weighting gives equal weight to individuals, whereas cluster balancing gives equal total representation
to each cluster. These schemes can define different estimands. Conventional observation-weighted inference can fail under ULC, whereas the \(Z\)-SACR framework supports inference under its stated conditions while accommodating arbitrary realised cluster sizes. The maintained common-zero restriction preserves the structural zero under reweighting. Indeed, in several applications, economic theory or the treatment assignment design motivates observation-level exogeneity conditions that centre each
individual score at a common structural parameter \citep[e.g.,][]{parente_quantile_2016,hagemann-2017-cluster}. Averaging within
clusters then gives the common cluster-level zero, so SACR reweighting
preserves that structural zero. Observation-level exogeneity is a
sufficient route; the generic theory permits the weaker cluster-average
restriction to be imposed directly. Separation of the reweighted
population moments establishes identification, while the moment and
regularity conditions support consistency and asymptotic normality.

Inverse-cluster-size estimating equations have antecedents in the literature on informative cluster size. \citet{hoffman-sen-weinberg-2001} motivates within-cluster resampling by distinguishing selection of an individual from the pooled population from selection of a cluster followed by selection of an individual within it. Cluster-weighted generalised estimating equations implement inverse-size weighting directly \citep{williamson-datta-satten-2003,benhin-rao-scott-2005}. Cluster-weighted marginal distributions, quantiles, and regression functionals have also been studied by \citet{nevalainen-datta-oja-2014}. The corresponding distinction between individual- and cluster-weighted treatment effects in cluster-randomised designs is discussed by \citet{su-ding-2021} and \citet{bugni_inference_2025}.

Building on these precedents, this paper develops pointwise and functional \(Z\)-estimation theory for independent, potentially non-identically distributed clusters when conventional cluster-size negligibility conditions fail. Primitive individual-level moment and score function class conditions in this paper support cluster-average laws of large numbers and empirical-process arguments, leading to consistency, asymptotic linearity, covariance-normalised pointwise normality, weak convergence, and functional multiplier-bootstrap inference. The common-zero restriction distinguishes preservation of a structural parameter from a change in population representation. The framework accommodates smooth and non-smooth estimating equations. Distribution Regression (DR), Quantile Regression (QR), and Expectile Regression (ER) serve as concrete examples.

The remainder of the paper is organised as follows. \Cref{sect:ULC_problem} formalises the ULC problem within the \(Z\)-estimation framework. \Cref{sect:theory_generic} develops the general asymptotic theory, and \Cref{sect:theory_concrete} verifies the theory for distribution, quantile, and expectile regression as concrete examples. \Cref{sect:MC} presents pointwise and uniform Monte Carlo evidence, and \Cref{sect:conclusion} concludes. The Supplementary Appendix contains proofs, simulation implementation details and additional experiments, and the empirical illustration.

\paragraph{Notation:} There are \(N\) observations in \(G\) clusters. \(\G:=\{1,\ldots,G\}\) denotes the set of cluster indices, \(\G_g\) is the set of observations in cluster \(g\), \(N_g:=|\G_g|\), and \(\bm W_g:=\{W_i:i\in\G_g\}\), where \(W_i\) is a vector of unit-\(i\)-specific observed data, with \(N=\sum_{g=1}^G N_g\). \(g(i)\) denotes the unique cluster containing observation \(i\), and \(p_\theta:=\dim(\theta)\) for a vector \(\theta\). As in empirical-process notation, \(\Prob f:=\int f\,d\Prob\) denotes integration under the marginal target population, while \(\Prob[A]\) denotes event probability. The operator \(\E_G\) denotes expectation under the stage-\(G\) triangular-array law; conditioning is displayed explicitly when needed. For individual-level functions \(f\) and cluster-level functions \(h\), define
\( \displaystyle 
\widehat{\Prob}_N f:=\frac1N\sum_{i=1}^N f(W_i),
\quad
\widehat{\Prob}_G h:=\frac1G\sum_{g=1}^G h(\bm W_g),
\quad \text{and} \quad
\Prob_Gh:=\E_G[\widehat{\Prob}_Gh].
\)
Thus \(\Prob_G\) is the cluster-averaged stage-\(G\) population operator. For functional parameters, \(\T\) is the index set, and \(\ell^\infty(\T)\) denotes the space of bounded functions on \(\T\) equipped with the uniform norm; for vector-valued paths, \( \displaystyle \|h\|_{\T}:=\sup_{\tau\in\T}\|h(\tau)\|\). \(\rightsquigarrow\) denotes weak convergence, and \(\rightsquigarrow_{\Prob}\) denotes conditional weak convergence in probability.

\section{Unignorably Large Clusters and Size-Adjusted \(Z\)-Estimation}\label{sect:ULC_problem}

This section formalises the Unignorably Large Cluster (ULC) problem within the \(Z\)-estimation framework and introduces the \(Z\)-SACR estimator as a remedy.

 \subsection{The ULC problem in \(Z\)-estimation}
Consider a population moment restriction of the generic form
\begin{equation}\label{eqn:pop_moment}
    \Prob \psi(W;\theta_o) = 0,
\end{equation}
where \(W\) is the vector of observable data, \(\theta_o \in \Theta \subset \R^{p_\theta}\) \emph{uniquely} solves the moment restriction, and \(\Theta\) is a compact parameter space. Given the moment restriction above and a sample of \(N\) observations, a natural \(Z\)-estimator of \(\theta_o\) is obtained as a solution to 
\[ 
\widehat{\Prob}_N \psi(W;\widehat \theta) = o_p(1). 
\] In addition to identification, a key ingredient for establishing consistency of \(\widehat \theta\) is a suitable uniform weak law of large numbers (UWLLN), 
\[ 
\sup_{\theta \in \Theta} \big\|(\widehat{\Prob}_N-\Prob)\psi(W;\theta)\big\| = o_p(1), 
\] for which pointwise convergence in probability of the estimating equations is necessary, namely
\[
(\widehat{\Prob}_N-\Prob)\psi(W;\theta) = o_p(1)
\]
for each \(\theta\in\Theta\). The sampling scheme becomes crucial at this point; it determines whether consistency, asymptotic normality, and valid inference can be established. Under independently sampled clusters, unknown within-cluster dependence, and unit-level marginal laws governed by \(\Prob\), suppose that
\( \displaystyle
\sup_{\theta \in \Theta} \Prob \|\psi(W;\theta)\|^2 \leq C < \infty,
\)
it follows that
\[
\Big\| \V\Big[ \widehat{\Prob}_N \psi(W;\theta_o) \Big] \Big\|
=
\Big\| \V\Big[ \frac{1}{N} \sum_{g=1}^{G} \Big( \sum_{i\in\G_g} \psi(W_i;\theta_o) \Big) \Big] \Big\|
\leq C \sum_{g=1}^{G} \Big( \frac{N_g}{N} \Big)^2.
\]
Under unrestricted within-cluster dependence, a sufficient size-negligibility condition for pointwise convergence of the observation-weighted moment is
\( \displaystyle
\sum_{g=1}^{G} \left( \frac{N_g}{N} \right)^2 \to 0,
\)
which is implied by
\( \displaystyle
\max_{1\leq g\leq G} \frac{N_g}{N} \to 0,
\)
since
\( \displaystyle
\sum_{g=1}^{G} \Big( \frac{N_g}{N} \Big)^2 \leq \max_{1\leq g\leq G} \frac{N_g}{N}.
\)
If this negligibility fails severely, the observation-weighted estimating equation may not even obey the pointwise WLLN needed for consistency.

Even when the law-of-large-numbers restriction is satisfied, Gaussian inference additionally requires control of cluster contributions relative to the covariance normalisation. A familiar sufficient size-negligibility benchmark is motivated by the bound
\( \displaystyle
\Big\| \V\Big[ \sqrt{N}\widehat{\Prob}_N\psi(W;\theta_o) \Big] \Big\|
\leq C \sum_{g=1}^{G} \frac{N_g^2}{N},
\)
which may diverge as \(N\to\infty\). The condition
\[
N^{-1} \max_{1\leq g\leq G} N_g^2 \to 0
\]
is a sufficient size-negligibility benchmark for standard CLT arguments under suitable moment and covariance conditions. When this condition fails, the estimator may still be consistent in favourable designs, but conventional Gaussian, cluster-robust variance, and bootstrap approximations can be invalid.

Large-sample justifications for clustered estimators typically rely on cluster-size restrictions of precisely this form. \citet[Assumption 2]{hansen2019asymptotic}, for example, imposes
\[
N^{-1} \Big( \sum_{g=1}^{G} N_g^r \Big)^{2/r} \leq C < \infty,
\text{ together with }
N^{-1} \max_{1\leq g\leq G} N_g^2 \to 0
\]
as \(N\to\infty\) for some \(2\leq r<\infty\). Together with suitable moment and covariance conditions, these restrictions support the CLT of \citet[Theorem~2]{hansen2019asymptotic}. Its WLLN uses the weaker maximum-share condition \( \displaystyle N^{-1} \max_{1\leq g\leq G} N_g \to 0 \). These are sufficient conditions whose violation does not itself imply failure of asymptotic normality, which also depends on within-cluster dependence and covariance normalisation.\footnote{\citet[Section~5, equations~(11)--(12)]{hansen2019asymptotic} explicitly identifies weaker covariance-adjusted sufficient conditions used in its CLT proof.} The ULC problem concerns settings where the size-negligibility benchmark fails and sufficiently large cluster scores can dominate the stochastic behaviour of the estimator. In such settings, consistency can fail in severe cases, limiting distributions may be non-normal, and conventional normal, Wald, and bootstrap approximations may be unreliable. More fundamentally, \citet{kojevnikov_impossibility_2023} shows that sufficiently large clusters can preclude consistent discrimination of the mean and consistent variance estimation under arbitrary within-cluster dependence. Under maintained heavy-tail assumptions, \citet[Theorem~6]{chiang-sasaki-wang-2025cluster} further shows that the wild cluster bootstrap is inconsistent.

\subsection{Target definition, cluster-level exogeneity, and invariance}

The primitive object remains the structural parameter \(\theta_o\) defined by the population moment condition \eqref{eqn:pop_moment}. This \(Z\)-estimation framework encompasses estimating equations, first-order conditions of loss functions, and economic moment restrictions without requiring an underlying objective function; see \citet[Section~2.2.5]{kosorok-2007} and \citet[Section E.1]{chernozhukov-fernandez-melly-2013inference}. The target population and its weighting are logically prior to identification and inference. To make this choice explicit, for nonnegative weights \(\{\pi_{g,G}\}_{g=1}^G\) satisfying \(\sum_{g=1}^G\pi_{g,G}=1\), define
\[
M_{\pi,G}(\theta)
:=
\sum_{g=1}^G\pi_{g,G}\E_G[\bar\psi(\bm W_g;\theta)]
\quad \text{and} \quad
\bar\psi(\bm W_g;\theta)
:=
\frac{1}{N_g}\sum_{i\in\G_g}\psi(W_i;\theta).
\]
The conventional observation-weighted moment uses \(\pi_{g,G}=N_g/N\),
whereas the cluster-balanced moment uses \(\pi_{g,G}=1/G\). The weighting
scheme determines whose outcomes the population moment represents.
Observation weighting corresponds to selecting an individual uniformly from
the pooled sample, so that cluster \(g\) is selected with probability
\(N_g/N\). Cluster balancing instead selects a cluster uniformly and then
a member uniformly within it, giving every cluster equal total
representation. Neither scheme intrinsically over- or under-represents clusters, as representation is assessed relative to the research question
\citep{nevalainen-datta-oja-2014,su-ding-2021,bugni_inference_2025}.
For example, student-weighted and municipality-balanced outcome
distributions describe different policy perspectives.

\citet{hoffman-sen-weinberg-2001} emphasises that informative cluster size
can make this representational choice consequential for the
estimand.\footnote{The term ``non-ignorable'' cluster size used there
refers to failure of conditional mean independence between outcomes and
cluster size given covariates. \citet[Definition~1]{nevalainen-datta-oja-2014}
extends noninformativeness to the conditional distribution. In this paper, ``unignorable'' concerns the
asymptotic influence of cluster sizes on observation-weighted estimating
equations. These conditions can occur together, but neither implies the
other.} When cluster size is related to outcome laws, changing the
relative representation of clusters can change the population moment and
its zero. Cluster balancing removes the additional representation of
larger clusters induced by pooled observation sampling; it does not
generally remove the underlying size--outcome dependence. The resulting
parameter is defined by the cluster-balanced moment, rather than generally
being an average of cluster-specific parameters.

The common-zero restriction maintained in this paper identifies a setting in which
this change in representation is innocuous; the structural zero is preserved, where every
cluster's expected score vanishes at the same \(\theta_o\). Without that
restriction, weighting-specific parameters can remain meaningful
scientific or causal targets under suitable identification conditions, but
reweighting need not preserve the parameter. ULC poses a separate
inferential question, i.e., even when observation weighting defines the desired
target, sufficiently influential clusters can invalidate conventional asymptotic approximations or even parameter estimator consistency.\footnote{\Cref{app:cluster_size_cases} develops this distinction and its implications for SACR in each of the four cases.}

The maintained economic condition in this paper is a common cluster-level zero,
\[
\E_G[\bar\psi(\bm W_g;\theta_o)]=0,
\qquad g=1,\ldots,G.
\]
This condition may be imposed directly as an economic identifying restriction or derived from the assignment design. For example, random assignment at the cluster level can imply cluster-level exogeneity for an appropriately constructed score. A separate sufficient route is the individual-level restriction \(\E_G[\psi(W_i;\theta_o)]=0\) for every \(i=1,\ldots,N\), e.g., \citet[Assumptions~2.2--2.3]{hagemann-2017-cluster}, which yields the displayed condition by averaging within clusters. Formally, \(M_{\pi,G}(\theta_o)=0\) for every deterministic weighting scheme \(\pi\); uniqueness remains a separate property of the selected population moment. The cluster-level restriction also supplies the centring used in the inference theorems below.\footnote{Kosorok's weighted bootstrap for independent non-identically distributed triangular arrays centres each summand by an estimated mean \(\widehat\mu_{ni}\); \citet[Theorem~11.18, condition~(G)]{kosorok-2007} requires \(\sup_{t\in T}\sum_i\{\widehat\mu_{ni}(t)-\mu_{ni}(t)\}^2=o_P(1)\). Without a common zero, heterogeneous cluster means generally obstruct consistent estimation of the sampling covariance. The uncentred analytical and multiplier-bootstrap score covariance estimators nevertheless coincide and can support conservative pointwise inference for a weighting-specific target; see \Cref{rem:bootstrap_centering}.}

The SACR adjustment operates by giving each cluster equal weight. Algebraically related inverse-size-weighted estimating equations and functionals arise in within-cluster resampling, cluster-weighted generalised estimating equations, and marginal distributional inference \citep{hoffman-sen-weinberg-2001,williamson-datta-satten-2003,benhin-rao-scott-2005,nevalainen-datta-oja-2014}; cluster averaging is used in this paper to obtain ULC-robust \(Z\)-inference. For a candidate value \(\theta\), define
\[
\widehat{\Prob}_G \psi^{SACR}(W;\theta)
:=
\frac{1}{G}\sum_{i=1}^N N_{g(i)}^{-1}\psi(W_i;\theta)
=
\frac{1}{G}\sum_{g=1}^G \bar\psi(\bm W_g;\theta).
\]
The corresponding \(Z\)-SACR estimator satisfies
\[
\widehat{\Prob}_G \psi^{SACR}(W;\widehat\theta^{SACR})=o_p(1),
\]
or the sharper \(o_p(G^{-1/2})\) rate when first-order asymptotic theory is invoked. The weight \(N_g^{-1}\) makes each cluster contribute one cluster-average score, so the empirical process is indexed by \(G\) rather than \(N\).

The following proposition records the resulting target-invariance implication for deterministic reweighting.

\begin{proposition}\label{prop:inv_id}
Let \(\theta_o \in \Theta\) be defined by the primitive population moment restriction \eqref{eqn:pop_moment}, and suppose that the cluster-level score satisfies \(\E_G[\bar\psi(\bm W_g;\theta_o)]=0\) for every \(g=1,\ldots,G\). Let \(\{\omega_g\}_{g=1}^G\) be any sequence of strictly positive deterministic weights.
Then, uniformly in \(G\geq 1\), the weighted population moment vanishes at \(\theta_o\), namely
\[
\E_G\!\left[\frac{1}{G}\sum_{i=1}^N\omega_{g(i)}\psi(W_i;\theta_o)\right]
=
\frac{1}{G}\sum_{g=1}^G \omega_g N_g\E_G[\bar\psi(\bm W_g;\theta_o)]
=
0.
\]
If, in addition, the weighted triangular-array population moment is separated away from zero outside every neighbourhood of \(\theta_o\), then point identification occurs at the same parameter value \(\theta_o\).
\end{proposition}

\Cref{prop:inv_id} separates the common-zero condition that gives target invariance from the separation condition that gives identification. Neither requires the variance restriction introduced next. Subject to finite second moments of individual scores, the SACR choice \(\omega_g=N_g^{-1}\) is a robust sufficient weighting rule under unknown within-cluster dependence. In the worst case, the variance of a cluster score \(\sum_{i\in\G_g}\psi(W_i;\theta)\) can scale as \(N_g^2\), so a sufficient condition for bounded variance under \(\sqrt G\)-normalisation is
\( \displaystyle
\frac{1}{G}\sum_{g=1}^{G}\omega_g^2N_g^2=\mathcal{O}(1)
\)
uniformly over the relevant parameter space. This condition is satisfied by \(\omega_g=N_g^{-1}\) without imposing restrictions on the growth rate of \(N_g\). Under stronger dependence restrictions, lighter reweighting may also be admissible, and the unweighted estimator can remain valid when cluster sizes are asymptotically negligible. Subject to the stated identification and moment conditions, the \(Z\)-SACR therefore provides a transparent cluster-balanced estimating equation that remains available under the ULC problem, where conventional unadjusted estimating equations may fail to converge.

\section{Asymptotic Theory for the \(Z\)-SACR}\label{sect:theory_generic}

Throughout this section, the asymptotic analysis uses a clustered sampling framework in which clusters are independent. The following assumption formalises this sampling structure.

\begin{assumption}[Sampling]\label{ass:sampling}
    The cluster vectors
    $\bm W_g:=\{W_i:i\in\mathcal{G}_g\}$,
    $g=1,\ldots,G$,
    are independent across clusters.
\end{assumption}

\noindent
\Cref{ass:sampling} requires only independence across clusters and permits arbitrary, potentially non-identically distributed dependence structures within clusters. Throughout, cluster sizes $\{N_g:g\leq G\}$ are treated as deterministic. Equivalently, all probability statements are interpreted conditionally on the realised cluster sizes. Under this conditional interpretation, the maintained independence, common-zero, moment and covariance conditions apply under the conditional law. As $G\to\infty$, the total size $N=N(G)$ and cluster sizes $N_g=N_g(G)$ may vary with $G$; this row dependence is suppressed in the notation.

\subsection{Pointwise consistency and asymptotic normality}

Define the cluster-average score
\[
\bar\psi(\bm W_g;\theta)
:=
\frac{1}{N_g}\sum_{i\in\G_g}\psi(W_i;\theta).
\]

The following running example and subsequent continuations thereof in this section use the just-identified linear instrumental variables (IV) model to translate the generic high-level conditions into familiar restrictions. Its ordinary least squares (OLS) special case recovers the size-adjusted least-squares estimator of \citet{chiang-sasaki-wang-2025cluster}.
\begin{runningexample}[Just-identified linear IV]
Consider the linear model \(\displaystyle Y_i=X_i'\theta_o+U_i\), where \(\displaystyle X_i,Z_i\in\mathbb R^{p_\theta}\), $X_i$ may contain endogenous regressors, and $Z_i$ is a vector of instruments. The just-identified IV score is \(\displaystyle \psi(W_i;\theta)=Z_i(Y_i-X_i'\theta)\), and the IV-SACR estimator solves \(\displaystyle G^{-1}\sum_{g=1}^G N_g^{-1}\sum_{i\in\G_g}Z_i(Y_i-X_i'\theta)=0\). Setting \(Z_i=X_i\) gives exactly the OLS-SACR normal equation of \citet{chiang-sasaki-wang-2025cluster}.
\end{runningexample}

The following assumptions, together with \Cref{ass:sampling}, are sufficient for consistency of the $Z$-SACR estimator. Define the unscaled centred cluster-average empirical process on \(\Theta\):
        \( \displaystyle
        \mathbb U_G(\theta)
        :=
        \frac{1}{G}
        \sum_{g=1}^G
        \left\{
        \bar\psi(\bm W_g;\theta)
        -
        \E_G[\bar\psi(\bm W_g;\theta)]
        \right\}.
        \)

\begin{assumption}[UWLLN]\label{ass:uwlln}\quad
    \begin{enumerate}[label=(\alph*)]
        \item\label{ass:uwlln_compact}
        \(\Theta\subset\mathbb R^{p_\theta}\) is compact.

        \item\label{ass:uwlln_bd}
        \( \displaystyle
        \lim_{M\to\infty}
        \sup_{G\geq1}
        \sup_{1\leq i\leq N}
        \sup_{\theta\in\Theta}
        \E_G\!\left[
        \|\psi(W_i;\theta)\|
        \indicator{\|\psi(W_i;\theta)\|>M}
        \right]
        =0.
        \)

        \item\label{ass:uwlln_se}
         For every \(\varepsilon>0\), 
        \( \displaystyle
        \lim_{\delta\downarrow0}
        \limsup_{G\to\infty}
        \Probbr{
        \sup_{\|\theta_1-\theta_2\|<\delta}
        \left\|
        \mathbb U_G(\theta_1)-\mathbb U_G(\theta_2)
        \right\|
        >\varepsilon
        }
        =0.
        \)
    \end{enumerate}
\end{assumption}

\noindent
\Cref{ass:uwlln} combines pointwise clustered triangular-array convergence, following the WLLN structure of \citet[Theorem~1]{hansen2019asymptotic}, with stochastic equicontinuity on a totally bounded parameter space. These conditions match the generic pointwise-to-uniform convergence result in \citet[Theorem~1(a)]{andrews-1992generic}. 

Define
\[
\widehat{\Prob}_G\bar\psi(\bm W_g;\theta)
:=
\frac{1}{G}\sum_{g=1}^{G}\bar\psi(\bm W_g;\theta),
\; \text{and} \;
\Prob_G\bar\psi(\bm W_g;\theta)
:=
\E_G\!\left[
\widehat{\Prob}_G\bar\psi(\bm W_g;\theta)
\right]
=
\frac{1}{G}\sum_{g=1}^{G}
\E_G[\bar\psi(\bm W_g;\theta)].
\]
The distinction between \(\Prob\), \(\E_G\), and \(\Prob_G\) deserves some discussion. The restriction
\(\Prob\psi(W;\theta_o)=0\)
defines the structural parameter in the target population. The triangular-array structure enters through cluster-specific expectations \(\E_G[\cdot]\) and their cluster average \(\Prob_G\). Economic theory or the assignment design supplies the identifying exogeneity condition, while the clustered sampling structure determines how the resulting scores are aggregated and how their sampling variation is analysed. The maintained array-level implication is the clusterwise zero \(\E_G[\bar\psi(\bm W_g;\theta_o)]=0\) for every \(g=1,\ldots,G\). An individual-level restriction \(\E_G[\psi(W_i;\theta_o)]=0\) for every sampled observation is sufficient, but not necessary, for this condition. Away from \(\theta_o\), the map
\(\theta\mapsto\Prob_G\bar\psi(\bm W_g;\theta)\)
need not coincide with
\(\theta\mapsto\Prob\psi(W;\theta)\),
so identification is imposed through array-level separation.

\begin{assumption}[Identification]\label{ass:pop_identification}\quad
    \begin{enumerate}[label=(\alph*)]

        \item\label{ass:pop_id_zero}
        The primitive population moment restriction
        \(
        \Prob\psi(W;\theta_o)=0
        \)
        holds, and the corresponding cluster-level score is centred along the triangular array:
        \[
        \E_G[\bar\psi(\bm W_g;\theta_o)]=0,
        \qquad g=1,\ldots,G.
        \]

        \item\label{ass:pop_id_sep}
        For every $\epsilon>0$,
        \( \displaystyle
        \liminf_{G\to\infty}
        \inf_{\theta\in\Theta:\|\theta-\theta_o\|\geq\epsilon}
        \big\|
        \Prob_G\bar\psi(\bm W_g;\theta)
        \big\|
        >0.
        \)

    \end{enumerate}
\end{assumption}

\noindent
\Cref{ass:pop_identification} ensures identification of the target parameter. Part~\ref{ass:pop_id_zero} gives the economically maintained cluster-level exogeneity condition and hence \(\Prob_G\bar\psi(\bm W_g;\theta_o)=0\). Part~\ref{ass:pop_id_sep} rules out approximate zeros of the triangular-array population moment away from \(\theta_o\).

The following theorem establishes a UWLLN and consistency of the $Z$-SACR estimator.
\begin{theorem}\label{Thm:uwlln}
Suppose that \Cref{ass:sampling,ass:uwlln} hold.
\begin{enumerate}[label=(\alph*)]
    \item\label{thm:uwlln_a}
    The cluster-average empirical process satisfies
    \begin{equation}
    \sup_{\theta \in \Theta} \big\rVert (\widehat{\Prob}_G-\Prob_G)\bar\psi(\bm{W}_g;\theta) \big\rVert \xrightarrow{p} 0.
    \label{eq:unfm_conv}
    \end{equation}
    \item\label{thm:uwlln_b}
    If, in addition, 
    \( \displaystyle 
    \widehat{\Prob}_G \bar\psi(\bm{W}_g;\widehat{\theta}^{SACR})=o_p(1)
    \)
    and \Cref{ass:pop_identification} holds,
    then \( \displaystyle \widehat{\theta}^{SACR} \xrightarrow{p} \theta_o \) as $G \rightarrow \infty$.
\end{enumerate}
\end{theorem}

The IV running example verifies the two ingredients of this result in turn.
\begin{runningexample}[continued: IV UWLLN]
For the IV score,
\(\psi(W_i;\theta)=Z_iU_i-Z_iX_i'(\theta-\theta_o)\).
Suppose that \(\|Z_iU_i\|\) and \(\|Z_iX_i'\|\) are uniformly
integrable over all observations and rows \(G\). Compactness of \(\Theta\) (\Cref{ass:uwlln}\ref{ass:uwlln_compact})
gives \(D_\Theta:=\sup_{\theta\in\Theta}\|\theta-\theta_o\|<\infty\), and
\[
\sup_{\theta\in\Theta}\|\psi(W_i;\theta)\|
\leq \|Z_iU_i\|+D_\Theta\|Z_iX_i'\|.
\]
The right-hand side is then uniformly integrable, verifying
\Cref{ass:uwlln}\ref{ass:uwlln_bd}.

Define
\(\bar Q_g:=N_g^{-1}\sum_{i\in\G_g}Z_iX_i'\). Cluster independence (\Cref{ass:sampling}) and uniform integrability (see \Cref{lem:cluster_average_limits}\ref{lem:cluster_average_wlln}), applied
to \(Z_iU_i\) and \(\operatorname{vec}(Z_iX_i')\), give
\(
\mathbb U_G(\theta_o)=o_p(1) \) and
\(
(\widehat{\Prob}_G-\Prob_G)\bar Q_g=o_p(1).
\)
The affine identity
\(\mathbb U_G(\theta)=\mathbb U_G(\theta_o)
-(\widehat{\Prob}_G-\Prob_G)\bar Q_g(\theta-\theta_o)\) yields
\[
\sup_{\substack{\theta_1,\theta_2\in\Theta\\
\|\theta_1-\theta_2\|<\delta}}
\|\mathbb U_G(\theta_1)-\mathbb U_G(\theta_2)\|
\leq\delta\|(\widehat{\Prob}_G-\Prob_G)\bar Q_g\|=o_p(1),
\]
for every fixed \(\delta>0\), which verifies \Cref{ass:uwlln}\ref{ass:uwlln_se}.
\end{runningexample}

\noindent The stated
primitive conditions above establish the IV UWLLN directly and verify the generic UWLLN assumptions.

The next example concerns primitives for IV identification.
\begin{runningexample}[continued: IV zero preservation and separation]
For the running IV model, cluster-level instrument validity (\Cref{ass:pop_identification}\ref{ass:pop_id_zero}) is \(\E_G[N_g^{-1}\sum_{i\in\G_g}Z_iU_i]=0\) for every \(g\). Individual-level restrictions \(\E_G[Z_iU_i]=0\), \(i=1,\ldots,N\), provide a familiar sufficient route. Define \(\displaystyle Q_G:=G^{-1}\sum_{g=1}^G\E_G[\bar Q_g]\), then \(\displaystyle \Prob_G\bar\psi(\bm W_g;\theta)=-Q_G(\theta-\theta_o)\). If \(\displaystyle \liminf_{G\to\infty}\lambda_{\min}(Q_G'Q_G)>0\), then
\[
\big\|\Prob_G\bar\psi(\bm W_g;\theta)\big\|^2
=(\theta-\theta_o)'Q_G'Q_G(\theta-\theta_o)
\geq
\lambda_{\min}(Q_G'Q_G)\|\theta-\theta_o\|^2,
\]
which verifies \Cref{ass:pop_identification}\ref{ass:pop_id_sep}. Thus cluster-level instrument validity preserves the structural zero, whereas uniform nonsingularity of the cluster-balanced first-stage cross-moment separates that zero from alternatives.
\end{runningexample}

The following assumption collects the additional first-order conditions used to derive the asymptotically linear representation and the normality of the Z-SACR estimator. Define
        \( \displaystyle
        H_G(\theta)
        :=
        \frac{\partial}{\partial\theta'}
        \Prob_G\bar\psi(\bm W_g;\theta)
        \) and 
        \( \displaystyle
        H_G:=H_G(\theta_o).
        \)
\begin{assumption}[Asymptotic linearisation]\label{ass:linearisation}\quad
    \begin{enumerate}[label=(\alph*)]
        \item\label{ass:lin_foc} The $Z$-SACR satisfies \(\widehat{\Prob}_G \bar\psi(\bm{W}_g;\widehat{\theta}^{SACR})=o_p(G^{-1/2})\).
        \item\label{ass:lin_int} $\theta_o$ is in the interior of $\Theta$.
        \item\label{ass:lin_cont} The array-level expected score \(\Prob_G \bar\psi(\bm W_g;\theta)\) is continuously differentiable in an open neighbourhood of \(\theta_o\), and
        \( \displaystyle
        \lim_{\delta\downarrow0}
        \limsup_{G\to\infty}
        \sup_{\|\theta-\theta_o\|\leq\delta}
        \|H_G(\theta)-H_G\|
        =0.
        \)
    \end{enumerate}
\end{assumption}

Define
\( \displaystyle
\Omega_G := \V_G\left[ \frac{1}{\sqrt{G}}\sum_{g=1}^G\bar\psi(\bm{W}_g;\theta_o) \right] =   G^{-1} \sum_{g=1}^G \E_G\big[ \bar\psi(\bm{W}_g;\theta_o) \bar\psi(\bm{W}_g;\theta_o)'\big].
\) The following supplies regularity conditions on the second moment of the score and the population Jacobian at the truth.
\begin{assumption}[Moments and nondegeneracy]\label{ass:norm_mmt_general}\quad
    \begin{enumerate}[label=(\alph*)]
        \item\label{ass:sup_finite}
        \( \displaystyle
        \lim_{M\to\infty}\sup_{G\geq1}\sup_{1\leq i\leq N}
        \E_G\!\left[
        \|\psi(W_i;\theta_o)\|^2
        \indicator{\|\psi(W_i;\theta_o)\|>M}
        \right]=0.
        \)
        \item\label{ass:nonsingular} The sequences $\Omega_G$ and $H_G$ satisfy
        \( \displaystyle
        \lambda_{\min}(\Omega_G)\geq\lambda
        \quad\text{and}\quad
        \lambda_{\min}(H_G'H_G)\geq\lambda^2
        \)
        for some $\lambda>0$.
    \end{enumerate}
\end{assumption}
\noindent A sufficient primitive condition for \Cref{ass:norm_mmt_general}\ref{ass:sup_finite} is $ \displaystyle \sup_{G\geq1}\sup_{1\leq i\leq N}\E_G[\|\psi(W_i;\theta_o)\|^{2+\delta}]<\infty$ for some $\delta>0$. \Cref{ass:norm_mmt_general}\ref{ass:sup_finite} is a primitive individual-level moment condition that yields corresponding cluster-average control by \Cref{lem:cluster_average_limits}\ref{lem:cluster_average_ui}. \Cref{ass:norm_mmt_general}\ref{ass:nonsingular} ensures $\Omega_G$ is positive definite and keeps the population Jacobian \(H_G\) uniformly invertible.

Define the first-order centred cluster-average empirical process by
\[
\mathbb G_G(\theta)
:=
\sqrt G\,\mathbb U_G(\theta)
=
\sqrt G(\widehat{\Prob}_G-\Prob_G)
\bar\psi(\bm W_g;\theta).
\]
The following assumption strengthens the preceding level-scale control to the \(\sqrt G\) scale locally around \(\theta_o\). Unlike the global condition in \Cref{ass:uwlln}\ref{ass:uwlln_se}, it is tailored to the empirical-process remainder in the asymptotic linearisation.
\begin{assumption}[Local first-order stochastic equicontinuity]\label{ass:stch_equ}
For every \(\varepsilon>0\),
\[
\lim_{\delta\downarrow0}
\limsup_{G\to\infty}\Probbr{
\sup_{\|\theta-\theta_o\|\leq\delta}
\left\|
\mathbb G_G(\theta)-\mathbb G_G(\theta_o)
\right\|>\varepsilon
}=0.
\]
\end{assumption}

\noindent
The sample score in a \(Z\)-estimation problem need not be continuously differentiable; regression quantiles are the canonical example \citep{koenker-bassett-1978}. The differentiability requirement in \Cref{ass:linearisation}\ref{ass:lin_cont} is therefore imposed only on the array-level expected score. To control the non-smooth sample component, \Cref{ass:stch_equ} imposes local stochastic equicontinuity on the first-order process \(\mathbb G_G\). This condition ensures that the empirical-process remainder is negligible at the \(G^{-1/2}\) scale in a neighbourhood of \(\theta_o\), even when the sample score is non-smooth.
 
Define \(\mathbf V_G:=H_G^{-1}\Omega_G(H_G^{-1})'\).
With consistency and the preceding regularity conditions in hand, the following establishes the asymptotic linearity and asymptotic normality of the $Z$-SACR.
\begin{theorem}\label{Thm:normality}
    Under \Cref{ass:sampling,ass:uwlln,ass:pop_identification,ass:linearisation,ass:norm_mmt_general,ass:stch_equ}, as $G \to \infty$,
    \begin{enumerate}[label=(\alph*)]
        \item\label{thm:normality_a} the $Z$-SACR admits the asymptotically linear representation
        \begin{equation}
        \sqrt{G}\big(\widehat{\theta}^{SACR}-\theta_o\big)
        = -H_G^{-1}\sqrt{G}\,\widehat{\Prob}_G \bar\psi(\bm{W}_g;\theta_o)+o_p(1);
        \end{equation}
        \item\label{thm:normality_b} and
        \begin{equation}
        \mathbf{V}_G^{-1/2}\sqrt{G} (\widehat{\theta}^{SACR} - \theta_o) \xrightarrow{d} \mathcal{N}(0, I).
        \end{equation}
    \end{enumerate}
\end{theorem}

\noindent
\Cref{Thm:normality}\ref{thm:normality_a} expresses $\sqrt{G}(\widehat{\theta}^{SACR}-\theta_o)$, to first order, as a scaled average of the mean-zero cluster-average scores $\bar\psi(\bm{W}_g;\theta_o)$. Since these summands are independent across clusters by \Cref{ass:sampling}, \Cref{Thm:normality}\ref{thm:normality_b} supplies their pointwise CLT. Importantly, $\mathbf{V}_G$ can be a drifting covariance, for which neither $\Omega_G$ nor $H_G$ need converge to a fixed matrix; the covariance sequence need only remain well behaved under \Cref{ass:norm_mmt_general}\ref{ass:nonsingular}.

For the IV running example, the remaining first-order checks take a particularly simple form.
\begin{runningexample}[continued: IV linearisation and covariance]
For the just-identified IV score, \(\displaystyle H_G=-Q_G\), \(\displaystyle \bar\psi(\bm W_g;\theta_o)=N_g^{-1}\sum_{i\in\G_g}Z_i U_i\), and \(\displaystyle \Omega_G=G^{-1}\sum_{g=1}^G\E_G[\bar\psi(\bm W_g;\theta_o)\bar\psi(\bm W_g;\theta_o)']\). Because the score is affine in \(\theta\), its population linearisation \(\displaystyle \Prob_G\bar\psi(\bm W_g;\theta)=\Prob_G\bar\psi(\bm W_g;\theta_o)-Q_G(\theta-\theta_o)\) is exact. The generic covariance matrix therefore reduces to the familiar just-identified IV sandwich \(\displaystyle \mathbf V_G=Q_G^{-1}\Omega_G(Q_G^{-1})'\). The relevance condition established in the running example's zero-preservation and separation discussion verifies the Jacobian lower bound in \Cref{ass:norm_mmt_general}\ref{ass:nonsingular} for all sufficiently large \(G\), which suffices for the asymptotic arguments. The separate lower bound on \(\lambda_{\min}(\Omega_G)\) prevents degeneracy of the cluster-average score variance.
\end{runningexample}

\begin{runningexample}[continued: IV stochastic equicontinuity]
Suppose additionally that
\[
\sup_{G\geq1}\sup_{i\leq N(G)}\E_G\|Z_iX_i'\|_F^2\leq C<\infty,
\]
where \(\|\cdot\|_F\) is the Frobenius norm.
The affine IV score satisfies \(\displaystyle \bar\psi(\bm W_g;\theta)-\bar\psi(\bm W_g;\theta_o)=-\bar Q_g(\theta-\theta_o)\). Consequently,
\begin{align*}
\sup_{\|\theta-\theta_o\|\leq\delta}
\left\|
\mathbb G_G(\theta)-\mathbb G_G(\theta_o)
\right\| &= \sup_{\|\theta-\theta_o\|\leq\delta}
\left\|
\sqrt G(\widehat{\Prob}_G-\Prob_G)\bar Q_g (\theta-\theta_o)
\right\| \\
&\leq
\delta\left\|
\sqrt G(\widehat{\Prob}_G-\Prob_G)\bar Q_g
\right\|_F=\Op(\delta),
\end{align*}
Indeed, cluster independence and Jensen's inequality give
\[
\E_G\big\|\sqrt G(\widehat{\Prob}_G-\Prob_G)\bar Q_g\big\|_F^2
\leq\frac1G\sum_{g=1}^G\frac1{N_g}\sum_{i\in\G_g}
\E_G\|Z_iX_i'\|_F^2\leq C.
\]
Markov's inequality therefore bounds the probability in
\Cref{ass:stch_equ} by \(C\delta^2/\varepsilon^2\), which vanishes
as \(\delta\downarrow0\).
\end{runningexample}

\begin{remark}[Structural role of SACR reweighting]\label{rem:sacr_structure}
The conventional observation-weighted moment can be written as
\[
\frac1N\sum_{g=1}^G\sum_{i\in\G_g}\psi(W_i;\theta)
=
\sum_{g=1}^G\frac{N_g}{N}\bar\psi(\bm W_g;\theta),
\]
so a large cluster receives weight \(N_g/N\) and may remain first-order influential. By contrast, the SACR moment \(G^{-1}\sum_{g=1}^G\bar\psi(\bm W_g;\theta)\) assigns every cluster the same deterministic weight \(1/G\). The effective triangular array therefore contains one independent cluster-average summand per cluster, with maximal deterministic weight \(1/G\). Treating each cluster average as one effective observation then makes the Hansen--Lee cluster-size restrictions \citep[Assumption 2]{hansen2019asymptotic} automatic. This is the structural reason SACR removes cluster-size dominance and yields the \(\sqrt G\)-linear representation in \Cref{Thm:normality}.
\end{remark}

The \(\sqrt G\)-normalisation does not by itself establish an efficiency cost relative to conventional estimation because under clustered dependence, the latter need not attain a \(\sqrt N\) rate. If all clusters have the same size, the SACR and conventional estimating equations differ only by a constant multiplier and have the same roots. With unequal sizes, cluster balancing can sacrifice efficiency when additional observations within a cluster provide substantial information, but can improve efficiency when strong within-cluster dependence makes that information largely redundant. There is therefore no general efficiency ranking; the comparison depends on cluster sizes, dependence, and the score and Jacobian structure.

The above representation motivates the natural plug-in estimator
\( \displaystyle
\widehat{\mathbf V}_G^{SACR}
 =
\widehat H_G^{-1}\widehat\Omega_G(\widehat H_G^{-1})',
\)
where
\( \displaystyle
\widehat H_G
:=
\frac{1}{G}\sum_{g=1}^G \widehat{\bar H}_g\) and 
\( \displaystyle
\widehat{\bar H}_g
:=
N_g^{-1}\sum_{i\in\G_g}
\frac{\partial}{\partial\theta'}
\psi(W_i;\widehat\theta^{SACR}),
\)
for smooth scores, with the usual model-specific population-Jacobian analogue used for non-smooth scores, and
\[
\widehat\Omega_G
:=
\frac{1}{G}\sum_{g=1}^G
\Big(\bar\psi(\bm W_g;\widehat{\theta}^{SACR})-\widehat\Prob_G \bar\psi(\bm W_g;\widehat\theta^{SACR})\Big)
\Big(\bar\psi(\bm W_g;\widehat{\theta}^{SACR})-\widehat\Prob_G \bar\psi(\bm W_g;\widehat\theta^{SACR})\Big)'.
\]
Under model-specific consistency conditions for the Jacobian and score-covariance estimators, this estimates the covariance of \(\sqrt G(\widehat\theta^{SACR}-\theta_o)\); the covariance of \(\widehat\theta^{SACR}\) itself is \(G^{-1}\widehat{\mathbf V}_G^{SACR}\). The \(Z\)-SACR modification enters through the cluster-size-adjusted score covariance, since \(\bar\psi(\bm W_g;\widehat\theta^{SACR})\) is already a cluster-average score.

\subsection{Functional consistency and weak convergence}
The ULC problem is increasingly important in clustered sampling settings involving economic parameters of interest such as distributional effects and tests of stochastic dominance, which are inherently functional in nature. It therefore becomes useful to extend the asymptotic theory from a finite-dimensional point parameter $\theta_o \in \mathbb{R}^{p_\theta}$ to a functional parameter $\theta_o(\tau)$, $\tau \in \mathcal{T}$.

Write the primitive estimating equation process as
\[
  \Psi(W;\theta) \;:=\; \bigl\{\psi_{\tau}(W;\theta) : \tau \in \T\bigr\},
\]
where $\mathcal{T}$ denotes a compact index set. The population moment restriction is $\Prob\Psi(W;\theta_o)=0$ in
$\ell^{\infty}(\T)$. At the cluster level, define the SACR process
\[
\bar\Psi(\bm W_g;\theta)
:=\left\{\bar\psi_\tau(\bm W_g;\theta(\tau)):\tau\in\T\right\}
\quad \text{where} \quad
\bar\psi_\tau(\bm W_g;\theta(\tau))
:=N_g^{-1}\sum_{i\in\G_g}\psi_\tau(W_i;\theta(\tau)).
\]
Thus,
$\widehat{\Prob}_G\bar\Psi(\bm W_g;\theta)$ and
$\Prob_G\bar\Psi(\bm W_g;\theta)$
are the functional analogues of the cluster-average moments used in
\Cref{Thm:normality}. \Cref{sect:theory_concrete} supplies the concrete DR, QR, and ER applications with the model-specific index notation.

The functional theory uses three related (score) function classes. The global class indexed jointly by \((\tau,t)\) delivers the functional UWLLN, the class evaluated at the true path delivers the functional CLT, and a localised increment class delivers first-order stochastic equicontinuity. Let \(\psi_{\tau,j}\) denote the \(j\)th coordinate of \(\psi_\tau\), and define
\[
\mathcal F_j:=\{\psi_{\tau,j}(\,\cdot\,;t):\tau\in\T,\ t\in\Theta\},
\qquad
\mathcal F_{o,j}:=\{\psi_{\tau,j}(\,\cdot\,;\theta_o(\tau)):\tau\in\T\},
\]
and, for \(\delta>0\),
\( \displaystyle 
\mathcal F_j^{\Delta}(\delta)
:=
\{\psi_{\tau,j}(\,\cdot\,;t)-\psi_{\tau,j}(\,\cdot\,;\theta_o(\tau)):
\tau\in\T,\ \|t-\theta_o(\tau)\|\leq\delta\}.
\)
For every coordinate \(j\), let \(F_j\) be a measurable envelope of
\(\mathcal F_j\), so that
\[
\sup_{\tau\in\T,\,t\in\Theta}|\psi_{\tau,j}(w;t)|\leq F_j(w),
\qquad
\bar F_{Gg,j}:=
\left\{N_g^{-1}\sum_{i\in\G_g}F_j(W_i)^2\right\}^{1/2},
\]
and define the normalised cluster array and its envelope by
\[
Z_{Gg,j}(\tau,t):=G^{-1/2}\bar\psi_{\tau,j}(\bm W_g;t),
\qquad F_{Gg,j}:=G^{-1/2}\bar F_{Gg,j}.
\]

\begin{namedassumption}{\ref{ass:uwlln}'}[Functional manageability and moments]
\label{ass:uwlln_F}\quad
\begin{enumerate}[label=(\alph*)]

\item\label{ass:functional_manageability}
For every coordinate \(j\), the array
\(\{Z_{Gg,j}(\tau,t):(\tau,t)\in\T\times\Theta, g\leq G\}\)
is manageable with respect to \(\{F_{Gg,j}:g\leq G\}\), uniformly in
\(G\). Each coordinate is measurable at every fixed index. The full array,
its truth-indexed restriction, and, for every \(\delta>0\), the local
difference array
\(Z_{Gg,j}(\tau,t)-Z_{Gg,j}(\tau,\theta_o(\tau))\) indexed by
\(\{(\tau,t)\in\T\times\Theta:\|t-\theta_o(\tau)\|\leq\delta\}\)
are almost measurable Suslin (AMS), with common Suslin approximating index
sets and jointly measurable parametrisations across coordinates.

\item\label{ass:functional_global_envelope}
For every coordinate \(j\),
the squared individual envelopes are uniformly integrable:
\[
\lim_{M\to\infty}
\sup_{G\geq1}\sup_{g\leq G}\sup_{i\in\G_g}
\E_G\!\left[
F_j(W_i)^2\indicator{F_j(W_i)^2>M}
\right]=0.
\]

\item\label{ass:functional_local_L2}
For every coordinate \(j\), 
\[
\lim_{\delta\downarrow0}\limsup_{G\to\infty}
\sup_{\substack{\tau\in\T,\ t\in\Theta\\
\|t-\theta_o(\tau)\|\leq\delta}}
\frac1G\sum_{g=1}^G\E_G\!\left[
\big(\bar\psi_{\tau,j}(\bm W_g;t)
-\bar\psi_{\tau,j}(\bm W_g;\theta_o(\tau))\big)^2
\right]=0.
\]

\end{enumerate}
\end{namedassumption}

\noindent
Manageability in \Cref{ass:uwlln_F}\ref{ass:functional_manageability}
is understood in the sense of \citet[Definition~7.9]{pollard-1990}.
The almost-measurable-Suslin condition is that of
\citet[Section~2]{kosorok-2003}, applied jointly to the finitely many
coordinates with the approximation error measured by the sum of squared differences over both clusters and coordinates. A sufficient verification is joint separability; see 
\citet[Lemma~11.15]{kosorok-2007}.

Let \(\theta_o(\cdot)\) denote the true functional parameter. The object of interest is the process
\[
\sqrt{G}\bigl(\widehat\theta-\theta_o\bigr)(\cdot),
\]
viewed as a random element of \(\ell^\infty(\T)\). For a parameter path \(\theta(\cdot)\), extend the first-order process notation pointwise in \(\tau\) by
\[
\mathbb G_G(\theta)(\tau)
:=
\sqrt G(\widehat{\Prob}_G-\Prob_G)
\bar\psi_\tau(\bm W_g;\theta(\tau)).
\]
Throughout this subsection, \(\Theta_{\T}\) denotes the admissible collection of parameter paths equipped with the norm
\( \displaystyle
\|\theta\|_{\T}
:=
\sup_{\tau\in\T}\|\theta(\tau)\|,
\)
with each admissible path satisfying \(\theta(\tau)\in\Theta\) uniformly over \(\tau\in\T\).

\begin{lemma}[Functional UWLLN and local first-order stochastic equicontinuity]\label{lem:functional_ep}
Under \Cref{ass:sampling,ass:uwlln_F},
\begin{enumerate}[label=(\alph*)]
\item\label{lem:functional_uwlln}
\( \displaystyle 
\sup_{\theta\in\Theta_{\T}}
\bigl\|(
\widehat{\Prob}_G-\Prob_G)
\bar\Psi(\bm W_g;\theta)\bigr\|_{\T}=o_p(1);
\) and
\item\label{lem:functional_se}
for every \(\varepsilon>0\),
\begin{equation}\label{eq:SE}
\lim_{\delta\downarrow0}\limsup_{G\to\infty}
\Probbr{
\sup_{\|\theta-\theta_o\|_{\T}<\delta}
\|\mathbb G_G(\theta)-\mathbb G_G(\theta_o)\|_{\T}
>\varepsilon}=0.
\end{equation}
\end{enumerate}
\end{lemma}

\noindent
Part~\ref{lem:functional_uwlln} is the functional counterpart of \Cref{Thm:uwlln}\ref{thm:uwlln_a}. Part~\ref{lem:functional_se} strengthens the pointwise local condition in \Cref{ass:stch_equ} by making its first-order control uniform over \(\tau\). 

The remaining functional identification requirements concern the population map itself.

\begin{namedassumption}{\ref{ass:pop_identification}'}[Functional identification and separation]
\label{ass:pop_identification_F}\quad
  \begin{enumerate}[label=(\alph*)]

    \item\label{ass:functional_zero}
    The primitive functional moment restriction is preserved through cluster-level centring along the triangular array:
    \( \displaystyle 
    \E_G[\bar\psi_\tau(\bm W_g;\theta_o(\tau))]=0,
    \qquad g=1,\ldots,G,\quad \tau\in\T.
    \)

    \item\label{ass:functional_sep}
    For every \(\epsilon>0\),
    \( \displaystyle 
    \liminf_{G\to\infty}
    \inf_{\theta\in\Theta_{\T}:
      \|\theta-\theta_o\|_{\T}\ge\epsilon}
    \bigl\|
    \Prob_G\bar\Psi(\bm W_g;\theta)
    \bigr\|_{\T}
    >
    0.
    \)

  \end{enumerate}
\end{namedassumption}

The following functional linearisation conditions enable the first-order transfer from the score process to the estimator.

\begin{namedassumption}{\ref{ass:linearisation}'}[Functional asymptotic linearisation]
\label{ass:linearisation_F}\quad
  \begin{enumerate}[label=(\alph*)]
    \item\label{ass:functional_lin_foc}
    The process estimator \(\widehat\theta(\cdot)\) satisfies
    \begin{equation}\label{eq:functional_foc_rate}
      \widehat{\Prob}_G \bar\Psi(\bm W_g;\widehat\theta)
      \;=\;
      o_p(G^{-1/2})
      \qquad\text{in } \ell^{\infty}(\T).
    \end{equation}

    \item\label{ass:functional_lin_int}
    The true path \(\theta_o(\cdot)\) is uniformly interior to \(\Theta\):
    \( \displaystyle
    \inf_{\tau\in\T}
    \operatorname{dist}\bigl(\theta_o(\tau),\partial\Theta\bigr)
    >0,
    \)
    where \(\operatorname{dist}(\cdot,\partial\Theta)\) denotes Euclidean distance to the boundary of \(\Theta\).

    \item\label{ass:functional_lin_cont}
    The map
    \(
    t\mapsto\Prob_G\bar\psi_\tau(\bm W_g;t)
    \)
    is continuously differentiable in a neighbourhood of \(\theta_o(\tau)\), uniformly over \(\tau\in\T\), with
    \( \displaystyle 
    H_G(\tau)
    :=
    \frac{\partial}{\partial t'}
    \Prob_G\bar\psi_\tau(\bm W_g;t)
    \bigg|_{t=\theta_o(\tau)},
    \)
    and
    \[
    \lim_{\delta\downarrow0}
    \limsup_{G\to\infty}
    \sup_{\tau\in\T}
    \sup_{\|t-\theta_o(\tau)\|\le\delta}
    \left\|
    \frac{\partial}{\partial t'}
    \Prob_G\bar\psi_\tau(\bm W_g;t)
    -
    H_G(\tau)
    \right\|
    =0.
    \]
    Moreover, \(H_G(\tau)\to H(\tau)\) uniformly on \(\T\), and, for some \(\lambda>0\),
    \( \displaystyle
    \inf_{\tau\in\T}\lambda_{\min}\big(H(\tau)'H(\tau)\big)\geq\lambda^2.
    \)
  \end{enumerate}
\end{namedassumption}


For the truth-indexed score process, write
\[
\Omega_G(\tau,\tau')
:=
\frac{1}{G}\sum_{g=1}^G
\E_G\!\left[
\bar\psi_\tau(\bm W_g;\theta_o(\tau))
\bar\psi_{\tau'}(\bm W_g;\theta_o(\tau'))'
\right].
\]
For \(a\in\R^{p_\theta}\), define the semimetric 
\[
\rho_{G,a}(\tau,\tau')
:=
\left[
\frac{1}{G}\sum_{g=1}^G
\E_G\!\left[
\Big(
a'\left[
\bar\psi_\tau(\bm W_g;\theta_o(\tau))
-
\bar\psi_{\tau'}(\bm W_g;\theta_o(\tau'))
\right]
\Big)^{2}
\right]
\right]^{1/2}.
\]
For any positive semidefinite matrix-valued kernel \(K\), let
\(\mathbb G_K\) denote a centred Gaussian process with covariance kernel \(K\), and define its semimetric by
\[
\rho_{K,a}(\tau,\tau')
:=
\left\|
a'\big(\mathbb G_K(\tau)-\mathbb G_K(\tau')\big)
\right\|_{L^2}
=
\left[
a'\big(
K(\tau,\tau)+K(\tau',\tau')
-K(\tau,\tau')-K(\tau',\tau)
\big)a
\right]^{1/2}.
\]

\begin{namedassumption}{\ref{ass:norm_mmt_general}'}[Functional covariance kernel]
\label{ass:norm_mmt_general_F}\quad
\begin{enumerate}[label=(\alph*)]

\item\label{cond:functional_cov_kernel} The covariance kernels
        \(\Omega_G\) converge pointwise on \(\T\times\T\) to a positive
        semidefinite kernel \(\Omega\).

        \item\label{cond:functional_cov_metric}
        For any deterministic sequences $\tau_G,\tau_G'\in\T$ and
        \(a_G\in\R^{p_\theta}\) with \(\|a_G\|\leq1\),
        \(\rho_{\Omega,a_G}(\tau_G,\tau_G')\to0\) implies
        \(\rho_{G,a_G}(\tau_G,\tau_G')\to0\).
    \end{enumerate}
\end{namedassumption}

\noindent
\Cref{ass:norm_mmt_general_F} supplies the covariance and metric conditions in \citet[Theorem~1]{kosorok-2003}, equivalently \citet[Theorem~11.16]{kosorok-2007}. It is intentionally stronger than the condition required for \Cref{Thm:normality}. The pointwise result permits a drifting covariance sequence and achieves asymptotic pivotality through studentisation. By contrast, weak convergence in $\ell^\infty(\T)$ to a fixed Gaussian process requires convergence of the covariance kernel itself.

\begin{remark}[A sufficient route for covariance-kernel convergence]\label{rem:cov_kernel_sufficient}
An illustrative sufficient case decomposes the individual score at the true path as
\(\psi_\tau(W_i;\theta_o(\tau))=A_g(\tau)+E_{gi}(\tau)\) for \(i\in\G_g\), where \(A_g\) is a cluster-wide shock and \(E_{gi}\) is an idiosyncratic component. Suppose these components have zero means, are mutually uncorrelated across components and individuals at every pair \((\tau,\tau')\), and have row- and cluster-invariant second-moment kernels
\(K_A(\tau,\tau'):=\E_G[A_g(\tau)A_g(\tau')']\) and
\(K_E(\tau,\tau'):=\E_G[E_{gi}(\tau)E_{gi}(\tau')']\), with \(K_E\not\equiv0\). Consequently,
\[
\Omega_G(\tau,\tau')
=K_A(\tau,\tau')+c_G K_E(\tau,\tau'),
\qquad
c_G:=\frac1G\sum_{g=1}^G\frac1{N_g}.
\]
If \(c_G\to c\), the limiting kernel is \(\Omega=K_A+cK_E\) and both parts of \Cref{ass:norm_mmt_general_F} hold.
\end{remark}

\Cref{ass:uwlln_F,ass:norm_mmt_general_F}, together with \Cref{ass:sampling}, provide the functional CLT for the cluster empirical process at \(\theta_o(\cdot)\). For a scalar-valued target path, such as a quantile function, distribution function, quantile or distribution treatment effect, let
\( \displaystyle
\eta_o(\tau)
:=
\varphi_\tau(\theta_o(\tau)),
\quad
\widehat\eta(\tau)
:=
\varphi_\tau(\widehat\theta(\tau)),
\)
and define the induced map
\[
\Phi_\varphi:
\ell^\infty(\T)^{p_\theta}
\to
\ell^\infty(\T),
\qquad
\Phi_\varphi(\theta)(\tau)
:=
\varphi_\tau(\theta(\tau)).
\]

\begin{theorem}[Weak Convergence]\label{Thm:weak_convergence}
Suppose that Assumptions \ref{ass:sampling}, \ref{ass:uwlln_F}, \ref{ass:pop_identification_F}, \ref{ass:linearisation_F}, and \ref{ass:norm_mmt_general_F} hold. Then
\begin{enumerate}[label=(\alph*)]
\item\label{thm:weak_path}
\( \displaystyle
\sqrt{G}\bigl(\widehat\theta-\theta_o\bigr)(\cdot)
\rightsquigarrow
\mathbb Z_\theta(\cdot):=-H(\cdot)^{-1}\mathbb G_0(\cdot)
\quad\text{in }\ell^\infty(\T),
\)
where \(\mathbb G_0\) is the tight mean-zero Gaussian process with covariance kernel \(\Omega(\tau,\tau')\) from \Cref{ass:norm_mmt_general_F}.

\item\label{thm:weak_scalar} Additionally, suppose \(\Phi_\varphi\) is Hadamard differentiable at \(\theta_o\), tangentially to a subspace containing the support of \(\mathbb Z_\theta\), with derivative \(\Phi'_{\varphi,\theta_o}\). Then
\( \displaystyle 
\sqrt{G}
(\widehat\eta-\eta_o)(\cdot)
\rightsquigarrow
\Phi'_{\varphi,\theta_o}
[\mathbb Z_\theta](\cdot)
\;
\text{in }\ell^\infty(\T).
\)
\end{enumerate}
\end{theorem}

\subsection{Multiplier bootstrap and uniform inference}\label{sec:boot}

The weak-convergence result in \Cref{Thm:weak_convergence} characterises the asymptotic distribution of the full parameter path
\(\tau\mapsto\widehat\theta(\tau)\).
Inference for such path-valued parameters is conveniently conducted via the bootstrap, e.g., \citet{chernozhukov-fernandez-melly-2013inference}. It is therefore imperative to establish bootstrap validity in the same functional setting.

Let \(\{\xi_g:g\geq1\}\) be an i.i.d.\ sequence of centred bootstrap multipliers with a fixed law, independent of the data, and normalised so that
\(\E[\xi_g]=0\) and \(\E[\xi_g^2]=1\).
Conditional on the sample, these equalities are also written
\(\E^*[\xi_g]=0\) and \(\E^*[\xi_g^2]=1\), where \(\E^*\) denotes conditional expectation given the sample; see  \citet[condition~F]{kosorok-2003}. When the estimator is computed from a weighted objective function or weighted estimating equation, it is often convenient to use nonnegative cluster weights \(\tilde\xi_g\) with
\(\E^*[\tilde\xi_g]=1\)
and
\(\V^*(\tilde\xi_g)=1\);
the centred multiplier entering the first-order representation is then
\(\xi_g=\tilde\xi_g-1\).

A cluster-level multiplier perturbation of the empirical estimating equation
is obtained by setting \(\tilde\xi_g=1+\xi_g\) and defining
\[
\begin{aligned}
\widehat\Psi_G^*(\theta)(\tau)
&:=
\widehat{\Prob}_G
\bar\psi_\tau(\bm W_g;\theta(\tau))
+
\frac{1}{G}\sum_{g=1}^G
\xi_g\bar\psi_\tau(\bm W_g;\theta(\tau))
=
\frac{1}{G}\sum_{g=1}^G
\tilde\xi_g\bar\psi_\tau(\bm W_g;\theta(\tau)),
\quad
\tau\in\T.
\end{aligned}
\]
For nonnegative \(\tilde\xi_g\), this is the estimating equation induced by
weighting each cluster's contribution to the original criterion or estimating
equation. For general centred \(\xi_g\), it is understood as a direct
multiplier perturbation of the \(Z\)-equation.
Let \(\widehat\theta^*(\cdot)\) denote a corresponding bootstrap \(Z\)-SACR estimator satisfying
\[
\widehat\Psi_G^*(\widehat\theta^*)
=o_{\Prob^*}(G^{-1/2})
\qquad\text{in }\ell^\infty(\T)
\]
in outer probability. Define the corresponding first-order multiplier moment process by
\[
\mathbb M_G^*(\theta)(\tau)
:=
G^{-1/2}\sum_{g=1}^G
\xi_g\bar\psi_\tau(\bm W_g;\theta(\tau)),
\qquad
\tau\in\T.
\]

\begin{remark}[Heterogeneous cluster means and score covariance estimation]\label{rem:bootstrap_centering}
Let \(s_g=\bar\psi(\bm W_g;\theta_G)\) and \(\mu_g=\E_G[s_g]\), where \(\theta_G\) solves the SACR population moment, so \(\sum_g\mu_g=0\). For independent, mean-zero, unit-variance multipliers,
\[
\widehat\Omega_G^{A}
=\frac1G\sum_gs_gs_g'
=\V^*\!\left(G^{-1/2}\sum_g\xi_gs_g\right)
=\widehat\Omega_G^{MB}.
\]
Both have population counterpart
\(
\Omega_G+\frac1G\sum_g\mu_g\mu_g',
\text{ and }
\Omega_G:=\frac1G\sum_g\V_G(s_g).
\)
Thus both share the same positive semidefinite excess and, under suitable consistency and normality conditions, yield conservative pointwise Wald inference for \(\theta_G\). Consistent estimation of the sampling covariance requires this excess to vanish or be removed; heterogeneous cluster means are not generally consistently estimable from a single realised cluster under unrestricted within-cluster dependence. The present inference theorems maintain the common zero.
\end{remark}

The following theorem establishes bootstrap validity for the functional \(Z\)-SACR estimator and its Hadamard differentiable map.

\begin{theorem}[Functional Bootstrap Validity]
\label{Thm:boot_const}
Suppose the conditions of \Cref{Thm:weak_convergence}\ref{thm:weak_path} hold, and let the chosen multiplier weights satisfy the conditional independence, centring, and unit-variance conditions described above. Let \(\widehat\eta^*(\tau):=\varphi_\tau(\widehat\theta^*(\tau))\). Then the following hold.
\begin{enumerate}[label=(\alph*)]
\item\label{thm:boot_path} Conditionally on the sample, \quad
\( \displaystyle
\sqrt{G}
(\widehat\theta^*-\widehat\theta)(\cdot)
\rightsquigarrow_{\Prob}
\mathbb Z_\theta(\cdot)
\quad
\text{in }\ell^\infty(\T).
\)

\item\label{thm:boot_scalar} Additionally, suppose \(\Phi_\varphi\) is Hadamard differentiable at \(\theta_o\), tangentially to a subspace containing the support of \(\mathbb Z_\theta\), with derivative \(\Phi'_{\varphi,\theta_o}\). Then, conditionally on the sample,
\( \displaystyle 
\sqrt{G}
(\widehat\eta^*-\widehat\eta)(\cdot)
\rightsquigarrow_{\Prob}
\Phi'_{\varphi,\theta_o}
[\mathbb Z_\theta](\cdot)
\;
\text{in }\ell^\infty(\T).
\)
\end{enumerate}
\end{theorem}

\noindent
Consequently, the conditional distributions of
\(
\sqrt{G}
(\widehat\theta^*-\widehat\theta)(\cdot)
\) and
\(
\sqrt{G}
(\widehat\eta^*-\widehat\eta)(\cdot)
\)
consistently approximate the Gaussian limits of
\(
\sqrt{G}
(\widehat\theta-\theta_o)(\cdot)
\) 
and
\(
\sqrt{G}
(\widehat\eta - \eta_o)(\cdot).
\)

\Cref{Thm:boot_const}\ref{thm:boot_scalar} supplies the theoretical framework for inference on Hadamard differentiable maps of SACR estimators. The following remark specialises to distributional functionals and treatment effects as maps of distribution functions, which are the functional parameters of interest in \Cref{sect:empirics}.

\begin{remark}[Distributional scalar functionals]\label{rem:dist_scalar_functionals}
When the underlying object is a pair of marginal or counterfactual
distribution functions \((F_1,F_0)\), the Distribution Treatment Effect (DTE) path
is generated by the continuous linear contrast: 
\(
\Phi_{\mathrm{DTE}}(F_1,F_0)(y)=F_1(y)-F_0(y),
\quad y\in\mathcal Y.
\)
Similarly, quantile effects use the difference of the two left-inverse
maps, whose Hadamard differentiability requires continuity and positive-density
conditions around the target quantiles \citep[Theorem~4.1(ii),
Corollary~4.1, and Appendix~D.2]{chernozhukov-fernandez-melly-2013inference}.
Expectile effects are differences of the paths treated in
\Cref{prop:expectile_map}, which verifies the indexed expectile \(Z\)-map
conditions explicitly using the Z-map framework of
\citet[Appendix E]{chernozhukov-fernandez-melly-2013inference}.
\end{remark}

\noindent
If \(\varphi_\tau\) is continuously differentiable in \(\theta\), uniformly over \(\tau\in\T\), then the derivative takes the pointwise form
\[
\Phi'_{\varphi,\theta_o}[h](\tau)
=
\dot\varphi_\tau(\theta_o(\tau))'h(\tau).
\]

\Cref{Thm:boot_const} establishes bootstrap validity for the functional \(Z\)-SACR estimator under ULC. The next corollary specialises the functional-regime result to fixed values of \(\tau\). Hence its limit uses the fixed functional limit \(H(\tau)\) from \Cref{Thm:weak_convergence}; the more general pointwise theory in \Cref{Thm:normality} allows drifting \(H_G\) and \(\Omega_G\).

\begin{corollary}[Pointwise Bootstrap Validity]
\label{Crly:boot_pointwise}
Under the conditions of \Cref{Thm:boot_const}, for each fixed \(\tau\in\T\),
\[
\sqrt{G}
\big(\widehat\theta^*(\tau)-\widehat\theta(\tau)\big)
\rightsquigarrow_{\Prob}
-H(\tau)^{-1}\mathbb G_0(\tau).
\]
\end{corollary}

For uniform inference on the scalar path \(\eta_o(\cdot)\), \Cref{Thm:boot_const}\ref{thm:boot_scalar} provides a bootstrap approximation to the Gaussian limit of
\(
\sqrt{G}
(\widehat\eta-\eta_o)(\cdot).
\)
Fix \(0<q_2<q_1<1\), let \(z_q\) denote the \(q\)-quantile of a standard normal variable, and let \(\widehat q_q^*(\tau)\) denote the conditional \(q\)-quantile of
\(
\sqrt G\{\widehat\eta^*(\tau)-\widehat\eta(\tau)\}
\).
Following \citet[Supplementary Appendix, Lemma SA.1]{chernozhukov-fernandez-melly-2013inference}, define the quantile-spread studentiser
\[
\widehat s(\tau)
:=
\frac{\widehat q_{q_1}^*(\tau)-\widehat q_{q_2}^*(\tau)}
{z_{q_1}-z_{q_2}}.
\]
The implementation uses \(q_1=0.75\) and \(q_2=0.25\), giving the normalised bootstrap interquartile range (IQR).
Uniform confidence bands can then be constructed using the conditional quantiles of
\[
T_G^*
:=
\sup_{\tau\in\T}
\left|
\widehat s(\tau)^{-1}
\sqrt{G}
\{\widehat\eta^*(\tau)-\widehat\eta(\tau)\}
\right|.
\]
Let \(c_{1-\alpha}^*\) denote the conditional \((1-\alpha)\)-quantile of \(T_G^*\). Then
\[
\left[
\widehat\eta(\tau)
-
G^{-1/2}
c_{1-\alpha}^*
\widehat s(\tau),
\;
\widehat\eta(\tau)
+
G^{-1/2}
c_{1-\alpha}^*
\widehat s(\tau)
\right],
\qquad
\tau\in\T,
\]
forms an asymptotically valid \((1-\alpha)\) uniform confidence band for the path \(\eta_o(\cdot)\). Under the conditions of \Cref{Thm:boot_const}\ref{thm:boot_scalar}, continuity of the limiting sample paths, and pointwise standard deviations bounded and bounded away from zero, \citet[Supplementary Appendix, Lemma SA.1]{chernozhukov-fernandez-melly-2013inference} gives uniform consistency of this studentiser and validity of the resulting band.

\section{Model-Specific Verification}
\label{sect:theory_concrete}

This section verifies the theoretical framework established in the preceding section for distribution regression, quantile regression, and expectile regression estimating equations. The model-specific results establish identification and global score-class manageability with suitable envelopes. Together with the remaining generic sampling and regularity conditions, these verifications support the pointwise and functional inference results developed above. The supplementary empirical illustration in \Cref{sect:empirics} also considers quantile and expectile effects obtained from estimated distributions, a construction requiring additional distribution-to-functional map conditions; see \Cref{Thm:boot_const}\ref{thm:boot_scalar} and \Cref{rem:dist_scalar_functionals}.

DR involves a smooth finite-dimensional score indexed by a threshold, QR introduces a non-smooth threshold score, and ER combines a threshold indicator with an unbounded residual envelope. These examples encompass binary response models, Least Absolute Deviations (LAD)/median regression, and OLS as special cases. The examples below are understood under the generic compactness and interiority requirements in \Cref{ass:uwlln,ass:linearisation,ass:linearisation_F}. The following assumption collects the remaining common linear-index restrictions used across the DR, QR, and ER examples. These conditions provide model-specific primitive counterparts to the rank and \(X\)-moment requirements entering \Cref{ass:norm_mmt_general}.

\begin{modelassumption}{LI}{1}{Common linear-index conditions}\label{ass:LI_common}\quad
\begin{enumerate}[label=(\alph*)]
    \item\label{ass:LI_rank}
    \( \displaystyle
    \frac{1}{G}\sum_{g=1}^G
    \E_G\!\left[
    N_g^{-1}\sum_{i\in\G_g}X_iX_i'
    \right]
    \)
    is uniformly positive definite.

    \item\label{ass:LI_moment}
    For some \(\delta>0\),
    \( \displaystyle
    \sup_{G\geq1}\sup_{1\le i\le N}\E_G[\|X_i\|^{2+\delta}]<\infty.
    \)
\end{enumerate}
\end{modelassumption}

\subsection{Distribution regression}

For distribution regression, the target is the conditional distribution function
\[
F_{Y\mid X}(y\mid x)
=
\Lambda\!\left(x'\theta_o(y)\right),
\]
where $\Lambda$ is a known link function, such as the logistic link, and $y$ ranges over a compact subset $\mathcal Y$ of the support of $Y$.\footnote{The generic theory uses \(\tau\in\T\) as a neutral functional index. In the model-specific notation, DR is indexed by the outcome threshold \(y\in\mathcal Y\), while QR and ER retain \(\tau\in\T\subset(0,1)\).} The model implies the conditional moment restriction

\[
\Prob\psi_y(W;\theta_o(y))
=
\E\!\left[
X\left(
\indicator{Y\leq y}
-
\Lambda\!\left(X'\theta_o(y)\right)
\right)
\right]
=0,
\qquad
y\in\mathcal Y.
\]
Accordingly, the primitive score process is
\( \displaystyle
\psi_y(W;\theta(y))
=
X\left(
\indicator{Y\leq y}
-
\Lambda\!\left(X'\theta(y)\right)
\right),
\quad
y\in\mathcal Y.
\)
At the cluster level, the corresponding SACR score is
\[
\bar\psi_y(\bm W_g;\theta(y))
=
N_g^{-1}
\sum_{i\in\G_g}
X_i
\left(
\indicator{Y_i\leq y}
-
\Lambda\!\left(X_i'\theta(y)\right)
\right).
\]
For each coordinate \(j\), define the full jointly indexed DR score class
\[
\mathcal F_j^{DR}
:=
\left\{
w\mapsto x_j\big(\indicator{u\leq y}-\Lambda(x't)\big):
(y,t)\in\mathcal Y\times\Theta
\right\},
\qquad w=(u,x),
\]
and its normalised cluster-average array, unnormalised envelope, and
corresponding normalised envelope by
\[
Z_{Gg,j}^{DR}(y,t)
:=
G^{-1/2}\bar\psi_{y,j}(\bm W_g;t),
\quad
\bar F_{Gg,j}^{DR}
:=
\left\{
N_g^{-1}\sum_{i\in\G_g}X_{ij}^2
\right\}^{1/2},
\quad \text{and} \quad
F_{Gg,j}^{DR}:=G^{-1/2}\bar F_{Gg,j}^{DR}.
\]
The corresponding Jacobian and covariance kernel are
\[
H_G^{DR}(y)
=
-\frac{1}{G}
\sum_{g=1}^G
\E_G\!\left[
N_g^{-1}
\sum_{i\in\G_g}
X_iX_i'
\Lambda'\!\left(X_i'\theta_o(y)\right)
\right],
\]
and
\[
\Omega_G^{DR}(y,y')
=
\frac{1}{G}
\sum_{g=1}^G
\E_G\!\left[
\bar\psi_y(\bm W_g;\theta_o(y))
\bar\psi_{y'}(\bm W_g;\theta_o(y'))'
\right].
\]

The next assumptions establish identification of the DR-SACR population moment and functional separation.

\begin{modelassumption}{DR}{1}{Identification and array preservation}\label{ass:DR}\label{ass:DR_identification}\quad
The link \(\Lambda:\mathbb R\to(0,1)\) is known and strictly increasing. 
\[
F_{Y\mid X}(y\mid X)=\Lambda\bigl(X'\theta_o(y)\bigr),
\qquad y\in\mathcal Y
\]
in the target population, and this restriction is preserved along the sampled triangular array:
\[
F_{Y_i\mid X_i}(y\mid X_i)
:=\E_G\bigl[\indicator{Y_i\leq y}\mid X_i\bigr]
=\Lambda\bigl(X_i'\theta_o(y)\bigr)
\qquad\text{almost surely}
\]
for every \(G\), every sampled observation \(i\leq N\), and every \(y\in\mathcal Y\). Moreover, the coefficient path \(y\mapsto\theta_o(y)\) is Borel measurable.
\end{modelassumption}

The following condition strengthens strict link monotonicity
in the cluster-balanced population and supports uniform functional separation.

\begin{modelassumption}{DR}{2}{Regularity}\label{ass:DR_regular}\quad
\begin{enumerate}[label=(\alph*)]
    \item\label{ass:DR_link} The derivative \(\Lambda'\) is bounded and uniformly continuous on \(\mathbb R\).
    \item\label{ass:DR_jacobian} For
    \( \displaystyle
    Q_G^{DR}(y,t,s)
    :=
    \frac{1}{G}\sum_{g=1}^G
    \E_G\!\left[
    N_g^{-1}\sum_{i\in\G_g}
    X_iX_i'
    \Lambda'\!\left(X_i'\{\theta_o(y)+s(t-\theta_o(y))\}\right)
    \right],
    \)
    there is a constant \(c_H>0\) such that
    \( \displaystyle
    \inf_G\inf_{y\in\mathcal Y}\inf_{t\in\Theta}\inf_{s\in[0,1]}
    \lambda_{\min}\!\left(Q_G^{DR}(y,t,s)\right)
    \geq c_H.
    \)
\end{enumerate}
\end{modelassumption}

The following supplies the DR functional identification result.
\begin{proposition}[DR-SACR identification]\label{prop:DR_identification}
Under \Cref{ass:DR_identification,ass:LI_common}, \(\theta_o(y)\) is the unique solution to the DR-SACR population moment
$$
\frac{1}{G}\sum_{g=1}^G N_g^{-1} \sum_{i \in \mathcal{G}_g} \mathbb{E}_G\big[X_i\big(\indicator{Y_i \leq y} - \Lambda(X_i'\theta(y))\big)\big] = 0 \quad \text{for each } y \in \mathcal{Y}.
$$
If, in addition, \Cref{ass:DR_regular} holds, the DR-SACR population moments satisfy the
functional separation condition in \Cref{ass:pop_identification_F}\ref{ass:functional_sep}.
\end{proposition}

The next result verifies the global complexity and envelope conditions for the DR score process.
\begin{proposition}[DR global-class verification]\label{prop:DR_verify}
Suppose that Assumptions
\ref{ass:uwlln}\ref{ass:uwlln_compact}, 
\ref{ass:DR_identification}, \ref{ass:LI_common}, and 
\ref{ass:DR_regular}\ref{ass:DR_link} hold. Then
\Cref{ass:uwlln_F}\ref{ass:functional_manageability} and
\ref{ass:functional_global_envelope} hold for the DR score process.
\end{proposition}

\noindent Together with the generic approximate-root, interiority, Jacobian, covariance, and remaining functional conditions, the preceding DR conditions yield the conclusions of \Cref{Thm:uwlln,Thm:normality,Thm:weak_convergence,Thm:boot_const}. 

\subsection{Linear quantile regression}

For a growing number of clusters, the closest process-bootstrap benchmark is \citet{hagemann-2017-cluster}, which establishes cluster-robust inference for the quantile-regression process under bounded-cluster-size restrictions. Inverse-cluster-size-weighted quantile estimating equations also appear in the censored-data setting of \citet{kim-choi-park-choi-bandyopadhyay-2024}, where cluster sizes are relatively small. The results below use cluster-average scores within the general functional Z-SACR framework with a growing number of arbitrarily sized clusters.

For linear quantile regression,
\[
Q_{Y\mid X}(\tau\mid X)
=
X'\theta_o(\tau),
\qquad
\tau\in\T\subset(0,1),
\]
and the model implies the conditional moment restriction
\[
\Prob\psi_\tau(W;\theta_o(\tau))
=
\E\!\left[
X\big(
\tau-\indicator{Y\leq X'\theta_o(\tau)}
\big)
\right]
=0,
\qquad
\tau\in\T.
\]
Accordingly, the primitive score process is
\( \displaystyle
\psi_\tau(W;\theta(\tau))
=
X\big(
\tau-\indicator{Y\leq X'\theta(\tau)}
\big),
\quad
\tau\in\T.
\)
At the cluster level, the corresponding SACR score is
\[
\bar\psi_\tau(\bm W_g;\theta(\tau))
=
N_g^{-1}
\sum_{i\in\G_g}
X_i
\big(
\tau-\indicator{Y_i\leq X_i'\theta(\tau)}
\big).
\]
For each coordinate \(j\), define the full jointly indexed QR score class
\[
\mathcal F_j^{QR}
:=
\left\{
w\mapsto x_j\big(\tau-\indicator{u\leq x't}\big):
(\tau,t)\in\T\times\Theta
\right\},
\qquad w=(u,x),
\]
and its normalised cluster-average array, unnormalised envelope, and
corresponding normalised envelope by
\[
Z_{Gg,j}^{QR}(\tau,t)
:=
G^{-1/2}\bar\psi_{\tau,j}(\bm W_g;t),
\quad
\bar F_{Gg,j}^{QR}
:=
\left\{
N_g^{-1}\sum_{i\in\G_g}X_{ij}^2
\right\}^{1/2},
\quad \text{and} \quad
F_{Gg,j}^{QR}:=G^{-1/2}\bar F_{Gg,j}^{QR}.
\]
The corresponding Jacobian and covariance kernel are
\[
H_G^{QR}(\tau)
=
-\frac{1}{G}
\sum_{g=1}^G
\E_G\!\left[
N_g^{-1}
\sum_{i\in\G_g}
X_iX_i'
f_{Y_i\mid X_i}
\!\left(
X_i'\theta_o(\tau)\mid X_i
\right)
\right],
\]
and
\[
\Omega_G^{QR}(\tau,\tau')
=
\frac{1}{G}
\sum_{g=1}^G
\E_G\!\left[
\bar\psi_\tau(\bm W_g;\theta_o(\tau))
\bar\psi_{\tau'}(\bm W_g;\theta_o(\tau'))'
\right].
\]

The next assumptions establish identification of the QR-SACR population moment and provide primitive conditions for verifying the remaining high-level $Z$-SACR requirements.

\begin{modelassumption}{QR}{1}{Identification and regularity}\label{ass:QR}\label{ass:QR_identification}\quad
\begin{enumerate}[label=(\alph*)]
    \item\label{ass:QR_ident}
    \(Q_{Y\mid X}(\tau\mid X)=X'\theta_o(\tau)\) almost surely for every
    \(\tau\in\T\) in the target population, and this conditional-quantile restriction is preserved along
    the sampled triangular array:
    \[
    F_{Y_i\mid X_i}\!\left(X_i'\theta_o(\tau)\mid X_i\right)=\tau
    \qquad \text{almost surely under the stage-}G\text{ law}
    \]
    \noindent for every
    \(G\), every sampled observation \(i\leq N\), and every \(\tau\in\T\).

    \item\label{ass:QR_density}
    The conditional distribution of \(Y_i\) given \(X_i\) is absolutely
    continuous, with a density \(f_{Y_i\mid X_i}(\cdot\mid X_i)\) bounded above by a
    common constant \(C_f<\infty\), almost surely in \(X_i\), uniformly
    over \(G\) and \(i\leq N\). For some \(r>0\) and a common
    nondecreasing modulus \(\omega(u)\to0\) as \(u\downarrow0\),
    \( \displaystyle
    |f_{Y_i\mid X_i}(z\mid X_i)-f_{Y_i\mid X_i}(z'\mid X_i)|
    \leq\omega(|z-z'|)
    \)
    whenever both \(z\) and \(z'\) lie within \(r\|X_i\|\) of
    \(X_i'\theta_o(\tau)\), uniformly over \(G,i,\tau\), almost surely
    in \(X_i\).

    \item\label{ass:QR_segment_rank}
    For
    \( \displaystyle
    Q_G^{QR}(\tau,t,s)
    :=
    \frac{1}{G}\sum_{g=1}^G
    \E_G\!\left[
    N_g^{-1}\sum_{i\in\G_g}
    X_iX_i'
    f_{Y_i\mid X_i}\!\left(
    X_i'\{\theta_o(\tau)+s(t-\theta_o(\tau))\}\mid X_i
    \right)
    \right],
    \)
    there is a constant \(c_f>0\) such that
    \( \displaystyle
    \inf_G\inf_{\tau\in\T}\inf_{t\in\Theta}\inf_{s\in[0,1]}
    \lambda_{\min}\!\left(Q_G^{QR}(\tau,t,s)\right)
    \geq c_f.
    \)
\end{enumerate}
\end{modelassumption}

The following supplies the QR functional identification result.
\begin{proposition}[QR-SACR identification]\label{prop:QR_identification}
Under \Cref{ass:QR_identification,ass:LI_common}, \(\theta_o(\tau)\) is the unique solution to the QR-SACR population moment
\[
\frac{1}{G}\sum_{g=1}^G
\E_G\!\Big[
N_g^{-1}
\sum_{i\in\G_g}
X_i
\big(
\tau-\indicator{Y_i\leq X_i'\theta(\tau)}
\big)
\Big]
=
0,
\qquad
\tau\in\T.
\]
The population moments also satisfy the functional separation condition
in \Cref{ass:pop_identification_F}\ref{ass:functional_sep}.
\end{proposition}

The next result verifies the global complexity and envelope conditions for the QR score process.
\begin{proposition}[QR global-class verification]\label{prop:QR_verify}
Suppose that Assumptions
\ref{ass:uwlln}\ref{ass:uwlln_compact}, 
\ref{ass:QR_identification}, and \ref{ass:LI_common} hold. Then
\Cref{ass:uwlln_F}\ref{ass:functional_manageability} and
\ref{ass:functional_global_envelope} hold for the QR score process.
\end{proposition}

\noindent Together with the generic approximate-root, interiority, Jacobian, covariance, and remaining functional conditions, the preceding QR conditions yield the
conclusions of
\Cref{Thm:uwlln,Thm:normality,Thm:weak_convergence,Thm:boot_const}.

\subsection{Linear expectile regression}

For linear expectile regression,
\[
\E_\tau(Y\mid X)
=
X'\theta_o(\tau),
\qquad
\tau\in\T\subset[\underline\tau,\bar\tau]\subset(0,1),
\]
where \(\E_\tau(Y\mid X)\) denotes the \(\tau\)th conditional expectile of \(Y\) given \(X\), the model implies the conditional moment restriction
\[
\Prob\psi_\tau(W;\theta_o(\tau))
=
\E\!\Big[
X\big(
Y-X'\theta_o(\tau)
\big)
\big|
\tau-\indicator{Y<X'\theta_o(\tau)}
\big|
\Big]
=0,
\qquad
\tau\in\T.
\]
Accordingly, the primitive score process is
\[
\psi_\tau(W;\theta(\tau))
=
X\big(
Y-X'\theta(\tau)
\big)
\left|
\tau-\indicator{Y<X'\theta(\tau)}
\right|,
\qquad
\tau\in\T.
\]
When \(0.5\in\T\), the score at that index satisfies
\(
\psi_{0.5}(W;\theta)=\tfrac12 X(Y-X'\theta)
\), which is one half of the OLS score obtained by setting \(Z=X\)
in the running IV example.

At the cluster level, the corresponding SACR score is
\[
\bar\psi_\tau(\bm W_g;\theta(\tau))
=
N_g^{-1}
\sum_{i\in\G_g}
X_i
\big(
Y_i-X_i'\theta(\tau)
\big)
\left|
\tau-\indicator{Y_i<X_i'\theta(\tau)}
\right|.
\]
For each coordinate \(j\), let
\( \displaystyle B_\Theta:=\sup_{t\in\Theta}\|t\|\) and define the full jointly indexed ER
score class and its measurable envelope by
\[
\mathcal F_j^{ER}
:=
\left\{
w\mapsto x_j(u-x't)\left|\tau-\indicator{u<x't}\right|:
(\tau,t)\in\T\times\Theta
\right\},
\qquad w=(u,x),
\]
\[
F_j^{ER}(W):=|X_j|\big(|Y|+B_\Theta\|X\|\big).
\]
Define the corresponding normalised cluster-average array, unnormalised
envelope, and normalised envelope by
\[
\begin{aligned}
Z_{Gg,j}^{ER}(\tau,t)
&:=G^{-1/2}\bar\psi_{\tau,j}(\bm W_g;t),
&
\bar F_{Gg,j}^{ER}
&:=\left\{
N_g^{-1}\sum_{i\in\G_g}\bigl\{F_j^{ER}(W_i)\bigr\}^2
\right\}^{1/2},\;\text{and}\;
F_{Gg,j}^{ER}
&:=G^{-1/2}\bar F_{Gg,j}^{ER}.
\end{aligned}
\]
The corresponding Jacobian and covariance kernel are
\[
H_G^{ER}(\tau)
=
-\frac{1}{G}
\sum_{g=1}^G
\E_G\!\left[
N_g^{-1}
\sum_{i\in\G_g}
X_iX_i'
\E_G\!\Big[
\left|
\tau-\indicator{Y_i<X_i'\theta_o(\tau)}
\right|
\Big| X_i
\Big]
\right],
\]
and
\[
\Omega_G^{ER}(\tau,\tau')
=
\frac{1}{G}
\sum_{g=1}^G
\E_G\!\left[
\bar\psi_\tau(\bm W_g;\theta_o(\tau))
\bar\psi_{\tau'}(\bm W_g;\theta_o(\tau'))'
\right].
\]

The next assumptions establish identification of the ER-SACR population moment and provide primitive conditions for verifying the remaining high-level \(Z\)-SACR requirements.

\begin{modelassumption}{ER}{1}{Identification and array preservation}\label{ass:cond_expectile}\label{ass:ER_identification}\quad
\( \displaystyle
\E_\tau(Y\mid X)=X'\theta_o(\tau)
\quad\text{almost surely}
\)
for every \(\tau\in\T\subset[\underline\tau,\bar\tau]\subset(0,1)\), in the target population, and the corresponding first-order restriction is preserved along the sampled
triangular array: 
\[
\E_G\!\Big[
\big(Y_i-X_i'\theta_o(\tau)\big)
\left|\tau-\indicator{Y_i<X_i'\theta_o(\tau)}\right|
\Big| X_i
\Big]=0
\qquad \text{almost surely under the stage-}G\text{ law}
\]
for every \(G\), every sampled observation \(i\leq N\), and
every \(\tau\in\T\).
\end{modelassumption}

\begin{modelassumption}{ER}{2}{Regularity}\label{ass:ER_regular}
For some \(\delta>0\),
\( \displaystyle
\sup_{G\geq1}\sup_{1\leq i\leq N}
\E_G\!\left[
\left(\|X_i\|\,|Y_i|\right)^{2+\delta}
+
\|X_i\|^{4+2\delta}
\right]
<\infty.
\)
\end{modelassumption}

The following supplies the ER functional identification result.
\begin{proposition}[ER-SACR identification]\label{prop:ER_identification}
Under \Cref{ass:ER_identification},
\Cref{ass:ER_regular}, and \Cref{ass:LI_common},
\(\theta_o(\tau)\) is the unique solution to the ER-SACR population moment
\[
\frac{1}{G}
\sum_{g=1}^G
\E_G\!\left[
N_g^{-1}
\sum_{i\in\G_g}
X_i
\big(
Y_i-X_i'\theta(\tau)
\big)
\left|
\tau-\indicator{Y_i<X_i'\theta(\tau)}
\right|
\right]
=
0,
\qquad
\tau\in\T.
\]
The population moments also satisfy the functional separation condition
in \Cref{ass:pop_identification_F}\ref{ass:functional_sep}.
\end{proposition}

The next result verifies the global complexity and envelope conditions for the ER score process.
\begin{proposition}[ER global-class verification]\label{prop:ER_verify}
Suppose that Assumptions
\ref{ass:uwlln}\ref{ass:uwlln_compact}, 
\ref{ass:ER_identification}, \ref{ass:LI_common}, and 
\ref{ass:ER_regular} hold. Then
\Cref{ass:uwlln_F}\ref{ass:functional_manageability} and
\ref{ass:functional_global_envelope} hold for the ER score process.
\end{proposition}

\noindent Together with the generic approximate-root, interiority, Jacobian, covariance, and remaining functional conditions, the preceding ER conditions yield the conclusions of \Cref{Thm:uwlln,Thm:normality,Thm:weak_convergence,Thm:boot_const}.

\section{Monte Carlo Evidence}\label{sect:MC}

This section studies the finite-sample performance of conventional cluster-robust and \(Z\)-SACR inference for both pointwise coefficients and distributional treatment-effect functions. The two experiments use the same no-covariate design, cluster-level treatment assignment, deterministic cluster-size arrays, cluster grid, and Monte Carlo settings. They differ only in whether the target is evaluated at a central index or over a functional domain. The data-generating processes (DGPs) impose common coefficient paths and centre the relevant scores within every cluster, so observation weighting and cluster weighting share the same structural target. The experiments therefore isolate the inferential consequences of cluster-size dominance rather than differences between estimands. Coverage and rejection comparisons concern complete inference procedures, which differ in estimator weighting, resampling scheme, and, where applicable, scale estimation; their differences cannot be attributed solely to SACR weighting. Point-estimate bias and root mean squared error (RMSE) comparisons assess the estimators independently of the bootstrap procedure.

\subsection{Simulation design and inference procedures}\label{sec:mc_design}
Let \(X_i=(1,T_g)'\) for \(i\in\G_g\), with
\(T_g=\indicator{g\equiv0\pmod 5}\). The DR design is
\(Y_i=X_i'\theta_o+\operatorname{logit}\{\Phi(U_{gi})\}\), while
QR and ER use \(Y_i=X_i'\theta_o+X_i'\gamma U_{gi}\), where
\(\theta_o=(-0.5,0.5)'\), \(\gamma=(0.2,1)'\), and
\(U_{gi}=\sqrt{0.96}Z_g+\sqrt{0.04}\varepsilon_{gi}\).
The common and individual shocks are mutually independent standard normals,
independent across clusters. The errors therefore have standard logistic
margins for DR and standard normal margins for QR and ER.

Cluster sizes are deterministic Pareto quantiles,
\(N_g=\lceil(1-(g-1/2)/G)^{-1/\beta}\rceil\).
For \(\beta=2.5\), \(\max_gN_g^2/N\to0\); for \(\beta=1\), this
condition fails although \(\max_gN_g/N\to0\). The latter regime isolates
failure of conventional Gaussian inference while retaining cluster-size
negligibility sufficient for a consistency argument. Both regimes use
\(G\in\{150,300,600\}\), \(1{,}000\) Monte Carlo replications, and
\(499\) bootstrap draws.

Analytical procedures use the relevant cluster-score sandwich, with a
kernel density estimate for conventional QR. Conventional bootstrap
procedures resample whole clusters in pairs; SACR bootstrap procedures
apply independent \(\operatorname{Exp}(1)\) cluster weights to the
size-adjusted estimating equations. \Cref{sec:mc_implementation} gives the
coefficient paths and implementation details. Supplementary experiments in \Cref{sec:mc_dependence,sec:mc_matched_exp}
examine weaker dependence and a common exponential-weight bootstrap. \Cref{sec:sup_mc} considers a single
dominant cluster.

\subsection{Pointwise coefficient inference}\label{sec:mc_pointwise}

The pointwise targets are the DR treatment coefficient at the cluster-balanced
population median threshold \(y=Q_Y(0.5)\), the QR coefficient at
\(\tau=0.5\), and the ER coefficient at \(\tau=0.5\).
The tables report signed bias, RMSE, average
standard error (Avg. SE), and rejection frequencies for nominal \(5\%\) two-sided
Wald tests using the critical value \(1.96\). SACR-A and CR-A use analytical
standard errors; SACR-B and CR-B use bootstrap standard errors. For both procedures, each coefficient's bootstrap standard error is the sample standard deviation of its bootstrap coefficient draws.

\begin{table}[!htbp]
\centering
\caption{Pointwise Monte Carlo results: asymptotically ignorable Pareto regime (\(\beta=2.5\))}
\label{tab:mc_point_beta25}
\begin{threeparttable}
\setlength{\tabcolsep}{2pt}
\footnotesize
\begin{tabular}{lcccccccccccc}
\toprule
& \multicolumn{4}{c}{DR: \(y=Q_Y(0.5)\)}
& \multicolumn{4}{c}{QR: \(\tau=0.5\)}
& \multicolumn{4}{c}{ER: \(\tau=0.5\)} \\
\cmidrule(lr){2-5}\cmidrule(lr){6-9}\cmidrule(lr){10-13}
Method
& Bias & RMSE & Avg. SE & Rej.
& Bias & RMSE & Avg. SE & Rej.
& Bias & RMSE & Avg. SE & Rej. \\
\midrule
\multicolumn{13}{l}{\textbf{$G = 150$}} \\ \cmidrule{1-1}
SACR-A & \multirow{2}{*}{0.0043} & \multirow{2}{*}{0.4050} & 0.3983 & 0.0460 & \multirow{2}{*}{$-0.0127$} & \multirow{2}{*}{0.2614} & 0.2566 & 0.0860 & \multirow{2}{*}{$-0.0023$} & \multirow{2}{*}{0.2161} & 0.2128 & 0.0620 \\
SACR-B &  &  & 0.4027 & 0.0450 &  &  & 0.2687 & 0.0620 &  &  & 0.2092 & 0.0690 \\
CR-A & \multirow{2}{*}{0.0034} & \multirow{2}{*}{0.4630} & 0.4383 & 0.0630 & \multirow{2}{*}{$-0.0181$} & \multirow{2}{*}{0.2914} & 0.2905 & 0.1040 & \multirow{2}{*}{$-0.0035$} & \multirow{2}{*}{0.2468} & 0.2348 & 0.0670 \\
CR-B &  &  & 0.4594 & 0.0480 &  &  & 0.3026 & 0.0650 &  &  & 0.2363 & 0.0660 \\
\midrule
\multicolumn{13}{l}{\textbf{$G = 300$}} \\ \cmidrule{1-1}
SACR-A & \multirow{2}{*}{$-0.0063$} & \multirow{2}{*}{0.2744} & 0.2801 & 0.0440 & \multirow{2}{*}{$-0.0067$} & \multirow{2}{*}{0.1805} & 0.1854 & 0.0540 & \multirow{2}{*}{0.0034} & \multirow{2}{*}{0.1478} & 0.1527 & 0.0410 \\
SACR-B &  &  & 0.2815 & 0.0410 &  &  & 0.1896 & 0.0500 &  &  & 0.1514 & 0.0440 \\
CR-A & \multirow{2}{*}{$-0.0081$} & \multirow{2}{*}{0.3153} & 0.3131 & 0.0570 & \multirow{2}{*}{$-0.0029$} & \multirow{2}{*}{0.2071} & 0.2109 & 0.0800 & \multirow{2}{*}{0.0046} & \multirow{2}{*}{0.1721} & 0.1703 & 0.0560 \\
CR-B &  &  & 0.3178 & 0.0510 &  &  & 0.2173 & 0.0670 &  &  & 0.1704 & 0.0550 \\
\midrule
\multicolumn{13}{l}{\textbf{$G = 600$}} \\ \cmidrule{1-1}
SACR-A & \multirow{2}{*}{0.0083} & \multirow{2}{*}{0.2030} & 0.1975 & 0.0580 & \multirow{2}{*}{0.0015} & \multirow{2}{*}{0.1310} & 0.1314 & 0.0650 & \multirow{2}{*}{0.0016} & \multirow{2}{*}{0.1087} & 0.1080 & 0.0620 \\
SACR-B &  &  & 0.1981 & 0.0600 &  &  & 0.1329 & 0.0690 &  &  & 0.1075 & 0.0640 \\
CR-A & \multirow{2}{*}{0.0088} & \multirow{2}{*}{0.2313} & 0.2262 & 0.0580 & \multirow{2}{*}{$-0.0039$} & \multirow{2}{*}{0.1530} & 0.1547 & 0.0660 & \multirow{2}{*}{0.0028} & \multirow{2}{*}{0.1286} & 0.1265 & 0.0560 \\
CR-B &  &  & 0.2270 & 0.0550 &  &  & 0.1563 & 0.0650 &  &  & 0.1258 & 0.0610 \\
\bottomrule
\end{tabular}
\begin{tablenotes}
\item The three column blocks report DR at \(y=Q_Y(0.5)\), QR at \(\tau=0.5\) (LAD), and ER at \(\tau=0.5\) (OLS), for \(G=150,300,600\). SACR-A and CR-A use analytical standard errors; SACR-B and CR-B use bootstrap sample standard deviations. Bias and RMSE are shared within each estimator's A/B pair because the two rows differ only in their inference procedure.
\end{tablenotes}
\end{threeparttable}
\end{table}

\begin{table}[!htbp]
\centering
\caption{Pointwise Monte Carlo results: ULC Pareto regime (\(\beta=1\))}
\label{tab:mc_point_beta1}
\begin{threeparttable}
\setlength{\tabcolsep}{2pt}
\footnotesize
\begin{tabular}{lcccccccccccc}
\toprule
& \multicolumn{4}{c}{DR: \(y=Q_Y(0.5)\)}
& \multicolumn{4}{c}{QR: \(\tau=0.5\)}
& \multicolumn{4}{c}{ER: \(\tau=0.5\)} \\
\cmidrule(lr){2-5}\cmidrule(lr){6-9}\cmidrule(lr){10-13}
Method
& Bias & RMSE & Avg. SE & Rej.
& Bias & RMSE & Avg. SE & Rej.
& Bias & RMSE & Avg. SE & Rej. \\
\midrule
\multicolumn{13}{l}{\textbf{$G = 150$}} \\ \cmidrule{1-1}
SACR-A & \multirow{2}{*}{$-0.0118$} & \multirow{2}{*}{0.4007} & 0.3950 & 0.0490 & \multirow{2}{*}{0.0039} & \multirow{2}{*}{0.2564} & 0.2575 & 0.0790 & \multirow{2}{*}{$-0.0044$} & \multirow{2}{*}{0.2093} & 0.2121 & 0.0580 \\
SACR-B &  &  & 0.4000 & 0.0460 &  &  & 0.2648 & 0.0690 &  &  & 0.2088 & 0.0610 \\
CR-A & \multirow{2}{*}{$-0.1739$} & \multirow{2}{*}{1.6398} & 0.8449 & 0.3780 & \multirow{2}{*}{$-0.0197$} & \multirow{2}{*}{1.1102} & 0.1098 & 0.9160 & \multirow{2}{*}{$-0.0158$} & \multirow{2}{*}{0.8286} & 0.2496 & 0.6750 \\
CR-B &  &  & 1.0816 & 0.1720 &  &  & 0.5501 & 0.2690 &  &  & 0.4365 & 0.2460 \\
\midrule
\multicolumn{13}{l}{\textbf{$G = 300$}} \\ \cmidrule{1-1}
SACR-A & \multirow{2}{*}{$-0.0027$} & \multirow{2}{*}{0.2749} & 0.2768 & 0.0420 & \multirow{2}{*}{$-0.0060$} & \multirow{2}{*}{0.1786} & 0.1861 & 0.0500 & \multirow{2}{*}{0.0001} & \multirow{2}{*}{0.1507} & 0.1526 & 0.0540 \\
SACR-B &  &  & 0.2785 & 0.0420 &  &  & 0.1878 & 0.0540 &  &  & 0.1512 & 0.0530 \\
CR-A & \multirow{2}{*}{$-0.1232$} & \multirow{2}{*}{1.5289} & 0.7939 & 0.3820 & \multirow{2}{*}{$-0.0162$} & \multirow{2}{*}{1.0700} & 0.1102 & 0.9090 & \multirow{2}{*}{$-0.0141$} & \multirow{2}{*}{0.7745} & 0.2457 & 0.6600 \\
CR-B &  &  & 0.9732 & 0.1920 &  &  & 0.5055 & 0.3480 &  &  & 0.3950 & 0.2940 \\
\midrule
\multicolumn{13}{l}{\textbf{$G = 600$}} \\ \cmidrule{1-1}
SACR-A & \multirow{2}{*}{0.0057} & \multirow{2}{*}{0.1963} & 0.1954 & 0.0540 & \multirow{2}{*}{0.0008} & \multirow{2}{*}{0.1328} & 0.1303 & 0.0730 & \multirow{2}{*}{0.0002} & \multirow{2}{*}{0.1109} & 0.1080 & 0.0570 \\
SACR-B &  &  & 0.1959 & 0.0550 &  &  & 0.1310 & 0.0750 &  &  & 0.1077 & 0.0600 \\
CR-A & \multirow{2}{*}{$-0.2057$} & \multirow{2}{*}{1.4058} & 0.7510 & 0.3640 & \multirow{2}{*}{0.0509} & \multirow{2}{*}{1.0810} & 0.1177 & 0.9100 & \multirow{2}{*}{0.0352} & \multirow{2}{*}{0.7500} & 0.2561 & 0.6270 \\
CR-B &  &  & 0.8929 & 0.1650 &  &  & 0.4960 & 0.3390 &  &  & 0.3825 & 0.2470 \\
\bottomrule
\end{tabular}
\begin{tablenotes}
\item The table has the same column blocks, targets, cluster grid, Monte Carlo replications, bootstrap draws, and inference procedures as \Cref{tab:mc_point_beta25}; only the Pareto tail index differs.
\end{tablenotes}
\end{threeparttable}
\end{table}

The paired tables isolate the effect of the cluster-size regime. Bias and RMSE are shared across each estimator's analytical and bootstrap rows because the inference procedure does not change the point estimates. In the asymptotically ignorable \(\beta=2.5\) regime, both estimators' RMSEs decline with \(G\), with lower RMSE for SACR. Both bootstrap procedures have rejection frequencies close to the nominal \(5\%\) level throughout the ignorable regime; the larger QR departures occur with analytical standard errors at \(G=150\). Under \(\beta=1\), SACR RMSE continues to fall, whereas conventional RMSE remains substantial. At \(G=600\), SACR-B rejects \(5.5\%\), \(7.5\%\), and \(6.0\%\) for DR, QR, and ER, respectively, all reasonably close to nominal. Its QR rejection rates remain reasonably close to nominal at all three cluster counts. Conventional analytical inference rejects \(36.4\%\), \(91.0\%\), and \(62.7\%\) at \(G=600\). The pairs bootstrap reduces these frequencies to \(16.5\%\), \(33.9\%\), and \(24.7\%\), but substantial size distortions remain.

\subsection{Uniform inference on distributional treatment effects}\label{sec:mc_functional}

The functional experiment studies
\( \displaystyle
\Delta_F(y)=F_{Y\mid T=1}(y)-F_{Y\mid T=0}(y),
\;
\Delta_Q(\tau)=Q_{Y\mid T=1}(\tau)-Q_{Y\mid T=0}(\tau),
\;
\text{and}
\;
\Delta_E(\tau)=\E_\tau(Y\mid T=1)-\E_\tau(Y\mid T=0).
\)
The Distribution Function (DF) is evaluated on the central unconditional population support
\( \displaystyle
\mathcal{Y}_{0.1,0.9}=[Q_Y(0.1),Q_Y(0.9)],
\)
while the Quantile Function (QF) and Expectile Function (EF) are evaluated over \(\tau\in[0.1,0.9]\). In this saturated binary-treatment design, the group DF, quantile, and expectile differences are equivalent to the corresponding fitted DR, QR, and ER treatment-effect functions, including under inverse-cluster-size weighting. The evaluation grids grow at rate \(G^{0.3}\).\footnote{See \Cref{sec:mc_grid_metrics} for a discussion on the calibration.} Uniform inference is bootstrap-based following \citet[Algorithm 3]{chernozhukov-fernandez-melly-2013inference}. Both conventional and SACR uniform bands use the normalised bootstrap IQR studentiser described in \Cref{sec:boot}.

Integrated bias (IBias) is the root mean squared displacement of the mean
estimated path from the true path across grid points; integrated RMSE
(IRMSE) also includes variation across Monte Carlo samples. Avg. Band is
the mean uniform-band half-width, and U-Rej. is the frequency with which
the band fails to cover the complete true path.\footnote{\Cref{sec:mc_grid_metrics}
provides the formulae.} The tables report SACR-B followed by CR-B; each
bootstrap draw is shared across the complete functional grid.

\begin{table}[!htbp]
\centering
\caption{Uniform Monte Carlo results for distributional treatment effects: asymptotically ignorable Pareto regime (\(\beta=2.5\))}
\label{tab:mc_unif_beta25}
\begin{threeparttable}
\setlength{\tabcolsep}{2pt}
\footnotesize
\begin{tabular}{lcccccccccccc}
\toprule
& \multicolumn{4}{c}{DF treatment effect}
& \multicolumn{4}{c}{QF treatment effect}
& \multicolumn{4}{c}{EF treatment effect} \\
\cmidrule(lr){2-5}\cmidrule(lr){6-9}\cmidrule(lr){10-13}
Method
& IBias & IRMSE & Avg. Band & U-Rej.
& IBias & IRMSE & Avg. Band & U-Rej.
& IBias & IRMSE & Avg. Band & U-Rej. \\
\midrule
\multicolumn{13}{l}{\textbf{$G = 150$}} \\ \cmidrule{1-1}
SACR-B & 0.0035 & 0.0810 & 0.2107 & 0.1250 & 0.0204 & 0.2909 & 1.8054 & 0.0470 & 0.0151 & 0.2376 & 0.5191 & 0.1260 \\
CR-B & 0.0050 & 0.0915 & 0.2250 & 0.1450 & 0.0290 & 0.3411 & 1.8382 & 0.0580 & 0.0229 & 0.2720 & 0.5870 & 0.1380 \\
\midrule
\multicolumn{13}{l}{\textbf{$G = 300$}} \\ \cmidrule{1-1}
SACR-B & 0.0012 & 0.0559 & 0.1513 & 0.0910 & 0.0076 & 0.2003 & 1.0373 & 0.0490 & 0.0069 & 0.1621 & 0.3749 & 0.0950 \\
CR-B & 0.0010 & 0.0638 & 0.1640 & 0.1110 & 0.0122 & 0.2352 & 1.0980 & 0.0690 & 0.0093 & 0.1884 & 0.4165 & 0.0970 \\
\midrule
\multicolumn{13}{l}{\textbf{$G = 600$}} \\ \cmidrule{1-1}
SACR-B & 0.0022 & 0.0413 & 0.1084 & 0.0780 & 0.0093 & 0.1445 & 0.6411 & 0.0460 & 0.0089 & 0.1157 & 0.2676 & 0.0820 \\
CR-B & 0.0017 & 0.0476 & 0.1203 & 0.0980 & 0.0083 & 0.1729 & 0.7027 & 0.0620 & 0.0068 & 0.1386 & 0.3049 & 0.0840 \\
\bottomrule
\end{tabular}
\begin{tablenotes}
\item The three column blocks report DF, QF, and EF treatment effects over the domains defined above. The grids use the curvature-based rule described in the text. For \(G=150,300,600\), respectively, the grid sizes are \(L_G^{(F)}=12,15,18\), \(L_G^{(Q)}=43,52,64\), and \(L_G^{(E)}=41,50,61\). Both procedures use normalised bootstrap IQR studentisation.
\end{tablenotes}
\end{threeparttable}
\end{table}

\begin{table}[!htbp]
\centering
\caption{Uniform Monte Carlo results for distributional treatment effects: ULC Pareto regime (\(\beta=1\))}
\label{tab:mc_unif_beta1}
\begin{threeparttable}
\setlength{\tabcolsep}{2pt}
\footnotesize
\begin{tabular}{lcccccccccccc}
\toprule
& \multicolumn{4}{c}{DF treatment effect}
& \multicolumn{4}{c}{QF treatment effect}
& \multicolumn{4}{c}{EF treatment effect} \\
\cmidrule(lr){2-5}\cmidrule(lr){6-9}\cmidrule(lr){10-13}
Method
& IBias & IRMSE & Avg. Band & U-Rej.
& IBias & IRMSE & Avg. Band & U-Rej.
& IBias & IRMSE & Avg. Band & U-Rej. \\
\midrule
\multicolumn{13}{l}{\textbf{$G = 150$}} \\ \cmidrule{1-1}
SACR-B & 0.0020 & 0.0795 & 0.2075 & 0.1210 & 0.0112 & 0.2804 & 1.6193 & 0.0460 & 0.0100 & 0.2295 & 0.5157 & 0.1250 \\
CR-B & 0.0070 & 0.2711 & 0.5444 & 0.1430 & 0.4191 & 1.0659 & 4.7908 & 0.1020 & 0.1859 & 0.7962 & 1.5596 & 0.1630 \\
\midrule
\multicolumn{13}{l}{\textbf{$G = 300$}} \\ \cmidrule{1-1}
SACR-B & 0.0014 & 0.0553 & 0.1490 & 0.0790 & 0.0066 & 0.1948 & 0.9410 & 0.0370 & 0.0063 & 0.1585 & 0.3739 & 0.0890 \\
CR-B & 0.0047 & 0.2636 & 0.4981 & 0.1610 & 0.3774 & 1.0417 & 4.1366 & 0.1170 & 0.1619 & 0.7756 & 1.3952 & 0.1760 \\
\midrule
\multicolumn{13}{l}{\textbf{$G = 600$}} \\ \cmidrule{1-1}
SACR-B & 0.0018 & 0.0392 & 0.1065 & 0.0410 & 0.0064 & 0.1386 & 0.5914 & 0.0410 & 0.0044 & 0.1134 & 0.2673 & 0.0700 \\
CR-B & 0.0075 & 0.2472 & 0.4548 & 0.1590 & 0.3531 & 0.9818 & 3.8343 & 0.1520 & 0.1519 & 0.7222 & 1.2629 & 0.1860 \\
\bottomrule
\end{tabular}
\begin{tablenotes}
\item The table has the same functional targets, supports, grids, and bootstrap procedures as \Cref{tab:mc_unif_beta25}; only the Pareto tail index differs.
\end{tablenotes}
\end{threeparttable}
\end{table}

The functional results reinforce the pointwise comparison. In the \(\beta=2.5\) regime, SACR-B QF rejection frequencies remain close to \(5\%\); DF and EF distortions at smaller \(G\) diminish as the number of clusters increases, reaching \(7.8\%\) and \(8.2\%\), respectively, at \(G=600\). Under \(\beta=1\), SACR-B integrated RMSE declines with \(G\), and its uniform rejection frequencies at \(G=600\) are \(4.1\%\), \(4.1\%\), and \(7.0\%\) for DF, QF, and EF, all reasonably close to nominal. The corresponding CR-B frequencies are \(15.9\%\), \(15.2\%\), and \(18.6\%\), alongside substantially larger integrated RMSE and wider bands. Thus size adjustment improves both estimation and uniform inference under severe cluster-size heterogeneity, with the remaining SACR distortions concentrated at smaller cluster counts.

A supplementary empirical illustration in \Cref{sect:empirics} revisits the municipal-school setting of
\citet{Akhtari2022}, where students are clustered within municipalities of
substantially unequal size. The empirical illustration reports pointwise coefficient
comparisons and distributional effects, together with the identifying
assumptions and implementation details. The application illustrates the
consequences of cluster weighting in practice.

\section{Conclusion}\label{sect:conclusion}

Unignorably large clusters can invalidate conventional inference for \(Z\)-estimators. This paper develops pointwise and functional inference for cluster-balanced estimating equations under independent clusters and unrestricted within-cluster dependence, subject to the stated conditions. Building on the SACR reweighting of \citet{chiang-sasaki-wang-2025cluster}, the theory covers smooth and non-smooth scores, with distribution, quantile, and expectile regression as applications. Pointwise normality permits a drifting covariance sequence, whereas functional weak convergence and bootstrap inference require a stable covariance kernel. A common cluster-level zero preserves the structural target under size adjustment, while separation of the reweighted moments ensures identification. Without that common zero, cluster balancing can change the estimand; the appropriate population representation then depends on the research question. Monte Carlo experiments illustrate pointwise and uniform inference when the competing moments share a common structural zero.

\medskip

\noindent \texttt{Acknowledgments:}
This paper grew out of the first author's master's thesis.

\medskip

\noindent \texttt{Declaration of AI use:}
During the preparation of this manuscript, the authors used OpenAI's ChatGPT and Codex for research assistance. The authors reviewed and edited all AI-assisted output and take full responsibility for the content of the manuscript.

\printbibliography[title={References}]
\end{refsection}

\clearpage
\hypersetup{pageanchor=false}
\begin{refsection}
\appendix
\setcounter{page}{1}
\setcounter{table}{0}
\setcounter{figure}{0}
\setcounter{equation}{0}
\renewcommand{\thepage}{S-\arabic{page}}
\renewcommand{\thetable}{S.\arabic{table}}
\renewcommand{\thefigure}{S.\arabic{figure}}
\renewcommand{\thesection}{S.\arabic{section}}
\renewcommand{\thesubsection}{\thesection.\arabic{subsection}}
\renewcommand{\theequation}{\thesection.\arabic{equation}}

\begin{center}
{\LARGE\bfseries
Supplement to ``\(Z\)-Estimation under Unignorably Large Clusters''\par}

\vspace{0.5cm}

{\large
Agossou Michel Yomakou\textsuperscript{1}\par
Abderrahim Taamouti\textsuperscript{2}\par
Emmanuel S. Tsyawo\textsuperscript{3}\par}

\vspace{0.5cm}

\textsuperscript{1}AIRESS \& FGSES, Universit\'e Mohammed VI Polytechnique,
Morocco; \href{mailto:michel.yomakou@um6p.ma}{michel.yomakou@um6p.ma}\par
\textsuperscript{2}Department of Economics, University of Liverpool Management
School; \href{mailto:abderrahim.taamouti@liverpool.ac.uk}{abderrahim.taamouti@liverpool.ac.uk}\par
\textsuperscript{3}Department of Economics, Finance and Legal Studies,
Culverhouse College of Business, University of Alabama;
\href{mailto:estsyawo@gmail.com}{estsyawo@gmail.com}
\end{center}

\thispagestyle{plain}

The Supplementary Appendix contains the target-invariance proof (\Cref{app:target_invariance}), proofs of the general \(Z\)-SACR theory (\Cref{app:proofs_gen}), and proofs and supporting lemmas for the model-specific results (\Cref{app:proofs_specific}). These are followed by the distribution-to-expectile map and its proof (\Cref{app:expectile_map}). \Cref{app:mc_supplement} collects Monte Carlo implementation details and supplementary experiments, while \Cref{sect:empirics} presents the empirical illustration. \Cref{app:cluster_size_cases} concludes with a classification distinguishing cluster-size informativeness from asymptotic size influence.

\section{Proof of \Cref{prop:inv_id}}\label{app:target_invariance}
By the maintained cluster-level exogeneity condition, for every cluster \(g\),
\[
\E_G[\bar\psi(\bm W_g;\theta_o)]
=0,
\]
and therefore \(\E_G[\sum_{i\in\G_g}\psi(W_i;\theta_o)]=N_g\E_G[\bar\psi(\bm W_g;\theta_o)]=0\). Since \(\omega_g\) is deterministic, summing the cluster-level equalities with these weights gives
\[
\E_G\!\left[\frac{1}{G}\sum_{i=1}^N\omega_{g(i)}\psi(W_i;\theta_o)\right]
=
\frac{1}{G}\sum_{g=1}^G\omega_gN_g\E_G[\bar\psi(\bm W_g;\theta_o)]
=0.
\]
This establishes the population zero for every deterministic weighting scheme. Since no common zero is imposed for \(\theta\ne\theta_o\), uniqueness away from the truth is supplied by the weighted triangular-array separation condition. \qed

\section{Proofs for the General \(Z\)-SACR Theory}\label{app:proofs_gen}
The following supplies three reusable Z-SACR-compatible results based on theoretical results established in \citet{hansen2019asymptotic}.
\begin{lemma}[Cluster-average moment transfer and pointwise limits]
\label{lem:cluster_average_limits}
For each \(G\), let \(X_{gi,G}\in\mathbb R^p\), \(i=1,\ldots,N_g\), and define
\( \displaystyle 
\bar X_{g,G}:=\frac{1}{N_g}\sum_{i=1}^{N_g}X_{gi,G}.
\)
For \(r\geq1\), suppose that
\begin{equation}\label{eq:cluster_average_ui_primitive}
\lim_{M\to\infty}
\sup_{G\geq1}\sup_{g\leq G}\sup_{i\leq N_g}
\E_G\!\left[
\|X_{gi,G}\|^r\indicator{\|X_{gi,G}\|>M}
\right]=0.
\end{equation}
Then
\begin{enumerate}[label=(\alph*)]
\item\label{lem:cluster_average_ui}
\( \displaystyle 
\lim_{M\to\infty}
\sup_{G\geq1}\sup_{g\leq G}
\E_G\!\left[
\|\bar X_{g,G}\|^r\indicator{\|\bar X_{g,G}\|>M}
\right]=0;
\)
\item\label{lem:cluster_average_wlln} 
\( \displaystyle 
\frac1G\sum_{g=1}^G
\{\bar X_{g,G}-\E_G[\bar X_{g,G}]\}
\xrightarrow{p}0;
\)
if the cluster vectors are independent with \(r=1\); and 
\item\label{lem:cluster_average_op}
for
\( \displaystyle 
\Omega_{X,G}:=\frac1G\sum_{g=1}^G\V_G(\bar X_{g,G}),
\)
one has \(\sup_G\|\Omega_{X,G}\|<\infty\) and
\( \displaystyle 
\frac1{\sqrt G}\sum_{g=1}^G
\{\bar X_{g,G}-\E_G[\bar X_{g,G}]\}=\Op(1).
\) 
if the cluster vectors are independent with \(r=2\). If, in addition, \(\lambda_{\min}(\Omega_{X,G})\geq\lambda>0\), then
\[
\Omega_{X,G}^{-1/2}\frac1{\sqrt G}\sum_{g=1}^G
\{\bar X_{g,G}-\E_G[\bar X_{g,G}]\}
\xrightarrow{d}\mathcal N(0,I_p).
\]
\end{enumerate}
\end{lemma}

\noindent
\textit{Proof.}
Part~\ref{lem:cluster_average_ui}: \qquad This part follows from the argument of
\citet[Lemma~1]{hansen2019asymptotic}, applied uniformly over the triangular array. For the remaining parts, treat each
\(\bar X_{g,G}\) as one observation in a singleton effective cluster. 

\noindent Part~\ref{lem:cluster_average_wlln}: \qquad
The
effective total sample size is \(G\), and every effective cluster size equals
one. Thus the cluster-size condition of \citet[Theorem~1]{hansen2019asymptotic}
is \(G^{-1}\to0\), which, together with
part~\ref{lem:cluster_average_ui} for \(r=1\), proves
part~\ref{lem:cluster_average_wlln}. 

\noindent Part~\ref{lem:cluster_average_op}: \qquad
For \(r=2\), the two conditions in
\citet[Assumption~2]{hansen2019asymptotic} reduce to
\[
\frac{(\sum_{g=1}^G1^2)^{2/2}}{G}=1
\quad \text{and} \quad
\max_{g\leq G}\frac{1^2}{G}=\frac1G\to0.
\]
Part~\ref{lem:cluster_average_ui} supplies the squared uniform-integrability
condition in that paper. It also implies uniformly bounded second moments of the cluster
averages, so \(\sup_G\|\Omega_{X,G}\|<\infty\) and \( \displaystyle 
\frac1{\sqrt G}\sum_{g=1}^G
\{\bar X_{g,G}-\E_G[\bar X_{g,G}]\}=\Op(1)
\) by Chebyshev's inequality. Under the stated lower eigenvalue
bound, \citet[Theorem~2]{hansen2019asymptotic} gives the CLT.
\qed

\subsection{Proof of \Cref{Thm:uwlln}}

\proofpart{Proof of \Cref{Thm:uwlln}\ref{thm:uwlln_a}}
Let
\[
\bar\psi(\bm{W}_g;\theta):=\frac{1}{N_g}\sum_{i\in\G_g}\psi(W_i;\theta).
\]
Then \(\widehat{\Prob}_G \bar\psi(\bm{W}_g;\theta)=G^{-1}\sum_{g=1}^G \bar\psi(\bm{W}_g;\theta)\) and \(\Prob_G \bar\psi(\bm{W}_g;\theta)=G^{-1}\sum_{g=1}^G\E_G[\bar\psi(\bm{W}_g;\theta)]\).

For each fixed \(\theta\in\Theta\), apply
\Cref{lem:cluster_average_limits}\ref{lem:cluster_average_wlln} to
\(X_{gi,G}=\psi(W_i;\theta)\). Independence follows from
\Cref{ass:sampling}, and the primitive uniform-integrability condition is
\Cref{ass:uwlln}\ref{ass:uwlln_bd}. Thus, pointwise convergence holds:
$$
\frac{1}{G}\sum_{g=1}^G \big( \bar\psi(\bm{W}_g;\theta) - \E_G[\bar\psi(\bm{W}_g;\theta)]\big) \xrightarrow{p} 0.
$$
Because \(\Theta\) is compact, it is totally bounded under the Euclidean metric. This pointwise convergence and the stochastic equicontinuity imposed in \Cref{ass:uwlln}\ref{ass:uwlln_se} therefore verify, respectively, P-WCON and SE in \citet[Theorem~1(a)]{andrews-1992generic}. Applying that result componentwise to the finite-dimensional process \(\mathbb U_G\) gives
$$
\sup_{\theta \in \Theta} \left\|\frac{1}{G}\sum_{g=1}^G \big( \bar\psi(\bm{W}_g;\theta) - \E_G[\bar\psi(\bm{W}_g;\theta)]\big)\right\| \xrightarrow{p} 0.
$$
Since \((\widehat{\Prob}_G-\Prob_G)\bar\psi(\bm{W}_g;\theta)=G^{-1}\sum_{g=1}^G\{\bar\psi(\bm{W}_g;\theta)-\E_G[\bar\psi(\bm{W}_g;\theta)]\}\), this establishes \Cref{eq:unfm_conv}.

\proofpart{Proof of \Cref{Thm:uwlln}\ref{thm:uwlln_b}}
Define the population cluster-average moment map
\[
\Psi_G(\theta):=\Prob_G\bar\psi(\bm{W}_g;\theta).
\]
By the triangle inequality,
\begin{align*}
\|\Psi_G(\widehat{\theta}^{SACR})\|
&=\big\|\Prob_G\bar\psi(\bm{W}_g;\widehat{\theta}^{SACR})\big\|\\
&\leq
\big\|(\widehat{\Prob}_G-\Prob_G)\bar\psi(\bm{W}_g;\widehat{\theta}^{SACR})\big\|
+\big\|\widehat{\Prob}_G\bar\psi(\bm{W}_g;\widehat{\theta}^{SACR})\big\|\\
&\leq
\sup_{\theta\in\Theta}\big\|(\widehat{\Prob}_G-\Prob_G)\bar\psi(\bm{W}_g;\theta)\big\|
+\big\|\widehat{\Prob}_G\bar\psi(\bm{W}_g;\widehat{\theta}^{SACR})\big\|.
\end{align*}
The first term is $o_p(1)$ by part \Cref{thm:uwlln_a} above, and the second term is $o_p(1)$ by the approximate estimating equation condition in part \ref{thm:uwlln_b} of the statement of the theorem. Hence
\[
\|\Psi_G(\widehat{\theta}^{SACR})\|=o_p(1).
\]

It remains to translate this approximate population zero into parameter consistency. Fix any $\epsilon>0$ and set
\[
c_\epsilon:=\liminf_{G\to\infty}\inf_{\theta\in\Theta:\|\theta-\theta_o\|\geq\epsilon}\|\Psi_G(\theta)\|.
\]
By the separation condition in \Cref{ass:pop_identification}\ref{ass:pop_id_sep}, $c_\epsilon>0$. Therefore, for all sufficiently large $G$,
\[
\inf_{\theta\in\Theta:\|\theta-\theta_o\|\geq\epsilon}\|\Psi_G(\theta)\|\geq c_\epsilon/2.
\]
On the event $\{\|\widehat{\theta}^{SACR}-\theta_o\|\geq\epsilon\}$, this implies
\(
\|\Psi_G(\widehat{\theta}^{SACR})\|\geq c_\epsilon/2.
\)
Consequently,
\[
\Probbr{\|\widehat{\theta}^{SACR}-\theta_o\|\geq\epsilon}
\leq
\Probbr{\|\Psi_G(\widehat{\theta}^{SACR})\|\geq c_\epsilon/2}\to 0.
\]
Since $\epsilon>0$ is arbitrary, $\widehat{\theta}^{SACR}\xrightarrow{p}\theta_o$.

\qed

\subsection{Proof of \Cref{Thm:normality}}
\proofpart{Proof of \Cref{Thm:normality}\ref{thm:normality_a}}
By \Cref{Thm:uwlln}\ref{thm:uwlln_b},
\(\widehat\theta^{SACR}\to_p\theta_o\). Define
\[
\mathcal E_G
:=
\sqrt G\,\widehat{\Prob}_G
\bar\psi(\bm W_g;\widehat\theta^{SACR})
=o_p(1)
\]
and, on the event that the line segment joining
\(\widehat\theta^{SACR}\) and \(\theta_o\) lies in the differentiability
neighbourhood, define
\[
\widetilde H_G
:=
\int_0^1
H_G\!\left(
\theta_o+u(\widehat{\theta}^{SACR}-\theta_o)
\right)\,du.
\]
The population zero in
\Cref{ass:pop_identification}\ref{ass:pop_id_zero} and the fundamental
theorem of calculus give, on the event that \(\widetilde H_G\) is
invertible, the exact identity
\begin{equation}\label{eq:pointwise_headline_decomp}
\sqrt G(\widehat\theta^{SACR}-\theta_o)
=-\widetilde H_G^{-1}\left[
\mathbb G_G(\theta_o)
+\underbrace{\{\mathbb G_G(\widehat\theta^{SACR})
-\mathbb G_G(\theta_o)\}}_{\mathcal R_G}
-\mathcal E_G
\right],
\end{equation}
which holds thanks to consistency and interiority in
\Cref{ass:linearisation}\ref{ass:lin_int} with
probability approaching one. In addition, local stochastic equicontinuity in
\Cref{ass:stch_equ} gives \(\mathcal R_G=o_p(1)\), while
\Cref{ass:linearisation}\ref{ass:lin_foc} gives
\(\mathcal E_G=o_p(1)\).

Consistency and
\Cref{ass:linearisation}\ref{ass:lin_cont} also give
\(\widetilde H_G-H_G=o_p(1)\). The uniform lower singular-value bound in
\Cref{ass:norm_mmt_general}\ref{ass:nonsingular} therefore makes
\(\widetilde H_G\) and \(H_G\) uniformly invertible with probability
approaching one, and the inverse identity gives
\(\widetilde H_G^{-1}-H_G^{-1}=o_p(1)\). Moreover,
\Cref{lem:cluster_average_limits}\ref{lem:cluster_average_op}, applied to
\(X_{gi,G}=\psi(W_i;\theta_o)\), gives
\(\mathbb G_G(\theta_o)=\Op(1)\). Consequently,
\[
\big(\widetilde H_G^{-1}-H_G^{-1}\big)\mathbb G_G(\theta_o)=o_p(1).
\]
Equation~\eqref{eq:pointwise_headline_decomp} therefore gives the
asymptotically linear representation
\begin{equation}\label{eqn:lin_rep}
	\sqrt{G}\big(\widehat{\theta}^{SACR}-\theta_o\big)
		= -H_G^{-1}\mathbb G_G(\theta_o)+o_p(1)
\end{equation}
as claimed.

\proofpart{Proof of \Cref{Thm:normality}\ref{thm:normality_b}}
Let \(Z_{g,G}:=\bar\psi(\bm W_g;\theta_o)\). After SACR reweighting, this array has one summand per cluster:
\[
\sqrt{G}\,\widehat{\Prob}_G \bar\psi(\bm{W}_g;\theta_o)
= \frac{1}{\sqrt{G}}\sum_{g=1}^G Z_{g,G}.
\]
Independence holds by \Cref{ass:sampling}, and the zero-mean condition follows because \Cref{ass:pop_identification}\ref{ass:pop_id_zero} implies $\E_G[Z_{g,G}]=\E_G[\bar\psi(\bm W_g;\theta_o)]=0$ for every $g$. The individual squared uniform-integrability condition in \Cref{ass:norm_mmt_general}\ref{ass:sup_finite}, together with the lower covariance bound in \Cref{ass:norm_mmt_general}\ref{ass:nonsingular}, verifies \Cref{lem:cluster_average_limits}\ref{lem:cluster_average_op} for \(X_{gi,G}=\psi(W_i;\theta_o)\). Therefore
\[
\Omega_G^{-1/2}\sqrt{G}\,\widehat{\Prob}_G \bar\psi(\bm{W}_g;\theta_o)
\xrightarrow{d} \mathcal{N}(0,I).
\]
To combine this standardised CLT with the asymptotically linear
representation \eqref{eqn:lin_rep}, define
\[
\mathbf{V}_G := H_G^{-1}\Omega_G(H_G^{-1})'.
\]
It remains only to justify standardisation by the drifting matrix sequence.
The uniform-integrability condition in
\Cref{ass:uwlln}\ref{ass:uwlln_bd} implies
\(\sup_{G,\theta}\|\Prob_G\bar\psi(\bm W_g;\theta)\|<\infty\).
Interiority permits a fixed \(r>0\) such that
\(\theta_o+rv\in\Theta\) for every unit vector \(v\), and
\Cref{ass:linearisation}\ref{ass:lin_cont} permits this \(r\) to be chosen
so that \(\sup_{\|\theta-\theta_o\|\leq r}
\|H_G(\theta)-H_G\|\leq1\) for all sufficiently large \(G\). Define \(M_G(\theta):=\Prob_G\bar\psi(\bm W_g;\theta)\).
Applying the fundamental theorem of calculus along
\(t\mapsto\theta_o+tv\) gives
\[
M_G(\theta_o+rv)-M_G(\theta_o)
=\int_0^r H_G(\theta_o+tv)v\,dt.
\]
Consequently,
\begin{align*}
    \|M_G(\theta_o+rv)-M_G(\theta_o)-rH_Gv\| &= \Big\| \int_0^r H_G(\theta_o+tv)v\,dt - \int_0^r H_Gv\,dt \Big\| \\
&\leq\int_0^r\|H_G(\theta_o+tv)-H_G\|\,dt
\leq r.
\end{align*}
Taking \(K<\infty\) to be a uniform bound on
\(\|M_G(\theta)\|\), the triangle inequality yields
\[
r\|H_Gv\|
\leq
\|\Prob_G\bar\psi(\bm W_g;\theta_o+rv)
-\Prob_G\bar\psi(\bm W_g;\theta_o)\|+r
\leq2K+r.
\]
Thus \(\sup_G\|H_G\|<\infty\). Moreover,
\( \displaystyle
\lambda_{\min}(\mathbf V_G)
\geq\frac{\lambda_{\min}(\Omega_G)}{\|H_G\|^2},
\)
so \(\inf_G\lambda_{\min}(\mathbf V_G)>0\) and
\(\sup_G\|\mathbf V_G^{-1/2}\|<\infty\). Let
\( \displaystyle
U_G:=\Omega_G^{-1/2}\sqrt G\,\widehat{\Prob}_G
\bar\psi(\bm W_g;\theta_o),
\quad \text{and} \quad
Q_G:=\mathbf V_G^{-1/2}H_G^{-1}\Omega_G^{1/2}.
\)
Then \(U_G\xrightarrow{d}\mathcal N(0,I)\) and \(Q_GQ_G'=I\).
The linear representation therefore yields
\begin{align*}
    \mathbf V_G^{-1/2}\sqrt G(\widehat\theta^{SACR}-\theta_o) 
    &= -\mathbf V_G^{-1/2}H_G^{-1}\Omega_G^{1/2}\Omega_G^{-1/2}\sqrt{G}\,\widehat{\Prob}_G \bar\psi(\bm{W}_g;\theta_o)+o_p(1) \\
    &=-Q_GU_G+o_p(1).
\end{align*}
Every \(Q_G\) is orthogonal, so rotational
invariance gives the studentised form
\[
\mathbf{V}_G^{-1/2}\sqrt{G}\big(\widehat{\theta}^{SACR}-\theta_o\big)
\xrightarrow{d}\mathcal{N}(0,I).
\]

\qed

\subsection{Proof of \Cref{lem:functional_ep}}
\noindent Part (a): \qquad
Fix a coordinate \(j\). By \Cref{ass:uwlln_F}\ref{ass:functional_manageability}, the normalised full cluster array is manageable with envelope \(F_{Gg,j}\). Its joint almost-measurable-Suslin parametrisation justifies the symmetrisation and interchange of expectations, as in \citet[Lemma~1]{kosorok-2003}. Pollard's maximal inequality \citep[equation~(7.10)]{pollard-1990}, with \(p=2\), and the cluster-envelope moment bound supplied by \Cref{lem:cluster_average_limits}\ref{lem:cluster_average_ui}, applied to the squared individual envelopes under \Cref{ass:uwlln_F}\ref{ass:functional_global_envelope}, give
\[
\E_G^*\!\left[
\sup_{\tau\in\T,\,t\in\Theta}
\left|
\sum_{g=1}^G
\big(Z_{Gg,j}(\tau,t)-\E_G Z_{Gg,j}(\tau,t)\big)
\right|^2
\right]
\leq C\sum_{g=1}^G\E_G F_{Gg,j}^2 = \frac{C}{G}\sum_{g=1}^G\E_G \bar F_{Gg,j}^2 = \mathcal{O}(1).
\]
Consequently, \( \displaystyle \sup_{\tau\in\T,\,t\in\Theta}
\bigg|
\frac{1}{G}\sum_{g=1}^G
\big\{ \bar\psi_{\tau,j}(\bm W_g;t) - \E_G \bar\psi_{\tau,j}(\bm W_g;t) \big\}
\bigg| = \Op(G^{-1/2}) = o_p(1) \). Taking the maximum over the finitely many
coordinates and observing that every admissible path is contained in the
joint index set proves part~\ref{lem:functional_uwlln}.

\noindent Part~\ref{lem:functional_se}:
Cluster-array manageability is imposed directly in
\Cref{ass:uwlln_F}\ref{ass:functional_manageability}.
Fix a coordinate \(j\), and write
\[
d_{Gg,j}(\tau,t)
:=\bar\psi_{\tau,j}(\bm W_g;t)
-\bar\psi_{\tau,j}(\bm W_g;\theta_o(\tau)).
\]
Define
\[
\mathcal I_\delta
:=\{(\tau,t)\in\T\times\Theta:
\|t-\theta_o(\tau)\|\leq\delta\},
\qquad
R_{G,j}^{2}(\delta)
:=\sup_{(\tau,t)\in\mathcal I_\delta}
\frac1G\sum_{g=1}^{G}d_{Gg,j}(\tau,t)^2.
\]
Manageability and the joint AMS property follow from
\Cref{ass:uwlln_F}\ref{ass:functional_manageability}, using stability of
manageability under differences. By \Cref{lem:cluster_average_limits}\ref{lem:cluster_average_ui}, applied to the squared individual envelopes, \Cref{ass:uwlln_F}\ref{ass:functional_global_envelope} implies uniform integrability of the squared cluster envelopes, supplying the required envelope moment and Lindeberg bounds. Applying the truncation
and squared-increment argument in the proof of
\citet[Theorem~10.6, equations~(10.8)--(10.9)]{pollard-1990} gives,
for each fixed \(\delta>0\),
\begin{equation}\label{eq:local_squared_increment_wlln}
\sup_{(\tau,t)\in\mathcal I_\delta}
\left|
\frac1G\sum_{g=1}^{G}
\big(d_{Gg,j}(\tau,t)^2
-\E_G[d_{Gg,j}(\tau,t)^2]\big)
\right|=o_p(1).
\end{equation}
Here Lindeberg truncation permits application of the squared-process
maximal inequality without imposing fourth moments on the original
envelope. Consequently,
\Cref{ass:uwlln_F}\ref{ass:functional_local_L2} implies
\begin{equation}\label{eq:local_empirical_radius}
\lim_{\delta\downarrow0}\limsup_{G\to\infty}
\Probbr{R_{G,j}(\delta)>\varepsilon}=0,
\qquad \varepsilon>0.
\end{equation}
The local maximal-inequality argument in the same proof of
\citet[Theorem~10.6]{pollard-1990} therefore yields
\[
\lim_{\delta\downarrow0}\limsup_{G\to\infty}
\Probbr{
\sup_{(\tau,t)\in\mathcal I_\delta}
\left|
\frac1{\sqrt G}\sum_{g=1}^{G}
\big(d_{Gg,j}(\tau,t)-\E_Gd_{Gg,j}(\tau,t)\big)
\right|>\varepsilon}=0.
\]
Taking the maximum over the finitely many coordinates and restricting the
supremum to admissible parameter paths proves
part~\ref{lem:functional_se}. In particular,
\(R_{G,j}(\delta_G)=o_p(1)\) for every deterministic sequence
\(\delta_G\downarrow0\).
\qed

\subsection{Proof of \Cref{Thm:weak_convergence}}

\noindent Part~\ref{thm:weak_path}:

The proof is organised around the exact functional counterpart of the
pointwise decomposition in \eqref{eq:pointwise_headline_decomp}. Define the scaled
approximate-zero process
\[
\mathcal E_G
:=
\sqrt G\,\widehat{\Prob}_G
\bar\Psi(\bm W_g;\widehat\theta)
=o_p(1)
\qquad\text{in }\ell^\infty(\T)
\]
and, whenever the line segments joining \(\theta_o(\tau)\) and
\(\widehat\theta(\tau)\) lie in the differentiability neighbourhood, let
\[
\widetilde H_G(\tau)
:=
\int_0^1
\frac{\partial}{\partial t'}
\Prob_G\bar\psi_\tau(\bm W_g;t)
\bigg|_{t=\theta_o(\tau)+u\{\widehat\theta(\tau)-\theta_o(\tau)\}}
\,du.
\]
The population zero in
\Cref{ass:pop_identification_F}\ref{ass:functional_zero} and the
fundamental theorem of calculus then give the exact identity
\[
\mathcal E_G
=
\mathbb G_G(\theta_o)
+\{\mathbb G_G(\widehat\theta)-\mathbb G_G(\theta_o)\}
+\widetilde H_G\sqrt G(\widehat\theta-\theta_o).
\]
Consequently, on the event that \(\widetilde H_G\) is uniformly invertible,
\begin{equation}\label{eq:functional_headline_decomp}
\sqrt G(\widehat\theta-\theta_o)
=-\widetilde H_G^{-1}\left[
\mathbb G_G(\theta_o)
+\underbrace{\{\mathbb G_G(\widehat\theta)-\mathbb G_G(\theta_o)\}}_{\mathcal R_G}
-\mathcal E_G
\right].
\end{equation}
The first two steps below verify the event and dispatch the two remainder
terms. The weak limit is established only afterwards for the surviving
first-order term.

\proofstep{1}{Establish uniform consistency of the estimator}
The argument is the sup-norm analogue of
\Cref{Thm:uwlln}\ref{thm:uwlln_b}. The functional UWLLN in
\Cref{lem:functional_ep}\ref{lem:functional_uwlln}, together with the
approximate empirical zero \eqref{eq:functional_foc_rate}, implies
\[
\big\|\Prob_G\bar\Psi(\bm W_g;\widehat\theta)\big\|_{\T}
\leq
\sup_{\theta\in\Theta_{\T}}
\big\|(\widehat{\Prob}_G-\Prob_G)\bar\Psi(\bm W_g;\theta)\big\|_{\T}
+\big\|\widehat{\Prob}_G\bar\Psi(\bm W_g;\widehat\theta)\big\|_{\T}
=o_p(1).
\]
By arguments analogous to the concluding part of the proof of \Cref{Thm:uwlln}\ref{thm:uwlln_b}, the functional separation condition in
\Cref{ass:pop_identification_F}\ref{ass:functional_sep}
therefore yields
\[
\|\widehat\theta-\theta_o\|_{\T}=o_p(1).
\]
Indeed, failure of this conclusion along a subsequence would place
\(\widehat\theta\), with non-vanishing probability, in a set on which
functional separation bounds the population moment uniformly away from
zero, contradicting the preceding display. 

\proofstep{2}{Validate the decomposition \eqref{eq:functional_headline_decomp} and dispatch its degenerate terms}
Uniform consistency, uniform interiority in
\Cref{ass:linearisation_F}\ref{ass:functional_lin_int}, and uniform
differentiability in
\Cref{ass:linearisation_F}\ref{ass:functional_lin_cont} imply that the
line segments used to define \(\widetilde H_G\) lie in the differentiability
neighbourhood with probability approaching one. Thus
\eqref{eq:functional_headline_decomp} is an exact identity on such an event.
The same conditions give
\[
\|\widetilde H_G-H\|_{\T}=o_p(1).
\]
The local stochastic equicontinuity result in
\Cref{lem:functional_ep}\ref{lem:functional_se}, evaluated at the uniformly
consistent path, dispatches the remainder in
\eqref{eq:functional_headline_decomp}:
\[
\|\mathcal R_G\|_{\T}
=o_p(1)
\]
The approximate-zero error satisfies
\(\|\mathcal E_G\|_{\T}=o_p(1)\) directly by
\eqref{eq:functional_foc_rate}. Finally, the uniform lower singular-value
bound on \(H\), together with
\(\|\widetilde H_G-H\|_{\T}=o_p(1)\) (\Cref{ass:linearisation_F}\ref{ass:functional_lin_cont}), ensures that
\(\widetilde H_G\) is uniformly invertible and
\(\|\widetilde H_G^{-1}\|_{\T}=\Op(1)\). Hence the exact decomposition reduces to
\begin{equation}\label{eq:functional_surviving_score}
\sqrt G(\widehat\theta-\theta_o)(\cdot)
=
-\widetilde H_G(\cdot)^{-1}\mathbb G_G(\theta_o)(\cdot)
+o_p(1)
\quad\text{in }\ell^\infty(\T).
\end{equation}

\proofstep{3}{Establish weak convergence of the surviving score process}
For each \(a\in\R^{p_\theta}\), apply \citet[Theorem~1]{kosorok-2003},
equivalently \citet[Theorem~11.16]{kosorok-2007}, to the scalar array
\[
X_{Gg}^{a}(\tau)
:=
G^{-1/2}a'\bar\psi_\tau(\bm W_g;\theta_o(\tau)).
\]
Finite sums and nonnegative rescalings of manageable arrays are manageable
by \citet[Section~5]{pollard-1990}, with envelope
\(\sum_j|a_j|F_{Gg,j}\). The truth-indexed vector array is almost
measurable Suslin by
\Cref{ass:uwlln_F}\ref{ass:functional_manageability}. Its common Suslin
approximating sets also serve each scalar projection, since
\(\sum_g|a'(z_g-\widetilde z_g)|^2
\leq\|a\|^2\sum_g\|z_g-\widetilde z_g\|^2\), and scalar projection
preserves joint measurability on those sets.
Condition B follows from
\Cref{ass:norm_mmt_general_F}\ref{cond:functional_cov_kernel},
condition E is
\Cref{ass:norm_mmt_general_F}\ref{cond:functional_cov_metric},
and the envelope bound and Lindeberg conditions C and D follow from
\Cref{ass:uwlln_F}\ref{ass:functional_global_envelope} via
\Cref{lem:cluster_average_limits}\ref{lem:cluster_average_ui}, applied to the squared individual envelopes, since the truth-indexed class is a subclass of the full class. Thus every scalar
projection is asymptotically tight.

For joint finite-dimensional convergence, fix
\(\tau_1,\ldots,\tau_k\in\T\) and stack the individual scores as
\[
X_{gi,G}:=
\bigl(
\psi_{\tau_1}(W_i;\theta_o(\tau_1))',\ldots,
\psi_{\tau_k}(W_i;\theta_o(\tau_k))'
\bigr)' .
\]
The corresponding cluster average \(\bar X_{g,G}\) stacks the cluster
scores at these indices. Since
\(\|X_{gi,G}\|^2\leq k\sum_{j=1}^{p_\theta}F_j(W_i)^2\),
\Cref{ass:uwlln_F}\ref{ass:functional_global_envelope} implies the
primitive squared uniform-integrability condition
\eqref{eq:cluster_average_ui_primitive} with \(r=2\). Moreover,
\Cref{ass:pop_identification_F}\ref{ass:functional_zero} gives
\(\E_G[\bar X_{g,G}]=0\), and \Cref{ass:sampling} gives independence
across \(g\). Let \(\Omega_{k,G}\) denote the covariance matrix of
\(G^{-1/2}\sum_g\bar X_{g,G}\). Its blocks are
\(\Omega_G(\tau_\ell,\tau_m)\), so
\Cref{ass:norm_mmt_general_F}\ref{cond:functional_cov_kernel} gives
\(\Omega_{k,G}\to\Omega_k\).

For any fixed \(b\in\R^{kp_\theta}\), if
\(b'\Omega_kb>0\), then \(b'\Omega_{k,G}b\) is eventually bounded away
from zero and \Cref{lem:cluster_average_limits}\ref{lem:cluster_average_op},
applied to the scalar projection \(b'X_{gi,G}\), gives the standardised
CLT. If \(b'\Omega_kb=0\), Chebyshev's inequality instead gives
\(G^{-1/2}\sum_g b'\bar X_{g,G}\xrightarrow{p}0\). The Cram\'er--Wold
device therefore yields convergence of the stacked cluster-score vector to
\(\mathcal N(0,\Omega_k)\). Because \(p_\theta\) is finite, coordinatewise tightness and joint
finite-dimensional convergence give
\begin{equation}\label{eq:functional_score_fclt}
\mathbb G_G(\theta_o)
\rightsquigarrow \mathbb G_0
\quad\text{in }\ell^\infty(\T),
\end{equation}
with covariance kernel \(\Omega\). For every scalar projection, the same
theorem also gives total boundedness in its limiting semimetric and a
uniformly continuous version of the projected Gaussian limit. This proves
weak convergence of the sole surviving first-order term in
\eqref{eq:functional_headline_decomp}.

\proofstep{4}{Conclude from the decomposition}
The score functional CLT in \eqref{eq:functional_score_fclt} implies
\(\|\mathbb G_G(\theta_o)\|_{\T}=\Op(1)\). The inverse identity and the
uniform convergence established in Step~2 therefore give
\[
\big\|
\big(\widetilde H_G^{-1}-H^{-1}\big)\mathbb G_G(\theta_o)
\big\|_{\T}
=o_p(1).
\]
It follows from \eqref{eq:functional_surviving_score} that
\[
\sqrt G(\widehat\theta-\theta_o)(\cdot)
=-H(\cdot)^{-1}\mathbb G_G(\theta_o)(\cdot)+o_p(1).
\]
The continuous mapping theorem applied to
\eqref{eq:functional_score_fclt} now yields
\[
\sqrt{G}\,\bigl(\widehat\theta-\theta_o\bigr)(\cdot)
\rightsquigarrow -H(\cdot)^{-1}\mathbb G_0(\cdot)
\quad\text{in }\ell^\infty(\T).
\]

\noindent Part~\ref{thm:weak_scalar}: Since \(\widehat\eta=\Phi_\varphi(\widehat\theta)\), the ordinary functional delta method of
\citet[Lemma~B.1]{chernozhukov-fernandez-melly-2013inference}, applied to
\Cref{Thm:weak_convergence}, gives
\[
\sqrt{G}(\widehat\eta-\eta_o)(\cdot)
\rightsquigarrow
\Phi_{\varphi,\theta_o}'[\mathbb Z_\theta](\cdot)
\quad\text{in }\ell^\infty(\T).
\]
\qed

\subsection{Proof of \Cref{Thm:boot_const}}

\noindent Part~\ref{thm:boot_path}:

Define
\[
\widehat\Psi_G(\theta,\tau)
:=
\widehat{\Prob}_G\bar\psi_\tau(\bm W_g;\theta(\tau)),
\qquad
\Psi_G(\theta,\tau)
:=
\Prob_G\bar\psi_\tau(\bm W_g;\theta(\tau)).
\]
The difference between the bootstrap and original approximate-zero errors is
\[
\mathcal E_G^*
:=
\sqrt G\big(\widehat\Psi_G^*(\widehat\theta^*)
-\widehat\Psi_G(\widehat\theta)\big)
=o_{\Prob^*}(1)
\qquad\text{in }\ell^\infty(\T)
\]
in outer probability. On the event that the line segments joining
\(\widehat\theta\) and \(\widehat\theta^*\) lie in the differentiability
neighbourhood, define
\[
\widetilde H_G^*(\tau)
:=
\int_0^1
\frac{\partial}{\partial t'}\Psi_G(t,\tau)
\bigg|_{t=\widehat\theta(\tau)
+u\{\widehat\theta^*(\tau)-\widehat\theta(\tau)\}}
\,du
\]
and
\[
\begin{aligned}
\mathcal R_G^*
:={}&
\sqrt G\big[
(\widehat\Psi_G-\Psi_G)(\widehat\theta^*)
-(\widehat\Psi_G-\Psi_G)(\widehat\theta)
\big]\\
&+\mathbb M_G^*(\widehat\theta^*)
-\mathbb M_G^*(\theta_o).
\end{aligned}
\]
Subtracting the original approximate estimating equation from its bootstrap
analogue and applying the fundamental theorem of calculus gives, on the
event that \(\widetilde H_G^*\) is uniformly invertible, the exact decomposition
\begin{equation}\label{eq:bootstrap_headline_decomp}
\sqrt G(\widehat\theta^*-\widehat\theta)
=-(\widetilde H_G^*)^{-1}
\left[\mathbb M_G^*(\theta_o)+\mathcal R_G^*-\mathcal E_G^*\right].
\end{equation}
The first three steps below validate this identity and dispatch its remainder
terms before the conditional limit of the surviving multiplier score is
established.

\proofstep{1}{Control the multiplier perturbation globally}
The same conditional maximal inequality, now applied to the full jointly
indexed class, yields
\[
\E^*\!\left[
\sup_{\tau,t}\left|
G^{-1/2}\sum_{g=1}^G\xi_g
\bar\psi_{\tau,j}(\bm W_g;t)
\right|^2
\right]
\leq
C\,G^{-1}\sum_{g=1}^G\bar F_{Gg,j}^{\,2}
=\Op(1).
\]
The inequality uses \(\E^*\xi_g^2=1\), and the last equality follows by
Markov's inequality from the cluster-envelope moment bound supplied by
\Cref{lem:cluster_average_limits}\ref{lem:cluster_average_ui} under
\Cref{ass:uwlln_F}\ref{ass:functional_global_envelope}. Dividing
by \(\sqrt G\), applying conditional Markov's inequality, maximising over
the finitely many coordinates, and
restricting to admissible paths gives
\begin{equation}\label{eq:boot_global}
\sup_{\theta\in\Theta_{\T}}
\left\|G^{-1}\sum_{g=1}^G\xi_g
\bar\Psi(\bm W_g;\theta)\right\|_{\T}
=o_{\Prob^*}(1)
\quad\text{in outer probability}.
\end{equation}

\proofstep{2}{Localise the bootstrap estimator}
The functional UWLLN in
\Cref{lem:functional_ep}\ref{lem:functional_uwlln} gives
\[
\sup_{\theta\in\Theta_{\T}}
\|\widehat\Psi_G(\theta)-\Psi_G(\theta)\|_{\T}
=o_p(1),
\]
whereas \eqref{eq:boot_global} gives
\[
\sup_{\theta\in\Theta_{\T}}
\|\widehat\Psi_G^*(\theta)-\widehat\Psi_G(\theta)\|_{\T}
=o_{\Prob^*}(1)
\]
in outer probability. The triangle inequality therefore yields
\[
\sup_{\theta\in\Theta_{\T}}
\|\widehat\Psi_G^*(\theta)-\Psi_G(\theta)\|_{\T}
=o_p(1)+o_{\Prob^*}(1)
\]
in outer probability. Since
\(\|\widehat\Psi_G^*(\widehat\theta^*)\|_{\T}
=o_{\Prob^*}(G^{-1/2})\), the bootstrap approximate-zero condition implies
\[
\|\Psi_G(\widehat\theta^*)\|_{\T}
\leq
\sup_{\theta\in\Theta_{\T}}
\|\widehat\Psi_G^*(\theta)-\Psi_G(\theta)\|_{\T}
+
\|\widehat\Psi_G^*(\widehat\theta^*)\|_{\T}
=o_{\Prob^*}(1)
\]
in outer probability. Functional separation in
\Cref{ass:pop_identification_F}\ref{ass:functional_sep} then gives
\[
\|\widehat\theta^*-\theta_o\|_{\T}
=o_{\Prob^*}(1)
\]
in outer probability. The corresponding consistency
\(\|\widehat\theta-\theta_o\|_{\T}=o_p(1)\) was established in the proof of
\Cref{Thm:weak_convergence}.

\proofstep{3}{Validate the decomposition and dispatch remainders}
Conditionally on the data, multiplication by \(|\xi_g|\) is a nonnegative
rescaling and therefore preserves manageability in Pollard's definition.
The full and local vector arrays have the common
almost-measurable-Suslin parametrisations required by
\Cref{ass:uwlln_F}\ref{ass:functional_manageability}.
Multiplication by the finitely many row multipliers preserves these
parametrisations on the product probability space, i.e., the approximation error
is bounded by \(\max_{g\leq G}|\xi_g|^2\) times its original value.
The empirical-radius conclusion \eqref{eq:local_empirical_radius}
established in \Cref{lem:functional_ep} also controls the
multiplier-weighted local radius. With \(d_{Gg,j}\) and
\(\mathcal I_\delta\) defined there, for any \(K<\infty\),
\[
\begin{aligned}
&\sup_{(\tau,t)\in\mathcal I_\delta}
\frac1G\sum_{g=1}^{G}\xi_g^2d_{Gg,j}(\tau,t)^2 \leq K^2R_{G,j}^2(\delta)
+\frac4G\sum_{g=1}^{G}
\xi_g^2\indicator{|\xi_g|>K}\bar F_{Gg,j}^{\,2}.
\end{aligned}
\]
Independence of the multipliers and data, together with the envelope
moment bound, bounds the expectation of the second term by
\[
C\E[\xi^2\indicator{|\xi|>K}]\longrightarrow0
\qquad\text{as }K\to\infty.
\]
Thus, first localising for fixed \(K\) and then letting \(K\to\infty\)
makes this weighted radius negligible. Applying the symmetrisation and
maximal-inequality argument used in the proof of
\citet[Theorem~2]{kosorok-2003}, followed by conditional Markov's
inequality, gives
\begin{equation}\label{eq:boot_SE}
\lim_{\delta\downarrow0}\limsup_{G\to\infty}
\Probbr{\Probstarbr{
\sup_{\|\theta-\theta_o\|_{\T}<\delta}
\|\mathbb M_G^*(\theta)-\mathbb M_G^*(\theta_o)\|_{\T}
>\varepsilon}>\eta}=0
\qquad (\varepsilon,\eta>0).
\end{equation}

The two consistency results from Step~2 and the uniform differentiability
condition in
\Cref{ass:linearisation_F}\ref{ass:functional_lin_cont} validate
\eqref{eq:bootstrap_headline_decomp} with probability approaching one and
give
\[
\|\widetilde H_G^*-H\|_{\T}=o_{\Prob^*}(1)
\]
in outer probability. The empirical local stochastic equicontinuity in
\Cref{lem:functional_ep}\ref{lem:functional_se} gives
\[
\sqrt G\left\|
(\widehat\Psi_G-\Psi_G)(\widehat\theta^*)
-(\widehat\Psi_G-\Psi_G)(\widehat\theta)
\right\|_{\T}
=o_{\Prob^*}(1),
\]
and \eqref{eq:boot_SE} gives
\[
\|\mathbb M_G^*(\widehat\theta^*)
-\mathbb M_G^*(\theta_o)\|_{\T}
=o_{\Prob^*}(1).
\]
Hence \(\|\mathcal R_G^*\|_{\T}=o_{\Prob^*}(1)\), while the two
approximate-zero conditions give
\(\|\mathcal E_G^*\|_{\T}=o_{\Prob^*}(1)\), all in outer probability.
The uniform lower singular-value bound on \(H\) now makes
\(\widetilde H_G^*\) uniformly invertible with probability approaching one.
Thus \eqref{eq:bootstrap_headline_decomp} reduces to
\begin{equation}\label{eq:bootstrap_surviving_score}
\sqrt G(\widehat\theta^*-\widehat\theta)
=-(\widetilde H_G^*)^{-1}\mathbb M_G^*(\theta_o)
+o_{\Prob^*}(1)
\quad\text{in }\ell^\infty(\T),
\end{equation}
in outer probability.

\proofstep{4}{Establish weak convergence of the surviving multiplier score}
Apply \citet[Theorem~2]{kosorok-2003}, corresponding to
\citet[Theorem~11.19]{kosorok-2007}, to every scalar projection of the
truth-indexed array used in the proof of \Cref{Thm:weak_convergence}. Its
conditions are the same manageability, envelope, Lindeberg, covariance,
metric, and measurability conditions already verified there, together with
the mean-zero, unit-variance multiplier condition. Moreover,
\Cref{ass:pop_identification_F}\ref{ass:functional_zero} gives
\(\E_G\bar\psi_\tau(\bm W_g;\theta_o(\tau))=0\) for every \(g\) and
\(\tau\), so the theorem applies with the true component means equal to
zero. Equivalently, \citet[Theorem~11.18]{kosorok-2007} applies with
\(\widehat\mu_{Gg}=\mu_{Gg}=0\), making its conditions G--I immediate.
Conditional joint finite-dimensional convergence follows by the
Cram\'er--Wold device, and finite coordinatewise tightness then yields
\begin{equation}\label{eq:boot_multiplier_truth}
\mathbb M_G^*(\theta_o)
\rightsquigarrow_{\Prob}\mathbb G_0
\qquad\text{in }\ell^\infty(\T).
\end{equation}
This is the sole surviving first-order term in
\eqref{eq:bootstrap_headline_decomp}.

\proofstep{5}{Conclude from the decomposition}
The conditional multiplier limit implies
\(\|\mathbb M_G^*(\theta_o)\|_{\T}=\Opstar(1)\) in outer
probability. The inverse identity and the uniform convergence established in Step~3 therefore give
\[
\left\|
\big((\widetilde H_G^*)^{-1}-H^{-1}\big)
\mathbb M_G^*(\theta_o)
\right\|_{\T}
=o_{\Prob^*}(1).
\]
Combining \eqref{eq:bootstrap_surviving_score} and
\eqref{eq:boot_multiplier_truth}, and applying conditional Slutsky, gives
\[
\sqrt{G}(\widehat\theta^*-\widehat\theta)(\cdot)
\rightsquigarrow_{\Prob}
-H(\cdot)^{-1}\mathbb G_0(\cdot)
\quad\text{in }\ell^\infty(\T),
\]
which proves part~\ref{thm:boot_path}.

\noindent Part~\ref{thm:boot_scalar}: For the bootstrap statement, the ordinary weak convergence in
\Cref{Thm:weak_convergence} and the conditional weak convergence in
\Cref{Thm:boot_const} meet the hypotheses of the bootstrap functional
delta method in
\citet[Lemma~B.3]{chernozhukov-fernandez-melly-2013inference}. Hence
\[
\sqrt{G}(\widehat\eta^*-\widehat\eta)(\cdot)
\rightsquigarrow_{\Prob}
\Phi_{\varphi,\theta_o}'[\mathbb Z_\theta](\cdot)
\quad\text{in }\ell^\infty(\T).
\]
\qed

\subsection{Proof of \Cref{Crly:boot_pointwise}}

For each fixed \(\tau\in\T\), the evaluation map
\(z\mapsto z(\tau)\) is a bounded linear map on \(\ell^\infty(\T)\).
The result follows from \Cref{Thm:boot_const} by the continuous mapping
theorem.
\qed

\section{Proofs for the Model-Specific Results}\label{app:proofs_specific}

For ease of reference, the coordinate-score classes studied in this section are
\[
\begin{aligned}
\mathcal F_j^{DR}
&=\left\{(u,x)\mapsto x_j\big(\indicator{u\le y}-\Lambda(x't)\big):
(y,t)\in\mathcal Y\times\Theta\right\},\\
\mathcal F_j^{QR}
&=\left\{(u,x)\mapsto x_j\big(\tau-\indicator{u\le x't}\big):
(\tau,t)\in\T\times\Theta\right\},\\
\mathcal F_j^{ER}
&=\left\{(u,x)\mapsto x_j(u-x't)
|\tau-\indicator{u<x't}|:
(\tau,t)\in\T\times\Theta\right\}.
\end{aligned}
\]
We first establish their manageability and its preservation under cluster averaging, then prove the model-specific identification and verification results.

\begin{lemma}[Manageability of DR, QR, and ER score classes]
\label{lem:concrete_manageable_components}
Let \(w=(u,x)\), let \(\Theta\subset\R^{p_\theta}\) and
\(\T\subset(0,1)\) be compact, and set
\(B_\Theta:=\sup_{t\in\Theta}\|t\|\). For a monotone link
\(\Lambda\) taking values in \([0,1]\), the DR, QR, and ER
coordinate-score classes defined above are manageable in the sense of
\citet[Definition~7.9]{pollard-1990}, with respective envelopes
\(
|x_j|,\quad |x_j|,\quad \text{and} \quad
|x_j|\{|u|+B_\Theta\|x\|\}.
\)
\end{lemma}

\noindent
\textit{Proof.}
The linear-index, residual, and constant classes, respectively,
\[
\{(u,x)\mapsto x't:t\in\Theta\},\qquad
\{(u,x)\mapsto u-x't:t\in\Theta\},\qquad
\{(u,x)\mapsto\tau:\tau\in\T\},
\]
are contained in finite-dimensional vector spaces of measurable functions and are therefore
Vapnik--\v{C}ervonenkis (VC) subgraph classes by \citet[Lemma~9.6]{kosorok-2007}. The threshold and
halfspace indicator classes are, respectively,
\(
\{(u,x)\mapsto\indicator{u\leq y}:y\in\mathbb R\},\quad
\{(u,x)\mapsto\indicator{u\leq x't}:t\in\Theta\},\quad \text{and} \quad
\{(u,x)\mapsto\indicator{u<x't}:t\in\Theta\}.
\)
Their defining sets are VC by \citet[Lemma~9.12(i)]{kosorok-2007},
using complements for strict boundaries. By
\citet[Lemma~9.8]{kosorok-2007}, the indicator classes are therefore
VC-subgraph. The monotone-link class
\[
\{(u,x)\mapsto\Lambda(x't):t\in\Theta\}
\]
is also VC-subgraph by the monotone-composition result in
\citet[Lemma~9.9(viii)]{kosorok-2007}. All these VC indices are bounded
independently of the observations and sample size. Hence
\citet[Theorem~9.3]{kosorok-2007} gives uniform polynomial covering bounds
for each component class \(\mathcal F\), with its indicated envelope \(F\):
\[
\sup_Q N\bigl(\varepsilon\|F\|_{Q,2},\mathcal F,L_2(Q)\bigr)
\leq A\varepsilon^{-v},\qquad 0<\varepsilon<1,
\]
where the supremum is over finitely supported probability measures with
\(\|F\|_{Q,2}>0\), and \(A,v\) are finite constants.

The singleton class \(\{x_j\}\) has envelope \(|x_j|\) and covering
number one. The covering inequalities in the proofs of
\citet[Lemma~9.13 and Theorem~9.15]{kosorok-2007} preserve polynomial
bounds under finite sums and products, with sum and product envelopes.
Restricting a product index set to the common parameters used in a score
does not increase its entropy bound, apart from a constant adjustment of
the covering radius. For ER, use
\[
|\tau-\indicator{u<x't}|
=\tau+(1-2\tau)\indicator{u<x't}.
\]
These operations give polynomial bounds for the DR, QR, and ER
coordinate-score classes with envelopes bounded by constant multiples of
\(|x_j|\), \(|x_j|\), and
\(|x_j|\{|u|+B_\Theta\|x\|\}\), respectively. The scores themselves are
bounded by the stated envelopes; absorbing the fixed envelope multiples
into \(A\) gives polynomial bounds relative to those envelopes.

To connect these bounds to \citet[Definition~7.9, p.~38]{pollard-1990}, fix
\(w_1,\ldots,w_n\) and any nonnegative rescaling vector
\(a=(a_1,\ldots,a_n)\). Define the rescaled evaluation set and its envelope norm by
\[
V_a:=\{(a_1f(w_1),\ldots,a_nf(w_n)):f\in\mathcal F\},
\qquad
E_a:=\left\{\sum_i a_i^2F(w_i)^2\right\}^{1/2}.
\]
If \(E_a=0\), all rescaled evaluation vectors coincide and the conclusion is immediate.
Otherwise, put \(Q_a:=\sum_i a_i^2\delta_{w_i}/\sum_i a_i^2\). For any \(f,g\in\mathcal F\),
\[
\frac{\|f-g\|_{Q_a,2}}{\|F\|_{Q_a,2}}
=
\frac{\{\sum_i a_i^2[f(w_i)-g(w_i)]^2\}^{1/2}}{E_a}.
\]
Thus separation at radius \(\varepsilon E_a\) in \(V_a\) is equivalent to separation at radius
\(\varepsilon\|F\|_{Q_a,2}\) in \(L_2(Q_a)\). Writing \(D\) for the packing number, the
covering--packing inequality \citep[p.~10, following Definition~3.3]{pollard-1990} gives
\[
\begin{aligned}
D(\varepsilon E_a,V_a)
&\leq N\!\left(\frac{\varepsilon}{2}\|F\|_{Q_a,2},
                  \mathcal F,L_2(Q_a)\right)\\
&\leq A(\varepsilon/2)^{-v}
=2^vA\varepsilon^{-v}.
\end{aligned}
\]
Absorbing \(2^v\) into \(C\) gives the bound \(C\varepsilon^{-v}\), uniformly in the
observations, \(a\), and \(n\).
Taking \(C\geq1\), the deterministic capacity bound
\(\lambda(\varepsilon)=C\varepsilon^{-v}\) satisfies
\(\int_0^1\sqrt{\log\lambda(\varepsilon)}\,d\varepsilon<\infty\).
Thus all the asserted classes are Euclidean, and hence manageable, with
the indicated envelopes. \qed

The following lemma transfers manageability of individual-level evaluation arrays to cluster-average arrays. Let \(\mathcal F\) be a class of measurable functions with nonnegative measurable envelope \(F\). For any deterministic cluster partition with \(N_g\geq1\), define
\[
\bar f_{Gg}:=N_g^{-1}\sum_{i\in\G_g}f(W_i),
\qquad
\bar F_{Gg}:=
\left\{N_g^{-1}\sum_{i\in\G_g}F(W_i)^2\right\}^{1/2}.
\]
\begin{lemma}[Manageability under cluster averaging]
\label{lem:cluster_manageability_transfer}
Suppose the individual evaluation arrays of \(\mathcal F\) are
manageable in the sense of \citet[Definition~7.9]{pollard-1990}, uniformly
over sample sizes, evaluation points, and nonnegative rescaling vectors,
with a deterministic capacity bound \(\lambda\). Then the cluster-average array
\(\{\bar f_{Gg}:f\in\mathcal F,\ g\leq G\}\)
is manageable with envelope \(\{\bar F_{Gg}:g\leq G\}\) and capacity
bound \(\lambda\), uniformly over cluster sizes. Consequently,
\(\{G^{-1/2}\bar f_{Gg}:f\in\mathcal F,\ g\leq G\}\)
is manageable with envelope \(\{G^{-1/2}\bar F_{Gg}:g\leq G\}\).
\end{lemma}

\noindent\textit{Proof.}
Jensen's inequality gives \(|\bar f_{Gg}|\leq\bar F_{Gg}\).
Fix the evaluation points and any nonnegative rescaling vector
\(\alpha=(\alpha_1,\ldots,\alpha_G)\). For \(f,h\in\mathcal F\),
clusterwise Jensen's inequality gives
\begin{equation}\label{eq:cluster_covering_contraction}
\sum_{g=1}^G\alpha_g^2(\bar f_{Gg}-\bar h_{Gg})^2
\leq
\sum_{g=1}^G\frac{\alpha_g^2}{N_g}
\sum_{i\in\G_g}(f(W_i)-h(W_i))^2.
\end{equation}
Set \(\beta_i:=\alpha_{g(i)}/\sqrt{N_{g(i)}}\). The right-hand side is
the squared distance between the individual evaluation vectors rescaled
by \(\beta\), and the envelope scales agree:
\[
\sum_{g=1}^G\alpha_g^2\bar F_{Gg}^2
=\sum_{i=1}^N\beta_i^2F(W_i)^2.
\]
Choose one function \(f\in\mathcal F\) for each vector in a packing of the rescaled cluster-average vectors. By \eqref{eq:cluster_covering_contraction}, the corresponding rescaled individual evaluation vectors are separated by at least the same distance, so they form a packing at the same radius. The assumed
individual capacity bound therefore also bounds the cluster packing
numbers at radius
\(\varepsilon\{\sum_g\alpha_g^2\bar F_{Gg}^2\}^{1/2}\).
If the envelope norm is zero, all rescaled vectors coincide and the
conclusion is immediate. Otherwise, the cluster-average array satisfies condition (ii) of \citet[Definition~7.9, p.~38]{pollard-1990} with the same capacity bound \(\lambda\). Since this function is unchanged, its integrability condition (i) also carries over.
Absorbing \(G^{-1/2}\) into the arbitrary weights \(\alpha_g\) proves
the normalised-array conclusion. 
\qed

\setcounter{subsection}{1}
\subsection{Proof of \Cref{prop:DR_identification}}
Fix \(y\in\mathcal Y\). By the law of iterated expectations and
\Cref{ass:DR_identification},
\[
\E_G\!\left[
X_i\big(\indicator{Y_i\leq y}-\Lambda(X_i'\theta_o(y))\big)
\right]
=
\E_G\!\left[
X_i\big( F_{Y_i\mid X_i}(y\mid X_i)-\Lambda(X_i'\theta_o(y))\big)
\right]
=0
\]
for every sampled observation. Averaging these equalities within each cluster gives clusterwise exogeneity; averaging again with the SACR weights \(N_g^{-1}\) gives zero for the DR-SACR population moment at \(\theta_o(y)\), so the cluster-size adjustment does not move the target zero.

It remains to show uniqueness. Suppose, for contradiction, that the DR-SACR population moment is also zero at some \(\theta(y)\neq\theta_o(y)\), and set \(\delta=\theta_o(y)-\theta(y)\), so \(\delta\neq0\). Subtracting the zero moment at \(\theta_o(y)\) gives
\[
\frac{1}{G}\sum_{g=1}^G N_g^{-1}\sum_{i\in\G_g}
\E_G\!\left[
X_i\big(\Lambda(X_i'\theta_o(y))-\Lambda(X_i'\theta(y))\big)
\right]=0.
\]
Premultiplying by \(\delta'\) yields
\[
\frac{1}{G}\sum_{g=1}^G N_g^{-1}\sum_{i\in\G_g}
\E_G\!\left[
(X_i'\delta)\big(\Lambda(X_i'\theta_o(y))-\Lambda(X_i'\theta(y))\big)
\right]=0.
\]
Indeed, for real numbers \(a\) and \(b\), strict monotonicity of \(\Lambda\) implies \((a-b)\{\Lambda(a)-\Lambda(b)\}\geq0\), with equality only if \(a=b\). Taking \(a=X_i'\theta_o(y)\) and \(b=X_i'\theta(y)\), each summand is nonnegative and vanishes only when \(X_i'\delta=0\). Hence the sum can be zero only if \(X_i'\delta=0\) almost surely under the cluster-balanced array. Therefore
\[
\frac{1}{G}\sum_{g=1}^G
\E_G\!\left[
N_g^{-1}\sum_{i\in\G_g}(X_i'\delta)^2
\right]=0,
\]
or equivalently,
\[
\delta'
\left\{
\frac{1}{G}\sum_{g=1}^G
\E_G\!\left[N_g^{-1}\sum_{i\in\G_g}X_iX_i'\right]
\right\}
\delta=0.
\]
The uniform positive definiteness condition in \Cref{ass:LI_common}\ref{ass:LI_rank} forces \(\delta=0\), a contradiction. Thus, the DR-SACR population moment has the same unique zero as the primitive DR moment for each fixed \(y\).

The functional strengthening follows directly from Assumption-DR~2\ref{ass:DR_jacobian}. For any \(y\in\mathcal Y\), \(t\in\Theta\), and \(\delta=\theta_o(y)-t\), the fundamental theorem of calculus gives
\[
\delta'\Prob_G\bar\psi_y(\bm W_g;t)
=
\int_0^1
\delta'Q_G^{DR}\!\left(y,t,1-s\right)\delta\,ds
\geq c_H\|\delta\|^2.
\]
By Cauchy--Schwarz,
\[
c_H\|\delta\|^2
\leq \delta'\Prob_G\bar\psi_y(\bm W_g;t)
\leq \|\delta\|\,\|\Prob_G\bar\psi_y(\bm W_g;t)\|.
\]
Dividing by \(\|\delta\|\) when \(\delta\neq0\) gives
\(\|\Prob_G\bar\psi_y(\bm W_g;t)\|
\geq c_H\|t-\theta_o(y)\|\); the same bound is immediate when
\(\delta=0\). If an admissible path \(\theta\) satisfies
\(\|\theta-\theta_o\|_{\mathcal Y}\geq\epsilon\), choose \(y\) such that
\(\|\theta(y)-\theta_o(y)\|\geq\epsilon/2\). The preceding bound then gives
\[
\bigl\|\Prob_G\bar\Psi(\bm W_g;\theta)\bigr\|_{\mathcal Y}
\geq c_H\epsilon/2
\]
uniformly in \(G\), which is the functional separation condition in \Cref{ass:pop_identification_F}.
\qed

\subsection{Proof of \Cref{prop:DR_verify}}
The individual coordinate-score class is manageable with envelope
\(|X_j|\) by \Cref{lem:concrete_manageable_components}.
\Cref{lem:cluster_manageability_transfer} therefore gives manageability
of the normalised cluster-average array \(Z_{Gg,j}^{DR}\)
with the corresponding normalised root mean square (RMS) envelope \(F_{Gg,j}^{DR}\). Joint measurability of
the DR score and Borel measurability of \(\theta_o(\cdot)\) under \Cref{ass:DR_identification} give jointly measurable full,
truth-indexed, and local-difference arrays on Borel subsets of Euclidean
spaces; hence these arrays are jointly AMS, taking the entire respective
index spaces as the Suslin approximating sets. Finally,
\Cref{ass:LI_common}\ref{ass:LI_moment} establishes uniform integrability
of the squared envelope. \qed

\subsection{Proof of \Cref{prop:QR_identification}}
Fix \(\tau\in\T\). By the law of iterated expectations,
\[
\E_G\big[X_i\big(\tau-\indicator{Y_i\leq X_i'\theta_o(\tau)}\big)\big]
=
\E_G\big[
X_i\big(\tau-F_{Y_i\mid X_i}(X_i'\theta_o(\tau)\mid X_i)\big)
\big]
=0,
\]
where the last equality follows from the array-preservation restriction in
\Cref{ass:QR_identification}\ref{ass:QR_ident}. Averaging these zero moments
with the SACR weights \(N_g^{-1}\) gives the displayed QR-SACR population
moment at \(\theta_o(\tau)\), and hence gives clusterwise exogeneity.

It remains to show uniqueness. Suppose, for contradiction, that some \(\theta(\tau)\in\Theta\) with \(\theta(\tau)\neq\theta_o(\tau)\) also solves the QR-SACR population moment, and set \(\delta=\theta_o(\tau)-\theta(\tau)\), so \(\delta\neq0\). Subtracting the zero moment at \(\theta_o(\tau)\) gives
\[
\frac{1}{G}\sum_{g=1}^G N_g^{-1}\sum_{i\in\G_g}
\E_G\!\left[
X_i\big(F_{Y_i\mid X_i}(X_i'\theta_o(\tau)\mid X_i)
-F_{Y_i\mid X_i}(X_i'\theta(\tau)\mid X_i)\big)
\right]=0.
\]
Premultiplying by \(\delta'\) yields
\[
\frac{1}{G}\sum_{g=1}^G N_g^{-1}\sum_{i\in\G_g}
\E_G\!\left[
\big(X_i'\delta\big)
\big(F_{Y_i\mid X_i}(X_i'\theta_o(\tau)\mid X_i)
-F_{Y_i\mid X_i}(X_i'\theta(\tau)\mid X_i)\big)
\right]=0.
\]
Using the integral representation of the conditional cumulative distribution function (CDF) inside the preceding display gives
\[
\delta'\Prob_G\bar\psi_\tau(\bm W_g;\theta(\tau))
=
\int_0^1
\delta'Q_G^{QR}\!\left(\tau,\theta(\tau),1-s\right)\delta\,ds
\geq c_f\|\delta\|^2
\]
by \Cref{ass:QR_identification}\ref{ass:QR_segment_rank}. A second zero would make the left-hand side zero, forcing \(\delta=0\), a contradiction. Hence \(\theta_o(\tau)\) is the unique QR-SACR population zero for each fixed \(\tau\).

The same inequality and Cauchy--Schwarz imply
\begin{equation}\label{eq:QR_identification_bound}
\|\Prob_G\bar\psi_\tau(\bm W_g;t)\|
\geq c_f\|t-\theta_o(\tau)\|
\end{equation}
uniformly over \(G\), \(\tau\in\T\), and \(t\in\Theta\). If an admissible path \(\theta\) satisfies \(\|\theta-\theta_o\|_{\T}\geq\epsilon\), choose \(\tau\) such that \(\|\theta(\tau)-\theta_o(\tau)\|\geq\epsilon/2\). It follows that
\[
\bigl\|\Prob_G\bar\Psi(\bm W_g;\theta)\bigr\|_{\T}
\geq c_f\epsilon/2
\]
uniformly in \(G\), proving the functional separation condition.

\qed

\subsection{Proof of \Cref{prop:QR_verify}}
The individual coordinate-score class is manageable with envelope
\(|X_j|\) by \Cref{lem:concrete_manageable_components}.
\Cref{lem:cluster_manageability_transfer} therefore gives manageability
of the normalised cluster-average array \(Z_{Gg,j}^{QR}\)
with the corresponding normalised RMS envelope \(F_{Gg,j}^{QR}\). Joint measurability of
the QR score and continuity of \(\theta_o(\cdot)\) from \Cref{lem:QR_ER_path_lipschitz} give jointly measurable full,
truth-indexed, and local-difference arrays on Borel subsets of Euclidean
spaces; hence these arrays are jointly AMS, taking the entire respective
index spaces as the Suslin approximating sets. Finally,
\Cref{ass:LI_common}\ref{ass:LI_moment} establishes uniform integrability
of the squared envelope. \qed

\subsection{Proof of \Cref{prop:ER_identification}}
Fix \(\tau\in\T\) and let \(B_\Theta:=\sup_{t\in\Theta}\|t\|<\infty\).
\Cref{ass:ER_regular} ensures integrability of
\(\|X_i\||Y_i|+B_\Theta\|X_i\|^2\), so the score expectations below
are well defined. The array-preservation restriction in
\Cref{ass:ER_identification} gives
\[
\E_G\left[
\big(Y_i-X_i'\theta_o(\tau)\big)
\bigl|\tau-\indicator{Y_i<X_i'\theta_o(\tau)}\bigr|
\mid X_i
\right]=0
\]
for every \(i\). This is the first-order characterisation of the conditional expectile; see \citet[Theorem~1]{newey_asymmetric_1987}. Multiplying by \(X_i\), applying the law of iterated expectations, and averaging with the SACR weights \(N_g^{-1}\) gives the displayed ER-SACR population moment at \(\theta_o(\tau)\), and hence gives clusterwise exogeneity.

	For uniqueness, define the conditional expectile score
	\[
	q_\tau(m\mid X_i)
	:=
\E_G\left[
(Y_i-m)
\bigl|\tau-\indicator{Y_i<m}\bigr|
	\mid X_i
	\right].
	\]
For fixed \(y\), the score \(s_\tau(y,m):=(y-m)|\tau-\indicator{y<m}|\) is continuous and piecewise linear, with slopes \(-\tau\) for \(m<y\) and \(-(1-\tau)\) for \(m>y\). Thus, for any \(m_1,m_2\),
\[
s_\tau(y,m_1)-s_\tau(y,m_2)
=-\int_{m_2}^{m_1}
\left[\tau\indicator{y\ge u}+(1-\tau)\indicator{y<u}\right]\,du.
\]
Taking conditional expectations given \(X_i\) and applying conditional Fubini, justified by the bounded integrand on the finite interval, gives
\[
q_\tau(m_1\mid X_i)-q_\tau(m_2\mid X_i)
=-\int_{m_2}^{m_1}
\left[
\tau\Probbr{Y_i\geq u\mid X_i}
+(1-\tau)\Probbr{Y_i<u\mid X_i}
\right]\,du.
\]
The bracket lies between \(\min\{\tau,1-\tau\}\) and
\(\max\{\tau,1-\tau\}\), and hence
\begin{equation}\label{eq:ER_score_slope}
(m_1-m_2)\{q_\tau(m_1\mid X_i)-q_\tau(m_2\mid X_i)\}
\leq-\min\{\tau,1-\tau\}(m_1-m_2)^2.
\end{equation}
Suppose, for contradiction, that another
\(\theta(\tau)\neq\theta_o(\tau)\) solves the ER-SACR population moment,
and set \(\delta=\theta_o(\tau)-\theta(\tau)\), so \(\delta\neq0\).
Write \(c_\tau:=\min\{\tau,1-\tau\}>0\), and let \(c_X>0\) denote the
uniform lower eigenvalue bound in
\Cref{ass:LI_common}\ref{ass:LI_rank}.
Subtracting the zero moment at \(\theta_o(\tau)\) gives
\[
\frac{1}{G}\sum_{g=1}^G N_g^{-1}\sum_{i\in\G_g}
\E_G\!\left[
X_i\big(q_\tau(X_i'\theta_o(\tau)\mid X_i)
-q_\tau(X_i'\theta(\tau)\mid X_i)\big)
\right]=0.
\]
Premultiplication by \(\delta'\), followed by
\eqref{eq:ER_score_slope}, gives the contradiction
\begin{align*}
0
&=
\frac{1}{G}\sum_{g=1}^G N_g^{-1}\sum_{i\in\G_g}
\E_G\!\left[
\big(X_i'\delta\big)
\big(q_\tau(X_i'\theta_o(\tau)\mid X_i)
-q_\tau(X_i'\theta(\tau)\mid X_i)\big)
\right]\\
&\leq
-c_\tau\frac{1}{G}\sum_{g=1}^G
\E_G\!\left[
N_g^{-1}\sum_{i\in\G_g}(X_i'\delta)^2
\right]\\
&=
-c_\tau\delta'
\left\{
\frac{1}{G}\sum_{g=1}^G
\E_G\!\left[N_g^{-1}\sum_{i\in\G_g}X_iX_i'\right]
\right\}\delta
\leq-c_\tau c_X\|\delta\|^2<0.
\end{align*}
Hence \(\theta_o(\tau)\) is the unique ER-SACR population zero for each fixed \(\tau\).

Finally, suppose that functional separation failed. Then for some \(\epsilon>0\) one could find admissible paths \(\theta_G(\cdot)\in\Theta_{\T}\) with \(\|\theta_G-\theta_o\|_{\T}\geq\epsilon\) and
\[
\bigl\|
\Prob_G\bar\Psi(\bm W_g;\theta_G)
\bigr\|_{\T}
\to0
\]
along a subsequence. Choose \(\tau_G\in\T\) with
\(\|\theta_G(\tau_G)-\theta_o(\tau_G)\|\geq\epsilon/2\), and set
\(\underline c_\tau:=\min\{\underline\tau,1-\bar\tau\}>0\).
For \(\delta=\theta_o(\tau)-t\), \eqref{eq:ER_score_slope} and
\Cref{ass:LI_common}\ref{ass:LI_rank} imply
\[
\delta'\Prob_G\bar\psi_\tau(\bm W_g;t)
\geq
\underline c_\tau\,
\delta'
\left\{\frac1G\sum_{g=1}^G
\E_G\!\left[N_g^{-1}\sum_{i\in\G_g}X_iX_i'\right]\right\}
\delta
\geq\underline c_\tau c_X\|\delta\|^2,
\]
where \(c_X>0\) is the uniform eigenvalue bound. Cauchy--Schwarz gives
\begin{equation}\label{eq:ER_identification_bound}
\|\Prob_G\bar\psi_\tau(\bm W_g;t)\|
\geq\underline c_\tau c_X\|t-\theta_o(\tau)\|.
\end{equation}
At \(\tau_G\), the right-hand side is at least
\(\underline c_\tau c_X\epsilon/2\), contradicting convergence to zero.
Thus the functional separation condition holds.

\qed

The following lemma establishes Lipschitz continuity of the QR and ER coefficient paths, supplying the path regularity used in the subsequent score-process verification proofs.

\begin{lemma}[Lipschitz continuity of QR and ER coefficient paths]\label{lem:QR_ER_path_lipschitz}
\begin{enumerate}[label=(\alph*)]
\item\label{lem:QR_path_lipschitz} Under \Cref{ass:QR_identification,ass:LI_common}, the QR coefficient path satisfies
\[
\|\theta_o(\tau')-\theta_o(\tau)\|\leq C_{QR}|\tau'-\tau|,
\qquad \tau,\tau'\in\T,
\]
for some finite constant \(C_{QR}\).
\item\label{lem:ER_path_lipschitz} Under \Cref{ass:ER_identification,ass:LI_common} and \Cref{ass:ER_regular}, the ER coefficient path satisfies
\[
\|\theta_o(\tau')-\theta_o(\tau)\|\leq C_{ER}|\tau'-\tau|,
\qquad \tau,\tau'\in\T,
\]
for some finite constant \(C_{ER}\).
\end{enumerate}
\end{lemma}

\noindent\textit{Proof.}
\noindent Part~\ref{lem:QR_path_lipschitz}:
To establish regularity of the coefficient path, define
\[
\mu_G:=\frac1G\sum_{g=1}^G
\E_G\!\left[N_g^{-1}\sum_{i\in\G_g}X_i\right].
\]
The QR score is affine in \(\tau\), and the population moment vanishes
at \(\theta_o(\tau')\) for index \(\tau'\). Consequently,
\[
\Prob_G\bar\psi_\tau(\bm W_g;\theta_o(\tau'))
=
\Prob_G\bar\psi_{\tau'}(\bm W_g;\theta_o(\tau'))
+(\tau-\tau')\mu_G
=(\tau-\tau')\mu_G.
\]
Applying the identification bound \eqref{eq:QR_identification_bound} at \(t=\theta_o(\tau')\) gives
\[
c_f\|\theta_o(\tau')-\theta_o(\tau)\|
\leq |\tau-\tau'|\,\|\mu_G\|.
\]
By \Cref{ass:LI_common}\ref{ass:LI_moment},
\(\sup_G\|\mu_G\|\leq\sup_{G,i}\E_G\|X_i\|<\infty\).
Thus \(C_{QR}:=c_f^{-1}\sup_G\|\mu_G\|\) is a finite Lipschitz constant,
uniform over \(\tau,\tau'\in\T\). This is the cluster-average analogue
of \citet[Supplementary Appendix, equation~(C.1)]{hagemann-2017-cluster},
obtained directly from identification without differentiating the path.

\noindent Part~\ref{lem:ER_path_lipschitz}:
Let \(B_\Theta:=\sup_{t\in\Theta}\|t\|\), and let \(c_X>0\) denote the uniform eigenvalue bound in \Cref{ass:LI_common}\ref{ass:LI_rank}. Set \(\underline c_\tau:=\inf_{\tau\in\T}\min\{\tau,1-\tau\}>0\). For Lipschitz continuity, the ER score satisfies the exact identity
\[
\psi_\tau(W;t)-\psi_{\tau'}(W;t)
=(\tau-\tau')X|Y-X't|.
\]
The zero moment at index \(\tau'\) therefore gives
\[
\Prob_G\bar\psi_\tau(\bm W_g;\theta_o(\tau'))
=(\tau-\tau')\frac1G\sum_{g=1}^G
\E_G\!\left[
N_g^{-1}\sum_{i\in\G_g}X_i|Y_i-X_i'\theta_o(\tau')|
\right].
\]
Define
\[
M_{ER}:=\sup_{G,i}\E_G\!\left[
\|X_i\|\,|Y_i|+B_\Theta\|X_i\|^2
\right]<\infty,
\]
where finiteness follows from
\Cref{ass:ER_regular}. Applying the identification bound \eqref{eq:ER_identification_bound} at \(t=\theta_o(\tau')\), followed by the triangle
inequality, yields
\[
\underline c_\tau c_X\|\theta_o(\tau')-\theta_o(\tau)\|
\leq
\|\Prob_G\bar\psi_\tau(\bm W_g;\theta_o(\tau'))\|
\leq M_{ER}|\tau-\tau'|.
\]
Thus \(C_{ER}:=M_{ER}/(\underline c_\tau c_X)\) is a finite uniform
Lipschitz constant.

\qed

\subsection{Proof of \Cref{prop:ER_verify}}
The individual coordinate-score class is manageable with envelope
\(F_j^{ER}\) by \Cref{lem:concrete_manageable_components}.
\Cref{lem:cluster_manageability_transfer} therefore gives manageability
of the normalised cluster-average array \(Z_{Gg,j}^{ER}\)
with the corresponding normalised RMS envelope \(F_{Gg,j}^{ER}\). Continuity of the ER
score in \((\tau,t)\), together with continuity of \(\theta_o(\cdot)\)
from \Cref{lem:QR_ER_path_lipschitz}, makes the full, truth-indexed, and
local-difference arrays separable; hence they are AMS by
\citet[Lemma~11.15]{kosorok-2007}. Finally,
\Cref{ass:ER_regular} gives uniformly bounded
\((2+\delta)\)th moments of \(F_j^{ER}\), establishing the required
squared-envelope uniform integrability. \qed

\section{Expectile paths from counterfactual distributions}\label{app:expectile_map}

The preceding verification concerns the direct ER estimating equation. For expectile treatment effects computed instead from estimated counterfactual distributions, the required scalar functional is the distribution-to-expectile map whose limiting behaviour is handled in this section using the Z-map framework of \citet{chernozhukov-fernandez-melly-2013inference}. Let \(\T=[\underline\tau,\overline\tau]\subset(0,1)\), and let \(F\) be a distribution with a finite first moment. Set \(w_\tau(y,e)=\tau\indicator{y\geq e}+(1-\tau)\indicator{y<e}\), equivalently, \(w_\tau(y,e)=\tau+(1-2\tau)\indicator{y<e}\), and write \(e_\tau(F)\) for the unique zero in \(e\in\mathbb R\) of
\( \displaystyle
\Psi_F(e,\tau)
:=\int w_\tau(y,e)(e-y)\,dF(y).
\) The following imposes a regularity condition on the distribution-to-expectile map.

\begin{expectilemapcondition}
\(F\) is continuous at \(e_\tau(F)\) for every \(\tau\in\T\).
\end{expectilemapcondition}
Write \(\mathcal E(F):=(e_\tau(F))_{\tau\in\T}\), and define the candidate derivative map \(\mathcal E'_F:L^1(\mathbb R)\to\ell^\infty(\T)\) by 
\[
\mathcal E'_F[h](\tau)
=-a_\tau(F)^{-1}\int w_\tau(y,e_\tau(F))h(y)\,dy,
\qquad
a_\tau(F):=\int w_\tau(y,e_\tau(F))\,dF(y)
\]
and the tangent set
\[
\mathbb D_F
:=\left\{h\in L^1(\mathbb R):
\begin{array}{l}
	\text{there exist }t_n\downarrow0\text{ and }h_n\to h\text{ in }L^1,\\
	\text{with }F+t_nh_n\text{ a finite-first-moment distribution function}
\end{array}
\right\}.
\]
\begin{proposition}[Expectile map for distribution paths]\label{prop:expectile_map}
Under the preceding expectile-map regularity condition:
\begin{enumerate}[label=(\alph*)]
\item\label{prop:expectile_map_diff}
With differences of distribution functions equipped with the \(L^1(\mathbb R)\) norm, \(\mathcal E\) is Hadamard differentiable at \(F\), tangentially to \(\mathbb D_F\), with derivative \(\mathcal E'_F\).
\item\label{prop:expectile_map_weak}
If \(\sqrt G(\widehat F-F)\rightsquigarrow\mathbb H\) in \(L^1(\mathbb R)\), then
\(
\sqrt G\{\mathcal E(\widehat F)-\mathcal E(F)\}
\rightsquigarrow\mathcal E'_F[\mathbb H]
\quad\text{in }\ell^\infty(\T).
\)
\item\label{prop:expectile_map_boot}
Under the premise of part~\ref{prop:expectile_map_weak}, if \(\sqrt G(\widehat F^*-\widehat F)\rightsquigarrow_{\Prob}\mathbb H\) in \(L^1(\mathbb R)\), then
\(
\sqrt G\{\mathcal E(\widehat F^*)-\mathcal E(\widehat F)\}
\rightsquigarrow_{\Prob}\mathcal E'_F[\mathbb H]
\quad\text{in }\ell^\infty(\T).
\)
The same conclusion holds for approximate bootstrap roots \(\widetilde e_\tau^*\) satisfying \(\sup_{\tau\in\T}|\Psi_{\widehat F^*}(\widetilde e_\tau^*,\tau)|=o_{\Prob^*}(G^{-1/2})\), in outer probability.
\end{enumerate}
\end{proposition}

Unlike quantiles, expectiles depend on integrated probabilities on both sides of the candidate expectile. The \(L^1(\mathbb R)\) premise in \Cref{prop:expectile_map} states the population condition for full-distribution inference and is not implied by compact-threshold DR convergence. The empirical illustration in \Cref{sect:empirics} therefore implements an approximation explicitly on the observed support of the outcome.

\subsection{Proof of \Cref{prop:expectile_map}}

\noindent\textit{Part~\ref{prop:expectile_map_diff}.}
The finite first moment makes \(\Psi_F(e,\tau)\) finite and continuous in
\((e,\tau)\) on \(\mathbb R\times\mathcal U\) for any open interval
\(\mathcal U\) containing \(\T\) and contained in \((0,1)\). For each
\(\tau\), \(e\mapsto\Psi_F(e,\tau)\) is strictly increasing, crosses zero,
and is the gradient of a convex antiderivative. Existence and uniqueness,
together with strict monotonicity of the expectile path, follow from
\citet[Theorem~1(i)--(ii)]{newey_asymmetric_1987}. Monotonicity places the
entire root path in the compact interval
\([e_{\underline\tau}(F),e_{\overline\tau}(F)]\). To verify continuity,
suppose to the contrary that \(\tau_n\to\tau\) while, for some
\(\varepsilon>0\),
\[
|e_{\tau_n}(F)-e_\tau(F)|\geq\varepsilon.
\]
The Bolzano--Weierstrass Theorem gives a subsequence satisfying
\(e_{\tau_{n_k}}(F)\to e^*\). Joint continuity and the root restrictions
give
\(0=\lim_k\Psi_F(e_{\tau_{n_k}}(F),\tau_{n_k})=\Psi_F(e^*,\tau)\),
so uniqueness requires \(e^*=e_\tau(F)\), a contradiction. Thus
\(\tau\mapsto e_\tau(F)\) is continuous. Because \(F\) is continuous at
every point in the image of this path, it is continuous on that compact
root interval. At each root,
\[
\partial_e\Psi_F(e_\tau(F),\tau)=a_\tau(F)
\geq \min\{\underline\tau,1-\overline\tau\}>0,
\qquad
\partial_\tau\Psi_F(e_\tau(F),\tau)
=-\int |e_\tau(F)-y|\,dF(y).
\]
Dominated convergence, using the compact root interval and the finite
first moment, makes both derivatives continuous along the root path.
These properties connect the construction to the indexed \(Z\)-map
framework of \citet[Appendix~E]{chernozhukov-fernandez-melly-2013inference}.
The next step is to establish differentiability in the stated \(L^1\) topology.

To make the derivative explicit, consider an admissible perturbation
\(F_t=F+t h_t\), where \(h_t\to h\) in \(L^1(\mathbb R)\), and write
\(e_t=e_\tau(F_t)\) and \(e=e_\tau(F)\). Since both are exact roots,

\begin{align}
0
&=\Psi_{F_t}(e_t,\tau)-\Psi_F(e,\tau) \nonumber \\
&=\underbrace{\{\Psi_{F_t}(e_t,\tau)-\Psi_{F_t}(e,\tau)\}}
_{\Delta_{e,t}(\tau)}
+\underbrace{\{\Psi_{F_t}(e,\tau)-\Psi_F(e,\tau)\}}
_{\Delta_{F,t}(\tau)}. \label{eq:expectile_map_decomp}
\end{align}
Define \(a_\tau(F_t,u):=\int w_\tau(y,u)\,dF_t(y)\). For fixed \(y\), the function \(u\mapsto w_\tau(y,u)(u-y)\) is continuous and piecewise linear, with derivative \(w_\tau(y,u)\) almost everywhere. Hence
\[
w_\tau(y,e_t)(e_t-y)-w_\tau(y,e)(e-y)
=\int_e^{e_t}w_\tau(y,u)\,du.
\]
Integrating against \(dF_t(y)\) and applying Fubini's theorem, justified by \(0\le w_\tau\le1\), gives
\[
\Delta_{e,t}(\tau)
=\int_e^{e_t}a_\tau(F_t,u)\,du
=\bar a_{t,\tau}(e_t-e),
\]
where \(\bar a_{t,\tau}\) denotes the average of
\(a_\tau(F_t,\cdot)\) between \(e\) and \(e_t\), with the natural value
\(a_\tau(F_t,e)\) when \(e_t=e\). Set
\(H_t:=F_t-F=t h_t\). Splitting at \(e\) and applying
Lebesgue--Stieltjes integration by parts, the distribution-perturbation
component \(\Delta_{F,t}(\tau)\) in \eqref{eq:expectile_map_decomp} satisfies
\[
\begin{aligned}
\Delta_{F,t}(\tau)
&=(1-\tau)\int_{(-\infty,e)}(e-y)\,dH_t(y)
+\tau\int_{[e,\infty)}(e-y)\,dH_t(y)\\
&=(1-\tau)\int_{-\infty}^{e}H_t(y)\,dy
+\tau\int_{e}^{\infty}H_t(y)\,dy\\
&=\int_{\mathbb R}w_\tau(y,e)H_t(y)\,dy\\
&=t\int w_\tau(y,e)h_t(y)\,dy.
\end{aligned}
\]
The terms at \(e\) vanish because \(e-y=0\) there. The tail terms vanish
because the finite first moments of \(F_t\) and \(F\) imply the corresponding
tail limits \(y\big(1-F_t(y)\big),y\big(1-F(y)\big)\to0\) as \(y\to\infty\) and
\(|y|F_t(y),|y|F(y)\to0\) as \(y\to-\infty\).
Substitution of these two expressions into
\Cref{eq:expectile_map_decomp} gives
\[
\frac{e_t-e}{t}
=-\bar a_{t,\tau}^{-1}
\int w_\tau(y,e)h_t(y)\,dy.
\]
Set \(c:=\min\{\underline\tau,1-\overline\tau\}>0\).
Since \(\bar a_{t,\tau}\geq c\) and \(0\leq w_\tau\leq1\), the same
identity yields the bound
\begin{equation}\label{eq:expectile_root_displacement}
\sup_{\tau\in\T}|e_\tau(F_t)-e_\tau(F)|
\leq c^{-1}\|F_t-F\|_{L^1}=O(|t|).
\end{equation}
Let \(I:=[e_{\underline\tau}(F),e_{\overline\tau}(F)]\), and put
\[
\omega_F(r):=\sup_{x\in I}\{F(x+2r)-F(x-2r)\}.
\]
Continuity of \(F\) at every point of the compact interval \(I\)
implies \(\omega_F(r)\to0\) as \(r\downarrow0\).
For fixed \(r>0\), \eqref{eq:expectile_root_displacement} places every
\(u\) between \(e_\tau(F_t)\) and \(e_\tau(F)\) within distance \(r\)
of \(e_\tau(F)\) for all sufficiently small \(|t|\).
Monotonicity of the CDFs, averaging over \([u-r,u]\) and \([u,u+r]\),
then gives
\[
|F_t(u-)-F(e_\tau(F))|
\leq r^{-1}\|F_t-F\|_{L^1}+\omega_F(r).
\]
Because \(a_\tau(F_t,u)=\tau+(1-2\tau)F_t(u-)\) and
\(a_\tau(F)=\tau+(1-2\tau)F(e_\tau(F))\), averaging this inequality
along the root segment, first letting \(t\to0\) and then \(r\downarrow0\),
proves
\[
\sup_{\tau\in\T}|\bar a_{t,\tau}-a_\tau(F)|\longrightarrow0.
\]
Finally, the integral operator is linear and bounded on
\(L^1(\mathbb R)\), uniformly in \((e,\tau)\), so \(h_t\to h\) in
\(L^1\) gives the asserted derivative uniformly over \(\T\). Hence
\[
\frac{e_t(\tau)-e_\tau(F)}{t}
\longrightarrow
-a_\tau(F)^{-1}
\int w_\tau\bigl(y,e_\tau(F)\bigr)h(y)\,dy
\]
in \(\ell^\infty(\T)\). The limit is \(\mathcal E'_F[h]\). Since the admissible perturbation sequence was arbitrary, and \(\|\mathcal E'_F[h]\|_{\T}\le c^{-1}\|h\|_{L^1}\), this proves the asserted Hadamard differentiability with a continuous linear derivative.

\noindent\textit{Part~\ref{prop:expectile_map_weak}.}
Applying the ordinary functional delta method to part~\ref{prop:expectile_map_diff} and the stated \(L^1\) weak convergence gives the conclusion; see \citet[Lemma~B.1]{chernozhukov-fernandez-melly-2013inference}.

\noindent\textit{Part~\ref{prop:expectile_map_boot}.}
Applying the bootstrap functional delta method to part~\ref{prop:expectile_map_diff}, under the premises of parts~\ref{prop:expectile_map_weak} and~\ref{prop:expectile_map_boot}, gives the conclusion; see \citet[Lemma~B.3]{chernozhukov-fernandez-melly-2013inference}.
The uniform slope lower bound also bounds the displacement of an approximate bootstrap root from the exact root by its absolute score residual divided by \(c\). Thus the stipulated uniformly negligible residuals do not affect the bootstrap limit. \qed

\section{Supplementary Monte Carlo Experiments and Implementation}\label{app:mc_supplement}

\subsection{Data-generating processes and inference implementation}\label{sec:mc_implementation}

The baseline DGPs, treatment assignment, cluster-size arrays, and simulation settings are given in \Cref{sec:mc_design}. Their true coefficient paths are
\[
\begin{aligned}
\theta_o^{DR}(y)&=(y+0.5,-0.5)',\\
\theta_o^{QR}(\tau)&=\theta_o+\gamma\Phi^{-1}(\tau),\\
\theta_o^{ER}(\tau)&=\theta_o+\gamma\E_\tau(U),
\end{aligned}
\]
where \(U\) is standard normal and \(\E_\tau(U)\) is its \(\tau\)-expectile. In particular, the DR conditional CDF is \(F_{Y\mid T}(y\mid T_g)=\Lambda(X_i'\theta_o^{DR}(y))\), with logistic link \(\Lambda\).

For the baseline Pareto arrays, \(\beta=2.5\) gives \(N\asymp G\) and
\(\max_gN_g^2/N\asymp G^{-0.2}\to0\). When \(\beta=1\), \(N\asymp G\log G\),
\(\max_gN_g^2/N\asymp G/\log G\), and \(\max_gN_g/N\asymp1/\log G\to0\).

For conventional QR inference, CR-A uses an observation-weighted Jacobian estimated by a normal-kernel residual-density plug-in with bandwidth \(h=1.06s_rN^{-1/5}\), where \(s_r\) is the QR residual standard deviation. The remaining inference procedures are described in \Cref{sec:mc_pointwise,sec:mc_functional}.

\subsection{Evaluation grids and performance measures}\label{sec:mc_grid_metrics}
Let \(R_{\mathrm{MC}}\) denote the number of Monte Carlo replications, \(\widehat\theta_s\) the estimate in replication \(s\), and \(\theta_o\) its true value. The pointwise measures are
\[
\operatorname{Bias}=R_{\mathrm{MC}}^{-1}\sum_{s=1}^{R_{\mathrm{MC}}}(\widehat\theta_s-\theta_o),
\qquad
\operatorname{RMSE}=\left\{R_{\mathrm{MC}}^{-1}\sum_{s=1}^{R_{\mathrm{MC}}}(\widehat\theta_s-\theta_o)^2\right\}^{1/2}.
\]

For each target \(f\in\{F,Q,E\}\), the evaluation grid contains
\( \displaystyle 
L_G^{(f)}
=1+\Big\lceil
\frac{5M_f^{1/2}S_fG^{0.3}}{2\sqrt2}
\Big\rceil
\)
equally spaced points, where \(S_f\) is the length of the evaluation set and \(M_f\) is its target-specific population curvature envelope.\footnote{The rule follows \citet[Lemma~2.1 and Theorems~3.1--3.2]{tsyawo-2026grid} with the known Monte Carlo DGP supplying \(M_F=0.096\), \(M_Q=41.609\), and \(M_E=38.434\).}
For a given target \(f\) and cluster count \(G\), write \(L=L_G^{(f)}\). For grid points \(z_1,\ldots,z_L\), the reported integrated bias and integrated RMSE are
\[
\begin{aligned}
\operatorname{IBias}
=\bigg(L^{-1}\sum_{\ell=1}^L
\big(\overline{\widehat\Delta}(z_\ell)-\Delta(z_\ell)\big)^2\bigg)^{1/2},\quad
\operatorname{IRMSE}
=\bigg((R_{\mathrm{MC}}L)^{-1}\sum_{s=1}^{R_{\mathrm{MC}}}\sum_{\ell=1}^L
\big(\widehat\Delta_s(z_\ell)-\Delta(z_\ell)\big)^2\bigg)^{1/2}.
\end{aligned}
\] 
Here \(\overline{\widehat\Delta}(z_\ell)=R_{\mathrm{MC}}^{-1}\sum_{s=1}^{R_{\mathrm{MC}}}\widehat\Delta_s(z_\ell)\). IBias measures systematic displacement of the estimated path, whereas IRMSE also incorporates its Monte Carlo variation. Avg. Band and U-Rej. are defined in \Cref{sec:mc_functional}.

\subsection{Sensitivity to within-cluster dependence}\label{sec:mc_dependence}

\Cref{tab:mc_dependence} repeats the pointwise comparison at latent correlations \(0.25\) and \(0.50\). In the ignorable \(\beta=2.5\) regime, SACR/conventional RMSE ratios range from \(0.94\) to \(1.01\) at \(\rho=0.25\), and from \(0.89\) to \(0.96\) at \(\rho=0.50\), indicating little efficiency cost in these designs. Under \(\beta=1\), SACR retains substantial RMSE and rejection-rate advantages. At \(G=600\), its bootstrap rejection rates across both correlations and all models lie between \(5.1\%\) and \(5.7\%\), compared with \(18.8\%\) to \(26.4\%\) conventionally. The principal smaller-sample distortion is ER at \(\rho=0.25\) and \(G=150\) in the ignorable regime, where SACR-B rejects \(9.3\%\) of the time.

\begin{table}[!htbp]
\centering
\caption{Pointwise sensitivity to within-cluster dependence}
\label{tab:mc_dependence}
\resizebox{\textwidth}{!}{%
\begin{tabular}{ccc*{3}{rrr}}
\toprule
 & & & \multicolumn{3}{c}{DR} & \multicolumn{3}{c}{QR} & \multicolumn{3}{c}{ER} \\
\cmidrule(lr){4-6}\cmidrule(lr){7-9}\cmidrule(lr){10-12}
$\rho$ & $\beta$ & $G$ & Rel. RMSE & SACR-B rej. & CR-B rej.
& Rel. RMSE & SACR-B rej. & CR-B rej.
& Rel. RMSE & SACR-B rej. & CR-B rej. \\
\midrule
0.25 & 2.5 & 150 & 1.0059 & 0.0600 & 0.0590 & 1.0097 & 0.0750 & 0.0850 & 0.9877 & 0.0930 & 0.0680 \\
0.25 & 2.5 & 300 & 0.9924 & 0.0560 & 0.0670 & 0.9739 & 0.0540 & 0.0580 & 0.9497 & 0.0470 & 0.0600 \\
0.25 & 2.5 & 600 & 0.9970 & 0.0660 & 0.0540 & 0.9582 & 0.0620 & 0.0650 & 0.9444 & 0.0500 & 0.0520 \\
\midrule
0.25 & 1 & 150 & 0.4342 & 0.0550 & 0.1760 & 0.3935 & 0.0660 & 0.2100 & 0.3552 & 0.0660 & 0.2200 \\
0.25 & 1 & 300 & 0.3233 & 0.0490 & 0.2140 & 0.3032 & 0.0470 & 0.2820 & 0.2691 & 0.0610 & 0.2720 \\
0.25 & 1 & 600 & 0.2565 & 0.0530 & 0.1880 & 0.2322 & 0.0520 & 0.2430 & 0.2075 & 0.0570 & 0.2390 \\
\midrule
0.50 & 2.5 & 150 & 0.9530 & 0.0620 & 0.0640 & 0.9647 & 0.0700 & 0.0740 & 0.9351 & 0.0780 & 0.0730 \\
0.50 & 2.5 & 300 & 0.9397 & 0.0570 & 0.0640 & 0.9170 & 0.0410 & 0.0460 & 0.9001 & 0.0480 & 0.0590 \\
0.50 & 2.5 & 600 & 0.9422 & 0.0570 & 0.0560 & 0.9227 & 0.0490 & 0.0550 & 0.8945 & 0.0580 & 0.0570 \\
\midrule
0.50 & 1 & 150 & 0.3321 & 0.0510 & 0.1910 & 0.2910 & 0.0540 & 0.2420 & 0.2894 & 0.0640 & 0.2290 \\
0.50 & 1 & 300 & 0.2388 & 0.0380 & 0.2040 & 0.2203 & 0.0490 & 0.3100 & 0.2200 & 0.0640 & 0.2690 \\
0.50 & 1 & 600 & 0.1874 & 0.0510 & 0.1920 & 0.1692 & 0.0540 & 0.2640 & 0.1694 & 0.0550 & 0.2480 \\
\bottomrule
\end{tabular}}
\par\smallskip
\begin{minipage}{\textwidth}\footnotesize
Relative RMSE is SACR RMSE divided by conventional RMSE; values above one
indicate an efficiency cost. Rejection frequencies concern nominal 5\%
pointwise tests. Both SACR-B and CR-B use bootstrap sample standard deviations. $\rho$ is the Gaussian latent error correlation; DR errors use the existing logistic transformation. Each cell uses 1,000 Monte Carlo replications and 499 bootstrap draws.
\end{minipage}
\end{table}

\subsection{Matched exponential-weight bootstrap}\label{sec:mc_matched_exp}

\Cref{tab:mc_matched_exp} holds the bootstrap scheme and scale construction common across the two estimators: each replication uses the same \(\operatorname{Exp}(1)\) cluster draws, and both procedures use bootstrap sample standard deviations. Under \(\beta=1\) at \(G=600\), conventional rejection rates are \(43.5\%\), \(75.1\%\), and \(60.6\%\) for DR, QR, and ER, compared with \(5.0\%\), \(6.5\%\), and \(5.8\%\) for SACR. SACR also has shorter average intervals in these cells. Thus its ULC advantage persists when the bootstrap draws and scale construction are aligned. In the ignorable regime, SACR rejection frequencies are close to nominal across models and cluster counts. This diagnostic supplements the pairs-bootstrap benchmark and does not assert conventional exponential-bootstrap validity under ULC.

\begin{table}[!htbp]
\centering
\caption{Pointwise inference with matched exponential cluster weights}
\label{tab:mc_matched_exp}
\resizebox{\textwidth}{!}{%
\begin{tabular}{cc*{3}{rrr}}
\toprule
 & & \multicolumn{3}{c}{DR} & \multicolumn{3}{c}{QR} & \multicolumn{3}{c}{ER} \\
\cmidrule(lr){3-5}\cmidrule(lr){6-8}\cmidrule(lr){9-11}
$\beta$ & $G$ & SACR rej. & CR rej. & Rel. length
& SACR rej. & CR rej. & Rel. length
& SACR rej. & CR rej. & Rel. length \\
\midrule
2.5 & 150 & 0.0450 & 0.0700 & 0.9287 & 0.0670 & 0.0830 & 0.9000 & 0.0650 & 0.0740 & 0.9301 \\
2.5 & 300 & 0.0380 & 0.0470 & 0.9122 & 0.0550 & 0.0730 & 0.8833 & 0.0480 & 0.0630 & 0.9046 \\
2.5 & 600 & 0.0530 & 0.0660 & 0.8870 & 0.0670 & 0.0730 & 0.8543 & 0.0570 & 0.0630 & 0.8768 \\
\midrule
1 & 150 & 0.0320 & 0.4560 & 0.4913 & 0.0600 & 0.6730 & 0.7851 & 0.0490 & 0.5780 & 0.7293 \\
1 & 300 & 0.0570 & 0.4230 & 0.3758 & 0.0740 & 0.7080 & 0.5653 & 0.0640 & 0.5830 & 0.5456 \\
1 & 600 & 0.0500 & 0.4350 & 0.2785 & 0.0650 & 0.7510 & 0.4006 & 0.0580 & 0.6060 & 0.4040 \\
\bottomrule
\end{tabular}}
\par\smallskip
\begin{minipage}{\textwidth}\footnotesize
SACR-EXP-SD and CR-EXP-SD use identical exponential cluster draws within
each Monte Carlo replication and bootstrap sample standard deviations.
Relative length is SACR average interval length divided by conventional
average interval length; values below one indicate shorter SACR intervals.
The procedures use inverse-cluster-size and individual weighting,
respectively. The latent correlation is \(0.96\), with 1,000 replications,
499 bootstrap draws, and nominal 95\% normal intervals. All cluster counts
and both Pareto regimes are reported; full diagnostics remain in the
replication outputs.
\end{minipage}
\end{table}

\subsection{Single-dominant-cluster experiments}\label{sec:sup_mc}

The single-dominant-cluster experiments retain the DGPs and inference settings of \Cref{sec:mc_pointwise,sec:mc_functional}, replacing only the cluster-size array. The Pareto \(\beta=1\) regime violates the negligibility condition for conventional \(\sqrt N\)-inference, but its largest cluster's sample share still vanishes, leaving a route to consistency. The share in this case, on the contrary, stays bounded away from zero, removing this negligibility route for the conventional estimator as well.

Let \(s_{\mathrm{dom}}=0.3\) denote the target sample share of the dominant cluster. Define
\[
N=\operatorname{round}\!\left(\frac{G-1}{1-s_{\mathrm{dom}}}\right),
\qquad
N_G=N-(G-1),
\qquad
N_g=1,\quad g=1,\ldots,G-1.
\]
The dominant cluster is placed at \(g=G\). It is therefore treated under the common assignment rule \(T_g=\indicator{g\equiv0\pmod 5}\), matching the treatment status of the largest cluster in the ordered Pareto designs. Integer rounding makes the realised dominant share approximately, rather than exactly, \(0.3\). Because \(N_G/N\to s_{\mathrm{dom}}\), the dominant cluster remains asymptotically non-negligible and \(\max_gN_g^2/N\) diverges at rate \(G\). Under SACR, however, its cluster-average score still receives weight \(1/G\), and its contribution to the cluster-balanced covariance kernel is therefore \(\mathcal{O}(G^{-1})\). The stable proportions of treated and control clusters among the remaining clusters preserve the rank and covariance limits required by the SACR theory.

\subsubsection{Pointwise coefficient inference}\label{sec:sup_mc_ds_point}

The pointwise single-dominant-cluster experiment described in \Cref{sec:sup_mc} follows the targets, procedures, and performance measures in \Cref{sec:mc_pointwise}.

\begin{table}[!htbp]
\centering
\caption{Pointwise Monte Carlo results under the single-dominant-cluster design (\(s_{\mathrm{dom}}=0.3\))}
\label{tab:mc_point_ds}
\begin{threeparttable}
\setlength{\tabcolsep}{2pt}
\footnotesize
\begin{tabular}{lcccccccccccc}
\toprule
& \multicolumn{4}{c}{DR: \(y=Q_Y(0.5)\)}
& \multicolumn{4}{c}{QR: \(\tau=0.5\)}
& \multicolumn{4}{c}{ER: \(\tau=0.5\)} \\
\cmidrule(lr){2-5}\cmidrule(lr){6-9}\cmidrule(lr){10-13}
Method
& Bias & RMSE & Avg. SE & Rej.
& Bias & RMSE & Avg. SE & Rej.
& Bias & RMSE & Avg. SE & Rej. \\
\midrule
\multicolumn{13}{l}{\textbf{$G = 150$}} \\ \cmidrule{1-1}
SACR-A & \multirow{2}{*}{0.0004} & \multirow{2}{*}{0.4250} & 0.4204 & 0.0420 & \multirow{2}{*}{$-0.0230$} & \multirow{2}{*}{0.2662} & 0.2732 & 0.0740 & \multirow{2}{*}{0.0002} & \multirow{2}{*}{0.2223} & 0.2132 & 0.0730 \\
SACR-B &  &  & 0.4268 & 0.0380 &  &  & 0.2866 & 0.0550 &  &  & 0.2096 & 0.0810 \\
CR-A & \multirow{2}{*}{$-0.1544$} & \multirow{2}{*}{1.5521} & 0.7540 & 0.4610 & \multirow{2}{*}{0.0242} & \multirow{2}{*}{1.1097} & 0.1082 & 0.9270 & \multirow{2}{*}{0.0182} & \multirow{2}{*}{0.8192} & 0.2297 & 0.7590 \\
CR-B &  &  & 0.9712 & 0.1290 &  &  & 0.4944 & 0.3240 &  &  & 0.3974 & 0.2520 \\
\midrule
\multicolumn{13}{l}{\textbf{$G = 300$}} \\ \cmidrule{1-1}
SACR-A & \multirow{2}{*}{$-0.0279$} & \multirow{2}{*}{0.2996} & 0.2958 & 0.0500 & \multirow{2}{*}{$-0.0103$} & \multirow{2}{*}{0.1847} & 0.1936 & 0.0540 & \multirow{2}{*}{$-0.0041$} & \multirow{2}{*}{0.1525} & 0.1538 & 0.0450 \\
SACR-B &  &  & 0.2985 & 0.0460 &  &  & 0.2008 & 0.0470 &  &  & 0.1525 & 0.0470 \\
CR-A & \multirow{2}{*}{$-0.2032$} & \multirow{2}{*}{1.5081} & 0.6957 & 0.5850 & \multirow{2}{*}{$-0.0051$} & \multirow{2}{*}{1.0757} & 0.0943 & 0.9490 & \multirow{2}{*}{$-0.0042$} & \multirow{2}{*}{0.7890} & 0.2145 & 0.8300 \\
CR-B &  &  & 0.8802 & 0.1610 &  &  & 0.4580 & 0.4060 &  &  & 0.3663 & 0.2800 \\
\midrule
\multicolumn{13}{l}{\textbf{$G = 600$}} \\ \cmidrule{1-1}
SACR-A & \multirow{2}{*}{$-0.0057$} & \multirow{2}{*}{0.2045} & 0.2083 & 0.0430 & \multirow{2}{*}{$-0.0061$} & \multirow{2}{*}{0.1347} & 0.1384 & 0.0570 & \multirow{2}{*}{0.0025} & \multirow{2}{*}{0.1061} & 0.1096 & 0.0470 \\
SACR-B &  &  & 0.2084 & 0.0410 &  &  & 0.1407 & 0.0550 &  &  & 0.1092 & 0.0510 \\
CR-A & \multirow{2}{*}{$-0.1490$} & \multirow{2}{*}{1.4965} & 0.6917 & 0.6300 & \multirow{2}{*}{$-0.0022$} & \multirow{2}{*}{1.0796} & 0.0874 & 0.9650 & \multirow{2}{*}{$-0.0002$} & \multirow{2}{*}{0.7924} & 0.2091 & 0.8550 \\
CR-B &  &  & 0.8747 & 0.0990 &  &  & 0.4422 & 0.4760 &  &  & 0.3544 & 0.2810 \\
\bottomrule
\end{tabular}
\begin{tablenotes}
\item Targets, cluster counts, and inference procedures are as in \Cref{tab:mc_point_beta25}; only the cluster-size array differs. Bias and RMSE are shared within each estimator's A/B pair.
\end{tablenotes}
\end{threeparttable}
\end{table}

The single-dominant-cluster results sharpen the contrast with the Pareto designs. Pointwise SACR RMSE declines as \(G\) rises. At \(G=600\), SACR-B rejection rates are \(4.1\%\), \(5.5\%\), and \(5.1\%\) for DR, QR, and ER, all close to nominal. Conventional RMSE remains large: CR-A rejects \(63.0\%\), \(96.5\%\), and \(85.5\%\). The pairs bootstrap reduces these rates but still rejects \(9.9\%\), \(47.6\%\), and \(28.1\%\), respectively.

\subsubsection{Uniform inference on distributional treatment effects}\label{sec:sup_mc_functional}

The functional single-dominant-cluster experiment described in \Cref{sec:sup_mc} follows the targets, grids, procedures, and performance measures in \Cref{sec:mc_functional}.

\begin{table}[!htbp]
\centering
\caption{Uniform Monte Carlo results for distributional treatment effects under the single-dominant-cluster design (\(s_{\mathrm{dom}}=0.3\))}
\label{tab:mc_unif_ds}
\begin{threeparttable}
\setlength{\tabcolsep}{2pt}
\footnotesize
\begin{tabular}{lcccccccccccc}
\toprule
& \multicolumn{4}{c}{DF treatment effect}
& \multicolumn{4}{c}{QF treatment effect}
& \multicolumn{4}{c}{EF treatment effect} \\
\cmidrule(lr){2-5}\cmidrule(lr){6-9}\cmidrule(lr){10-13}
Method
& IBias & IRMSE & Avg. Band & U-Rej.
& IBias & IRMSE & Avg. Band & U-Rej.
& IBias & IRMSE & Avg. Band & U-Rej. \\
\midrule
\multicolumn{13}{l}{\textbf{$G = 150$}} \\ \cmidrule{1-1}
SACR-B & 0.0017 & 0.0847 & 0.2270 & 0.1530 & 0.0359 & 0.2995 & 2.6907 & 0.0340 & 0.0128 & 0.2304 & 0.5338 & 0.1210 \\
CR-B & 0.0035 & 0.2608 & 0.5191 & 0.1940 & 0.4174 & 1.0670 & 6.1238 & 0.0570 & 0.1776 & 0.7875 & 1.5282 & 0.1160 \\
\midrule
\multicolumn{13}{l}{\textbf{$G = 300$}} \\ \cmidrule{1-1}
SACR-B & 0.0018 & 0.0595 & 0.1633 & 0.0930 & 0.0172 & 0.2123 & 1.5278 & 0.0420 & 0.0069 & 0.1599 & 0.3832 & 0.0890 \\
CR-B & 0.0066 & 0.2575 & 0.4791 & 0.1850 & 0.4189 & 1.0451 & 5.5245 & 0.0640 & 0.1788 & 0.7642 & 1.4204 & 0.0960 \\
\midrule
\multicolumn{13}{l}{\textbf{$G = 600$}} \\ \cmidrule{1-1}
SACR-B & 0.0007 & 0.0423 & 0.1170 & 0.0710 & 0.0092 & 0.1519 & 0.9085 & 0.0340 & 0.0029 & 0.1176 & 0.2726 & 0.0770 \\
CR-B & 0.0026 & 0.2566 & 0.4403 & 0.2060 & 0.4131 & 1.0453 & 5.2497 & 0.0780 & 0.1743 & 0.7635 & 1.3531 & 0.1290 \\
\bottomrule
\end{tabular}
\begin{tablenotes}
\item Targets, supports, grids, and bootstrap procedures are as in \Cref{tab:mc_unif_beta25}; only the cluster-size array differs.
\end{tablenotes}
\end{threeparttable}
\end{table}

For the treatment-effect paths, SACR-B integrated RMSE also declines with \(G\). At \(G=600\), its uniform rejection frequencies are \(7.1\%\), \(3.4\%\), and \(7.7\%\) for DF, QF, and EF, compared with \(20.6\%\), \(7.8\%\), and \(12.9\%\) for CR-B. The conventional QF rejection rate is reasonably close to nominal, but this alone does not indicate comparable performance, because its average band half-width is \(5.2497\), against \(0.9085\) for SACR-B, while its integrated RMSE is \(1.0453\), against \(0.1519\). Thus conventional bootstrap bands can be extremely wide without resolving the substantial estimation error induced by the dominant treated cluster.

\section{Empirical Illustration}\label{sect:empirics}

The illustration revisits \citet{Akhtari2022}'s study of political turnover and student achievement in Brazilian municipal schools, replicating the fourth- and eighth-grade specifications in Table~3, Panels A and B, Columns 3--6. Observations are clustered by municipality, with substantially unequal cluster sizes. Pointwise OLS and LAD comparisons maintain common-coefficient RD specifications; distributional effects use municipality- and student-weighted cut-off distributions.\footnote{\citet{mackinnon-nielsen-webb-2023fast} and \citet[version~3]{chiang-sasaki-wang-2023subsampling} also use this empirical application to highlight similar issues related to cluster-size variation.}

\subsection{Empirical design, identification, and cluster-size diagnostics}

The replicated specification is
\[
\begin{aligned}
\text{score}_{gi,t+1}
= {} & \kappa
+
\theta_1\indicator{\text{IVM}_{gt} < 0}
+
\theta_2\text{IVM}_{gt}
+
\theta_3\indicator{\text{IVM}_{gt} < 0}\times\text{IVM}_{gt}
\\ & {}+\theta_4\text{score}^{\mathrm{base}}_{gi,t}
+C_{gi,t}'\Gamma
+
U_{gi,t},
\end{aligned}
\]
where \(g,t,i\) index municipalities, election cycles, and students. The outcome is the post-election test score; \(\text{score}^{\mathrm{base}}_{gi,t}\) is the baseline school mean, and \(C_{gi,t}\) contains the remaining controls. The incumbent party's vote margin \(\text{IVM}_{gt}\) is the running variable, with negative margins indicating turnover. Each municipality cluster contains all included student--cycle observations, permitting dependence across students and election cycles. Columns 3--4 use the fixed window \(0.07\), and Columns 5--6 use \(0.11\); the second specification in each pair adds full school and student controls and an election-cycle indicator. These windows provide a limited bandwidth-sensitivity comparison. The coefficient \(\theta_1\) represents the common local effect of turnover.

The identifying logic is the sharp regression-discontinuity (RD) comparison at the zero vote-margin cut-off. Let \(R\) denote the incumbent party's vote margin, let \(D=\indicator{R<0}\) denote political turnover, and let \(X=(X_{\mathrm{base}},C')'\) collect the predetermined baseline score, the remaining school and student characteristics, and the election-cycle indicator. Continuity at \(R=0\) of the potential-outcome distributions and the distribution of \(X\), together with consistency and the absence of precise manipulation of the cut-off, identifies the two side-specific potential-outcome distributions at the cut-off. In the maintained additive specification, the conditional mean discontinuity is \(\theta_1\).

For bandwidth \(h\), let \(V_{gti}(h)\) denote the regressor vector in the specification above. The maintained implication required by the Z-SACR theory is the municipality-level score restriction
\[
\E_G\!\left[\frac{1}{N_g}\sum_{(t,i)\in\G_g}V_{gti}(h)U_{gi,t}\right]=0
\qquad(g=1,\ldots,G).
\]
The RD design identifies the cut-off comparison, while the exact score restriction is the modelling condition that represents it through a common coefficient. Provided the corresponding cluster-balanced second-moment matrix is nonsingular, deterministic cluster-size adjustment preserves this coefficient.

At \(\tau=0.5\), ER coincides with OLS and QR with LAD. For LAD, the maintained additive local-linear conditional-median specification and continuity of potential-outcome conditional medians at the cut-off identify \(\theta_1\) as a conditional-median RD contrast common across covariates, distinct from the marginal Quantile Treatment Effect (QTE) at the median obtained by inverting cut-off distributions.

The conventional OLS coefficients and observation and municipality counts reproduce the original specifications to the reported precision. Pointwise inference follows the pairs and exponential-weight cluster-bootstrap procedures in \Cref{sec:mc_pointwise}, using \(499\) draws and bootstrap SD standard errors rather than the original study's variance calculation.

Across specifications, \(\max_gN_g^2/N\) ranges from \(10.4\) to \(39.7\), indicating substantial cluster-size concentration, while the largest municipality shares range from \(0.80\%\) to \(1.16\%\). \Cref{tab:results_expec,tab:results_quant} report the squared-size diagnostics by specification.

\subsection{Pointwise coefficient evidence}

\begin{table}[H]
  \centering
  \caption{OLS estimates across fixed-bandwidth specifications}
  \label{tab:results_expec}
  \scriptsize
  \begin{minipage}{\textwidth}\centering
    \resizebox{0.92\textwidth}{!}{%
    \begin{tabular}{ccccccccccccc}
      \toprule
      & & & & & & & \multicolumn{3}{c}{Z-SACR-B} & \multicolumn{3}{c}{Standard CR-B} \\
      \cmidrule(lr){8-10} \cmidrule(lr){11-13}
      Grade & Col. & Band & Controls & $N$ & $G$ & $\max_g N_g^2/N$ & Coef. & SE & $p$-value & Coef. & SE & $p$-value \\
      \midrule

4 & 3 & 0.07 & No & 295,170 & 1,538 & 39.7 & $-0.1066$ & 0.0315 & 0.0007 & $-0.0907$ & 0.0307 & 0.0031 \\
4 & 4 & 0.07 & Yes & 295,170 & 1,538 & 39.7 & $-0.0912$ & 0.0292 & 0.0018 & $-0.0748$ & 0.0269 & 0.0054 \\
4 & 5 & 0.11 & No & 429,979 & 2,101 & 27.2 & $-0.0829$ & 0.0260 & 0.0014 & $-0.0668$ & 0.0235 & 0.0045 \\
4 & 6 & 0.11 & Yes & 429,979 & 2,101 & 27.2 & $-0.0661$ & 0.0248 & 0.0077 & $-0.0554$ & 0.0219 & 0.0114 \\
8 & 3 & 0.07 & No & 126,855 & 965 & 10.4 & $-0.0303$ & 0.0359 & 0.3992 & $-0.0496$ & 0.0298 & 0.0959 \\
8 & 4 & 0.07 & Yes & 126,855 & 965 & 10.4 & $-0.0226$ & 0.0344 & 0.5115 & $-0.0458$ & 0.0305 & 0.1335 \\
8 & 5 & 0.11 & No & 191,169 & 1,335 & 13.3 & $-0.0311$ & 0.0293 & 0.2899 & $-0.0589$ & 0.0241 & 0.0145 \\
8 & 6 & 0.11 & Yes & 191,169 & 1,335 & 13.3 & $-0.0205$ & 0.0290 & 0.4797 & $-0.0488$ & 0.0230 & 0.0335 \\
\bottomrule
    \end{tabular}
    }
    \par\smallskip{\raggedright
      \footnotesize
      \noindent\textit{Notes:} The table reports ER estimates at \(\tau=0.5\), which coincide with OLS estimates under the linear conditional mean specification.
    \par}
  \end{minipage}
\end{table}

\begin{table}[H]
	\centering
	\caption{LAD estimates across fixed-bandwidth specifications}
	\label{tab:results_quant}
	\scriptsize
	\begin{minipage}{\textwidth}\centering
		\resizebox{0.92\textwidth}{!}{%
			\begin{tabular}{ccccccccccccc}
				\toprule
				& & & & & & & \multicolumn{3}{c}{Z-SACR-B} & \multicolumn{3}{c}{Standard CR-B} \\
				\cmidrule(lr){8-10} \cmidrule(lr){11-13}
				Grade & Col. & Band & Controls & $N$ & $G$ & $\max_g N_g^2/N$ & Coef. & SE & $p$-value & Coef. & SE & $p$-value \\
				\midrule

4 & 3 & 0.07 & No & 295,170 & 1,538 & 39.7 & $-0.1130$ & 0.0321 & 0.0004 & $-0.0898$ & 0.0325 & 0.0057 \\
4 & 4 & 0.07 & Yes & 295,170 & 1,538 & 39.7 & $-0.1011$ & 0.0323 & 0.0017 & $-0.0732$ & 0.0278 & 0.0084 \\
4 & 5 & 0.11 & No & 429,979 & 2,101 & 27.2 & $-0.0936$ & 0.0278 & 0.0008 & $-0.0667$ & 0.0248 & 0.0072 \\
4 & 6 & 0.11 & Yes & 429,979 & 2,101 & 27.2 & $-0.0735$ & 0.0265 & 0.0055 & $-0.0567$ & 0.0235 & 0.0157 \\
8 & 3 & 0.07 & No & 126,855 & 965 & 10.4 & $-0.0441$ & 0.0380 & 0.2454 & $-0.0490$ & 0.0308 & 0.1110 \\
8 & 4 & 0.07 & Yes & 126,855 & 965 & 10.4 & $-0.0333$ & 0.0350 & 0.3412 & $-0.0464$ & 0.0331 & 0.1607 \\
8 & 5 & 0.11 & No & 191,169 & 1,335 & 13.3 & $-0.0481$ & 0.0318 & 0.1306 & $-0.0651$ & 0.0272 & 0.0165 \\
8 & 6 & 0.11 & Yes & 191,169 & 1,335 & 13.3 & $-0.0351$ & 0.0314 & 0.2641 & $-0.0571$ & 0.0287 & 0.0467 \\
\bottomrule
			\end{tabular}
		}
		\par\smallskip{\raggedright
			\footnotesize
				\noindent\textit{Notes:} The table reports QR estimates at \(\tau=0.5\), that is, LAD estimates under the maintained local-linear conditional-median specification.
		\par}
	\end{minipage}
\end{table}

The pointwise results in \Cref{tab:results_quant,tab:results_expec} show that cluster-size adjustment can change both magnitudes and inferential conclusions under a potentially ULC setting where the conventional observation-weighted inference may be misleading. For fourth-grade scores, Z-SACR produces more negative LAD and OLS estimates than the corresponding standard estimates across all fixed-bandwidth specifications. For eighth-grade scores, the adjustment moves the estimates in the opposite direction: the negative effects are attenuated, and none of the Z-SACR estimates is significant at \(5\%\); conventional estimates remain significant in the wider-bandwidth specifications.

\subsection{Distributional effects and uniform inference}

Following the distributional RD approach of \citet{frandsen-frolich-melly-2012}, effects are constructed from side-specific conditional cut-off CDFs standardised over a common covariate distribution.

The distributional analysis uses the fully specified fourth-grade Column~6 sample with the original study's uniform bandwidth \(h=0.11\). Write \(Y(d)\) for the potential score under political-turnover state \(d\), with covariates \(X\) defined above. The causal standardisation below requires these covariates to be determined before turnover or unaffected by it. For each threshold \(y\) and covariate value \(x\), define the two conditional cut-off distributions separately under weighting scheme \(m\in\{S,C\}\). Since political turnover is defined by \(D=\indicator{R<0}\), the treated distribution uses the left-hand limit \(r\uparrow0\), and the control distribution uses the right-hand limit \(r\downarrow0\).
\[
F_1^{0,m}(y\mid x)
=\lim_{r\uparrow 0}\Prob_m[Y\leq y\mid R=r,X=x],
\quad
F_0^{0,m}(y\mid x)
=\lim_{r\downarrow 0}\Prob_m[Y\leq y\mid R=r,X=x].
\]
Here \(\Prob_S\) and \(\Prob_C\) denote municipality- and student-weighted laws. Under the RD continuity conditions, the marginal potential-outcome cut-off CDFs are obtained using the common limiting covariate distribution \(H_0^m\):
\[
F_d^{0,m}(y)=\int F_d^{0,m}(y\mid x)\,dH_0^m(x),
\qquad d\in\{0,1\}.
\]

All four local RD effects are functionals of this pair of marginal cut-off distributions: 
\( \Delta_F^{0,m}(y) = F_1^{0,m}(y)-F_0^{0,m}(y), \) \quad \( \Delta_\mu^{0,m} = \E_m[Y(1)-Y(0)\mid R=0]\allowbreak
=-\int_{\mathbb R}\Delta_F^{0,m}(y)\,dy, \) \quad \( \Delta_Q^{0,m}(\tau) = \bigl(F_1^{0,m}\bigr)^{-1}(\tau)
-\bigl(F_0^{0,m}\bigr)^{-1}(\tau), \) \quad and \quad \( \Delta_E^{0,m}(\tau) = e_\tau\bigl(F_1^{0,m}\bigr)
-e_\tau\bigl(F_0^{0,m}\bigr), \)
where \(e_\tau(F)\) denotes the \(\tau\)-expectile of \(F\). With finite first moments, \(\Delta_E^{0,m}(0.5)=\Delta_\mu^{0,m}\). QTE and the Expectile Treatment Effect (ETE) are functionals of the marginal cut-off distributions, rather than averages of conditional effects or direct QR and ER fits. Their map conditions for weak convergence and bootstrap validity are discussed in \Cref{rem:dist_scalar_functionals} and \Cref{prop:expectile_map}.

Within \(|R|\leq0.11\), separate DR fits on each side include a linear vote-margin term and controls \(X\). Predictions at \(R=0\) are averaged over the same pooled within-bandwidth covariate sample. This approximates \(H_0^m\) as the bandwidth shrinks; at the fixed bandwidth, it remains a finite-sample standardisation. Standard CR uses observation weights; Z-SACR uses inverse municipality-size weights in both fitting and covariate averaging.\footnote{The SACR covariate integrating measure is the empirical analogue, applied to \(X\), of the cluster-balanced distribution functional in \citet[Section~3.2]{nevalainen-datta-oja-2014}: it selects a municipality uniformly and then a student uniformly within that municipality in the pooled within-bandwidth sample.} The fitted marginal cut-off distributions are monotonically rearranged before inversion or integration \citep{chernozhukov-fernandezval-galichon-2010}.

Each of the \(499\) municipality-level bootstrap draws repeats the side-specific DR fits, pooled-covariate standardisation, rearrangement, and distribution-to-functional transformations. Pointwise intervals use bootstrap SDs; uniform bands and uniform homogeneity tests use normalised bootstrap IQRs. \Cref{sec:emp_grid} describes the grids and integration. These distribution-derived results illustrate implementation; their full-pipeline asymptotic validity is not established by the direct score-process results here.\footnote{Full-pipeline validity additionally requires joint weak convergence and bootstrap validity for the conditional distributions and covariate integrating measure, with appropriate centring under cluster heterogeneity; see \citet[Theorems~4.1--4.2]{chernozhukov-fernandez-melly-2013inference}.}

\begin{figure}[H]
  \centering
  \includegraphics[width=0.92\textwidth]{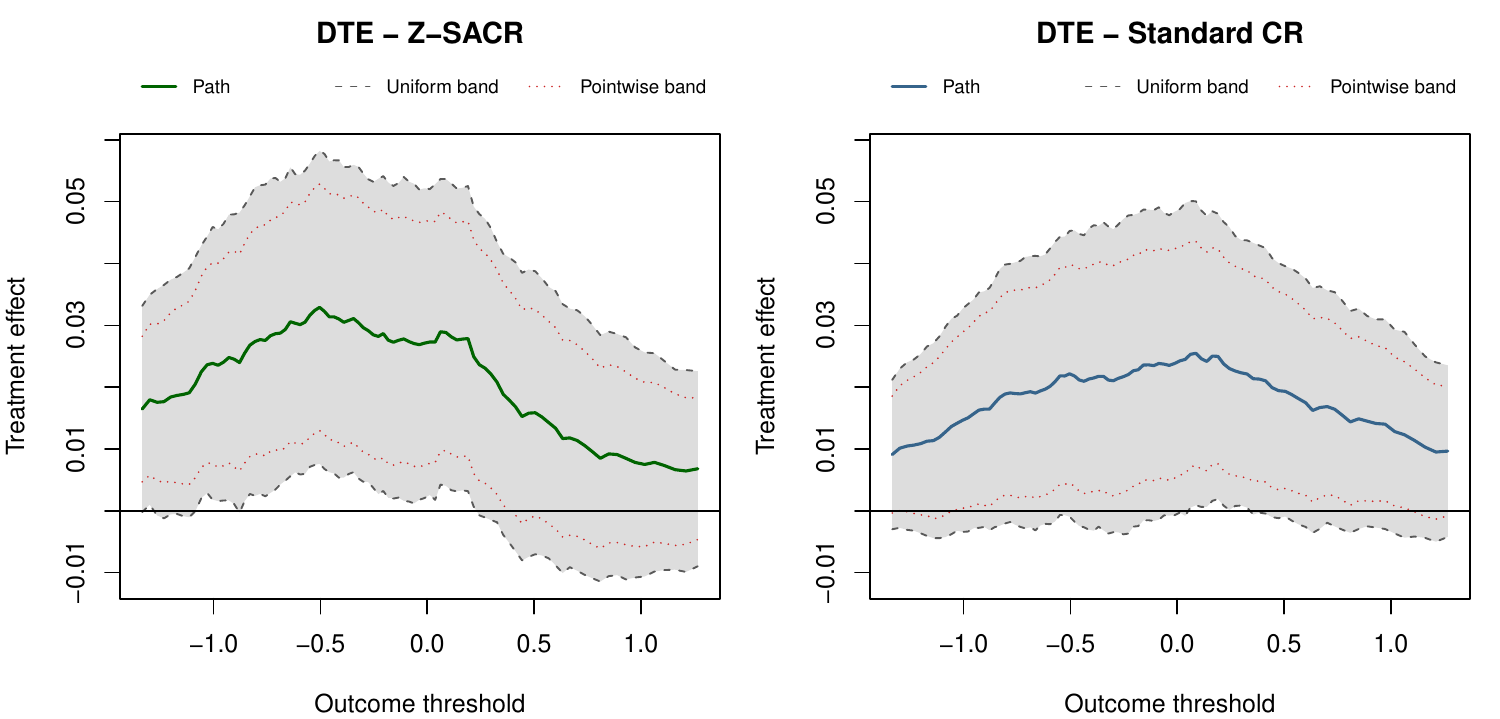}
  \caption{Local RD DTE path at the vote-margin cut-off: Z-SACR (left) and Standard CR (right)}
  \label{fig:emp_unif_dte}
\end{figure}

\begin{figure}[H]
  \centering
  \includegraphics[width=0.92\textwidth]{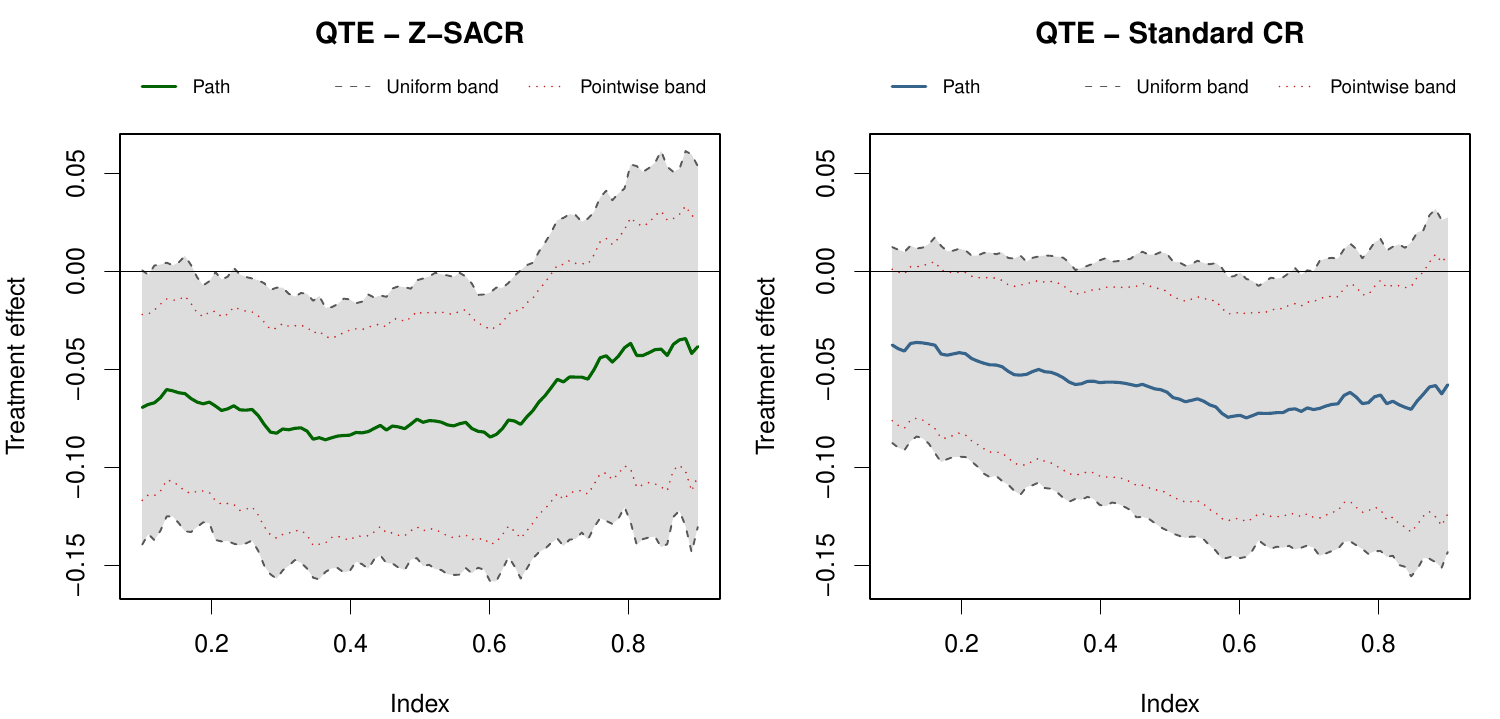}
  \caption{Local RD QTE path at the vote-margin cut-off: Z-SACR (left) and Standard CR (right)}
  \label{fig:emp_unif_qte}
\end{figure}

\begin{figure}[H]
  \centering
  \includegraphics[width=0.92\textwidth]{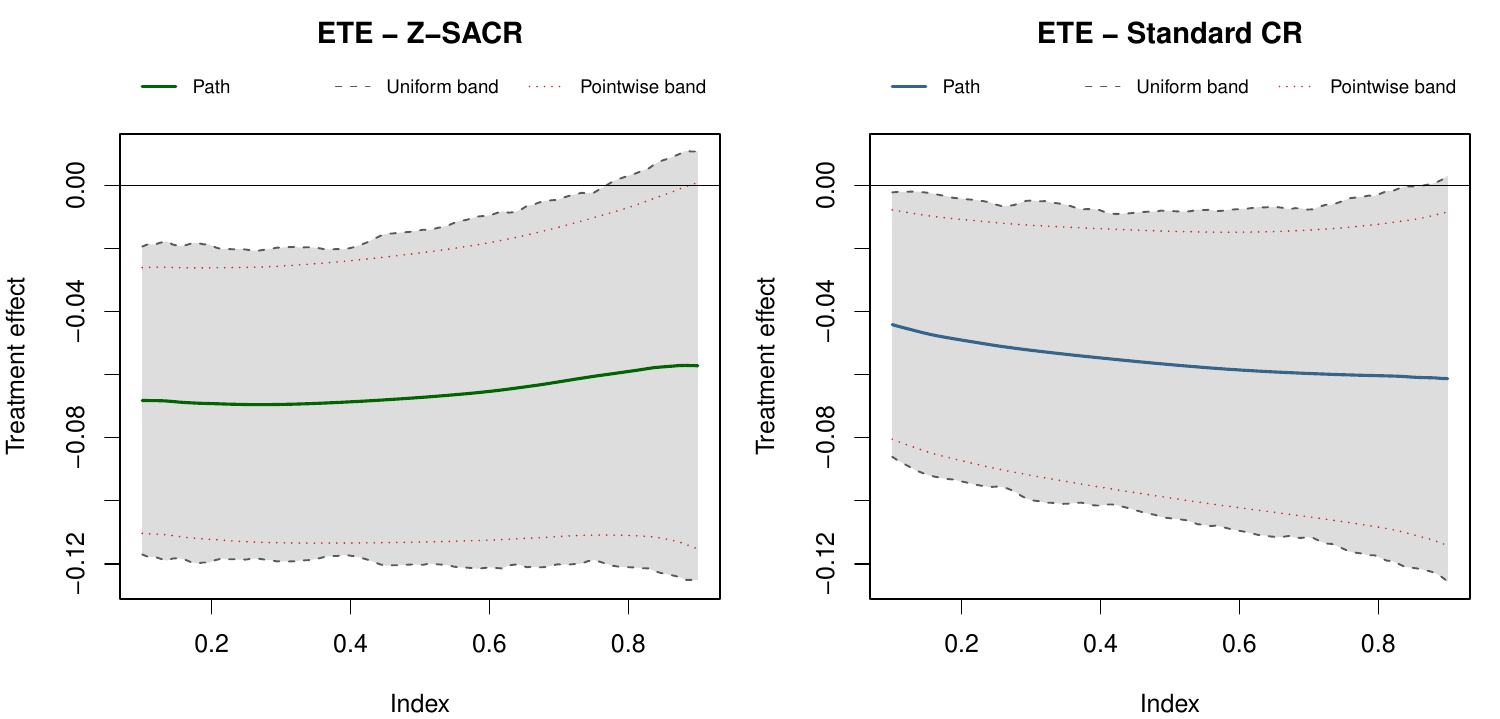}
  \caption{Local RD ETE path at the vote-margin cut-off: Z-SACR (left) and Standard CR (right)}
  \label{fig:emp_unif_ete}
\end{figure}

\begin{table}[H]
  \centering
  \caption{Local RD treatment-effect summaries at the vote-margin cut-off}
  \label{tab:emp_unif_summary}
  \footnotesize
  \begin{threeparttable}
  \begin{tabular}{llcccccc}
    \toprule
    Functional & Method & \(L\) & Avg. effect & Min. effect & Max. effect & \shortstack{Zero path\\excluded} & Homog. \(p\) \\
    \midrule
DTE & Z-SACR & 92 & 0.0224 & 0.0064 & 0.0329 & Yes & 0.0942 \\
DTE & Standard & 92 & 0.0186 & 0.0091 & 0.0255 & Yes & 0.1283 \\
QTE & Z-SACR & 92 & $-0.0673$ & $-0.0859$ & $-0.0342$ & Yes & 0.4770 \\
QTE & Standard & 92 & $-0.0588$ & $-0.0747$ & $-0.0362$ & Yes & 0.2184 \\
ETE & Z-SACR & 92 & $-0.0654$ & $-0.0695$ & $-0.0571$ & Yes & 0.5391 \\
ETE & Standard & 92 & $-0.0555$ & $-0.0613$ & $-0.0441$ & Yes & 0.3307 \\
    \bottomrule
  \end{tabular}
  \begin{tablenotes}[flushleft]
    \footnotesize
    \item \textit{Notes:} All effects are obtained from the two separately fitted cut-off CDFs; inference uses \(499\) municipality-level bootstrap draws. Zero path excluded means that the zero function is outside the simultaneous band. The homogeneity column gives the bootstrap \(p\)-value for a constant treatment-effect path. Positive DTE values mean a higher probability of scoring at or below the stated threshold under political turnover.
  \end{tablenotes}
  \end{threeparttable}
\end{table}

\Cref{fig:emp_unif_dte,fig:emp_unif_qte,fig:emp_unif_ete,tab:emp_unif_summary} report the distributional results. Both DTE paths are positive throughout the central grid, but neither simultaneous band is positive throughout, so the bands do not establish first-order stochastic dominance. Both QTE and ETE paths are negative. None of the homogeneity tests rejects a constant path at \(5\%\), although the Z-SACR DTE test has \(p=0.0942\).

Integrating the fitted cut-off DTE gives a DR-implied mean effect of \(-0.0672\) for Z-SACR, with a pointwise 95\% bootstrap confidence interval of \([-0.1131,-0.0214]\). The corresponding conventional effect is \(-0.0568\), with an interval of \([-0.0991,-0.0145]\). By construction, these integrals equal the fitted expectile differences at \(\tau=0.5\) to numerical precision, providing an internal check on the common cut-off-distribution construction.

\subsection{Distributional grid construction}\label{sec:emp_grid}

The grid construction fits a normal distribution and two- and three-component Gaussian mixtures to the outcome data and computes their implied curvature envelopes, i.e., the absolute diagonal element of the Hessian matrix of the corresponding function, on a dense grid. For each target, it uses the largest resulting grid size. The grid rule described in \Cref{sec:mc_grid_metrics} gives target-specific sizes of \(29\), \(92\), and \(87\) for DTE, QTE, and ETE; their maximum, \(92\), is used for all three reported paths. The reported evaluation ranges cover the central \(0.1\)--\(0.9\) region. A separate internal DR grid of \(92\) points spans the empirical \(0.1\)st--\(99.9\)th percentiles, on which each cut-off CDF after rearrangement is linearly interpolated for quantile inversion and expectile integration. Remaining lower- and upper-tail masses are assigned to the corresponding endpoints as a maintained tail specification.

\section{Cluster-Size Informativeness and Asymptotic Influence}
\label{app:cluster_size_cases}

Cluster size has two distinct dimensions. Conditional noninformativeness
means that the conditional outcome distribution, or the particular attribute
of interest, does not vary with cluster size (given covariates)
\citep{hoffman-sen-weinberg-2001,nevalainen-datta-oja-2014}.
Size negligibility concerns asymptotic influence: under unrestricted
within-cluster dependence, \(\max_gN_g^2/N\to0\) is a benchmark for
conventional Gaussian inference. Its failure defines the ULC regime
considered in this paper, but does not imply failure of Gaussian inference under
every dependence structure. The following classification separates these
questions.

\begin{center}
\small
\renewcommand{\arraystretch}{1.15}
\begin{tabular}{@{}p{0.20\textwidth}p{0.13\textwidth}p{0.59\textwidth}@{}}
\toprule
Size informativeness & Size regime & Implication \\
\midrule
Noninformative & Negligible & Conventional inference may be valid; marginal targets can still differ through covariate representation. \\
\hline
Informative & Negligible & Population representation matters even when conventional inference is available. \\
\hline
Noninformative & ULC & A common conditional outcome law does not prevent failure of conventional inference under strong dependence. \\
\hline
Informative & ULC & Both target representation and conventional inferential validity require attention. \\
\bottomrule
\end{tabular}
\end{center}

Conditional noninformativeness does not guarantee invariance of marginal targets, since cluster size may remain associated with covariates. Conversely, distributional informativeness does not rule out a common zero for a particular score. As \Cref{prop:inv_id} establishes, a common cluster-level score zero preserves the structural zero under deterministic reweighting; identification additionally requires separation, and inference requires the remaining assumptions. SACR limits size-driven influence by using cluster-average scores, with its inferential conclusions supplied by \Cref{sect:theory_generic}. Without a common zero, equal cluster representation can define a different meaningful target, but the present inference results cannot automatically be invoked for it. An aggregate zero alone does not ensure consistent estimation of the sampling covariance, although the coincident analytical and multiplier-bootstrap score covariance estimators can support conservative pointwise inference for the weighting-specific target; see \Cref{rem:bootstrap_centering}. There is no general efficiency ranking between observation-weighted and cluster-balanced estimators.

\printbibliography[title={References}]
\clearpage 
\end{refsection}

\end{document}